\documentclass[useAMS,usenatbib]{mn2e}
\usepackage{deluxetable}
\usepackage{graphics,epsfig}
\usepackage{graphicx}
\usepackage{amssymb}
\usepackage{txfonts}
\usepackage{times}
\usepackage{ulem}
\usepackage{array}

\newcommand{\ctwostar}{C{\sc ii}*$\lambda$1335.7}
\newcommand{\mgtwo}{Mg{\sc ii}~$\lambda$2796}
\newcommand{\mgone}{Mg{\sc i}~$\lambda$2852}
\newcommand{\sitwo}{Si{\sc ii}~$\lambda$1526}
\newcommand{\kms}{km\,s$^{-1}$} 
\newcommand{\ts}{\ensuremath{{\rm T_s}}}
\newcommand{\tr}{\ensuremath{{\rm T_{s, nCNM}}}}
\newcommand{\tk}{\ensuremath{{\rm T_k}}}

\newcommand{\nhi}{\ensuremath{\rm N_{HI}}}

\newcommand{\lsb}{\ensuremath{\left[}}
\newcommand{\rsb}{\ensuremath{\right]}}

\newcommand{\cm}{cm$^{-2}$}
\newcommand{\hi}{H{\sc i}}
\newcommand{\hii}{H{\sc i} 21cm}

\title[The spin temperature of high-$z$ DLAs]{The spin temperature of high-redshift damped Lyman-$\alpha$ systems}
\author[N. Kanekar et al.]{N. Kanekar$^{1}$\thanks{E-mail: nkanekar@ncra.tifr.res.in~(NK); Ramanujan Fellow.}, 
J.~X.~Prochaska$^2$, A. Smette$^3$, S.~L.~Ellison$^4$, E.~V.~Ryan-Weber$^5$, E.~Momjian$^6$,
\newauthor
F.~H.~Briggs$^7$, W.~M.~Lane$^8$, J.~N.~Chengalur$^1$, T.~Delafosse$^2$, J.~Grave$^2$, D.~Jacobsen$^{4}$, 
\newauthor and A.~G.~de~Bruyn$^{9,10}$\\
$^{1}$National Centre for Radio Astrophysics, TIFR, Post Bag 3, Ganeshkhind, Pune 411 007, India;
$^2${}UCO/Lick Observatory, UC Santa Cruz, Santa Cruz, \\ CA 95064, USA;
$^3${}European Southern Observatory, Alonso de Cordova 3107, Casilla 19001, Vitacura, Santiago, Chile; 
$^4${}Department of Physics and \\ Astronomy, University of Victoria, B.C., V8P 1A1, Canada; 
$^5${}Swinburne University of Technology, Mail H30, P.O. Box 218, Hawthorn, 3122 VIC, Australia; \\
$^6${}National Radio Astronomy Observatory, 1003 Lopezville Road, Socorro, NM 87801, USA;
$^7${}RSAA, The Australian National University, Mount Stromlo \\ Observatory, ACT 2611, Australia;
$^8${}Naval Research Laboratory, Code 7213, 4555 Overlook Ave SW, Washington, DC 20375, USA; 
$^9${}Kapteyn Astronomical \\ Institute, University of Groningen, PO Box 800, 9700 AV Groningen, The Netherlands; 
$^{10}${}ASTRON, PO Box 2, 7990 AA Dwingeloo, The Netherlands}

\begin{document}
\date{Accepted yyyy month dd. Received yyyy month dd; in original form yyyy month dd}


\maketitle

\label{firstpage}

\begin{abstract}
We report results from a program aimed at investigating the temperature of neutral gas in 
high-redshift damped Lyman-$\alpha$ absorbers (DLAs). This involved (1)~H{\sc i}\,21cm 
absorption studies of a large sample of DLAs towards radio-loud quasars, (2)~very long baseline 
interferometric studies to measure the low-frequency quasar core fractions, and (3)~optical/ultraviolet 
spectroscopy to determine DLA metallicities and the velocity widths of low-ionization metal lines.

Including literature data, our sample consists of 37 DLAs with estimates of the harmonic-mean 
spin temperature $\ts$. We find a statistically significant ($4\sigma$) difference between the $\ts$ distributions 
in the high-$z$ ($z > 2.4$) and low-$z$ ($z < 2.4$) DLA samples. The high-$z$ sample contains more 
systems with high spin temperature, $\ts \gtrsim 1000$~K. The $\ts$ distributions in DLAs and the Galaxy
are also significantly ($\approx 6\sigma$) different, with more high-$\ts$ sightlines in 
DLAs than in the Milky Way. The high $\ts$ values in the high-$z$ DLAs of our sample arise due to 
low fractions of the cold neutral medium (CNM). Only two of 23 DLAs at $z > 1.7$ have $\ts$ 
values indicating CNM fractions $> 20$\%, comparable to the median value ($\approx 27$\%) in the Galaxy.

We tested whether the H{\sc i} column density measured towards the optical quasar might
be systematically different from that towards the radio core by comparing the H{\sc i}
column densities inferred from H{\sc i}\,21cm emission studies at different spatial resolutions 
($\approx 15$~pc~$- 1$~kpc) in the Large Magellanic Cloud. The high-resolution $N_{\rm HI}$ values 
are, on average, larger than the smoothed ones for $N_{\rm HI} > 10^{21}$~cm$^{-2}$, but lower 
than the smoothed $N_{\rm HI}$ estimates for $N_{\rm HI} < 10^{21}$~cm$^{-2}$. Since there are 
far more DLAs with low $N_{\rm HI}$ values than high ones, the use of the optical $N_{\rm HI}$ 
value for the radio sightline results in a statistical tendency to under-estimate DLA spin temperatures. 

For 29 DLAs with metallicity estimates, we confirm the presence of an anti-correlation between 
$\ts$ and metallicity [Z/H], at $3.5\sigma$ significance via a non-parametric Kendall-tau test. 
This result was obtained with the assumption that the DLA covering factor is equal to the 
core fraction. However, Monte Carlo simulations show that the significance of the result is only 
marginally decreased if the covering factor and the core fraction are uncorrelated, or if there
is a random error in the inferred covering factor.

We also find statistically significant evidence for redshift evolution in DLA spin temperatures 
even for the DLA sub-sample at $z > 1$. Since all DLAs at $z > 1$ have angular diameter distances
comparable to or larger than those of their background quasars, they have similar efficiency 
in covering the quasars. We conclude that low covering factors in high-$z$ DLAs cannot account for 
the observed redshift evolution in spin temperatures.

\end{abstract}


\section{Introduction}
\label{sec:intro}

Quasar absorption spectra offer the possibility of selecting galaxies by their 
absorption signatures, and thus obtaining samples of high-$z$ galaxies without 
a bias towards the most luminous systems. The highest \hi\ column density absorbers 
detected in quasar spectra are the damped Lyman-$\alpha$ systems (DLAs). With 
\hi\ column densities $\nhi \ge 2 \times 10^{20}$~\cm\ \citep{wolfe05}, similar 
to values seen in sightlines through the Milky Way and nearby gas-rich galaxies, 
DLAs have long been identified as the high-redshift counterparts of normal galaxies 
in the local Universe. The nature of galaxies that give rise to DLAs at different 
redshifts, and their typical size, mass, kinematic structure and physical conditions,
are all important ingredients for understanding galaxy evolution.

The Sloan Digital Sky Survey (SDSS; \citealp{sdss7}) has resulted in the detection of a 
vast number of DLAs at high redshifts, with nearly seven thousand candidate absorbers now known at 
$z > 2.2$ \citep[e.g.][]{prochaska05,prochaska09,noterdaeme09,noterdaeme12b}. Unfortunately, 
contamination from the background quasars has made it very difficult to 
identify the host galaxy of the DLAs in optical images (e.g. \citealp{warren01}; but see 
\citealt{fumagalli10}). The low sensitivity of today's radio telescopes has meant that 
one cannot image the DLA hosts in the standard radio \hii\ and CO emission lines that have been used 
for detailed studies of the kinematics and dynamics of nearby galaxies. And, even following 
a number of recent studies, only a handful of high-$z$ DLAs have been detected in H$\alpha$ 
or Ly$\alpha$ emission \citep[e.g.][]{moller04,fynbo10,fynbo11,peroux12,noterdaeme12}, with 
typical star formation rates (SFRs) $\lesssim few \: M_\odot$ per year \citep{peroux12}, 
even for high-metallicity absorbers. Thus, despite much effort over the last three decades, 
relatively little information has so far been gleaned from emission studies of DLAs.  

Detailed absorption studies remain our primary source of information about the absorbers. 
Around two hundred DLAs have measured metallicities, elemental abundances and kinematics, 
from high-resolution optical echelle spectroscopy \citep[e.g.][]{pettini94,pettini97,pettini99,pettini08,prochaska03b,prochaska07,dessauges04,khare04,ledoux06,petitjean08,penprase10,cooke11,battisti12}. 
These studies have yielded interesting results. For example, mean DLA metallicities have been 
shown to increase with decreasing redshift, as expected from models of galaxy evolution, 
although low-metallicity DLAs are quite common even at low redshifts 
\citep{prochaska03a,kulkarni05,kulkarni10,rafelski12}. A positive correlation has been found 
between the metallicity and both the velocity spreads of low-ionization metal lines \citep{wolfe98,ledoux06},
and the rest equivalent width of the Si{\sc ii}$\lambda$1526 line \citep{prochaska08}. This 
has been interpreted as evidence for a mass-metallicity relation in DLAs, similar to that 
seen in emission-selected high-$z$ galaxies \cite[e.g.][]{tremonti04,savaglio05,erb06,neeleman13}.
Molecular hydrogen (H$_2$) absorption, along with C{\sc i} absorption, has been detected 
in about a dozen DLAs, with strong upper limits on the molecular fraction in 
$\sim 80$\% of the observed systems \citep[e.g.][]{levshakov85,ge97,ge01,ledoux03,noterdaeme08,milutinovic10,jorgenson10}. 
This has provided information on local conditions in the molecular phase, including estimates 
of the number density, temperature and strength of the ultraviolet (UV) radiation field 
\citep[e.g.][]{srianand05}, along with measurements of the microwave background temperature 
at different redshifts \citep[e.g.][]{noterdaeme11}.

While absorption spectra have yielded many successes in DLA studies, the fact that 
they only trace a pencil beam through each galaxy implies that many of the major
questions in the field remain to be answered. For example, the asymmetric wings seen 
in low-ionization metal line profiles were originally thought to imply an origin in 
rapidly-rotating thick disks \citep{prochaska97}, but have also been shown to 
arise naturally in hierarchical merging scenarios (\citealp{haehnelt98}; but see 
\citealp{prochaska10}), dwarf galaxy ejecta \citep{nulsen98}, and outflows 
from starburst galaxies \citep{schaye01}. The star formation histories of $z > 2$ DLAs, 
deduced from their elemental abundances, indicate an unambiguous origin in dwarf 
irregular galaxies for a few DLAs \citep{dessauges06}, but also suggest that 
all the absorbers do not arise in a single galaxy class \citep{dessauges07}.  
On the other hand, the large velocity spreads seen in a number of DLAs and the curious 
paucity of systems with narrow velocity spreads ($\Delta V \lesssim 30$~\kms) are 
difficult to reconcile with an origin in dwarf galaxies
\citep[e.g.][]{prochaska97,wolfe05}. The complications of interpreting absorption spectra 
have meant that even basic questions pertaining to the typical size and structure 
of DLAs, and their redshift evolution, remain issues of controversy.

\hii\ emission studies of high redshift DLAs are currently not possible owing to the 
weakness of the \hii\ line and the low sensitivity of today's radio telescopes; 
the highest redshift at which \hii\ emission has been detected is $z \approx 0.26$ 
\citep{catinella08}. However, \hii\ absorption studies of DLAs lying towards radio-loud 
quasars provide an important tool both to study physical conditions in the neutral gas in the 
absorbers and to probe fundamental constant evolution.
For DLAs lying towards extended background radio sources, \hii\ absorption studies provide 
a measure of the spatial extent of the neutral gas as well as the opacity-weighted velocity 
field, although such studies have only so far been possible in a couple of DLAs \citep[e.g.][]{briggs89,kanekar04}. 
Conversely, for DLAs towards compact radio sources, a comparison between the \hi\ column density 
measured from the Lyman-$\alpha$ absorption profile and the integrated \hii\ optical depth yields the 
spin temperature $\ts$ of the absorbing gas. In addition, for compact targets, comparisons between 
the redshifts of \hii\ absorption lines with those of other transitions \citep[e.g. resonant UV lines, 
OH Lambda-doubled lines, rotational lines;][]{wolfe76,drinkwater98,chengalur03} 
can be used to test for changes in the fundamental constants of physics.

The spin temperature contains information on the distribution of neutral gas between the different 
temperature phases of the interstellar medium (ISM) in the absorber, specifically on the fractions of 
warm and cold gas. The fact that the spin temperature appeared systematically higher in high-$z$ DLAs 
than in the Galaxy has made $\ts$ a quantity of interest in DLA studies for the last three decades 
\cite[e.g.][]{wolfe79,wolfe81}. However, until the recent advent of new radio 
telescopes like the Giant Metrewave Radio Telescope (GMRT) and the Green Bank 
Telescope (GBT) with excellent low frequency coverage, only a few $\ts$ 
estimates were available in high-$z$ DLAs 
\citep[e.g.][]{wolfe79,wolfe85,debruyn96,carilli96,briggs97,kanekar97}. Over the 
last decade, we have used the GBT, the GMRT and the Westerbork Synthesis Radio 
Telescope (WSRT) to carry out sensitive searches for redshifted \hii\ absorption 
in a large sample of DLAs at high redshifts, $z \gtrsim 2$. We have also used 
the Very Long Baseline Array (VLBA) to measure the low-frequency core fractions of 
the background quasars for most of the absorbers of the sample. Finally, we have obtained 
the metallicity, dust depletion and kinematic structure in the low ionization metal lines 
for a significant fraction of the DLAs in our sample, either from our own observations 
or from the literature. In this paper, we report results from these studies and 
their implications for physical conditions in the neutral gas.\footnote{When 
required, we will use the standard $\Lambda$-CDM cosmology, with H$_0 = 67.4$~\kms~Mpc$^{-1}$,
$\Omega_m = 0.315$ and $\Omega_\Lambda = 0.685$ \citep{planck13}.}

\section{Observations, data analysis and spectra}
\label{sec:obs}

\subsection{\hii\ absorption line searches}
\label{sec:21cm}

Our searches for redshifted \hii\ absorption were carried out with the GMRT, the GBT 
and the WSRT, targetting 26 DLAs at $0.68 < z < 3.42$. In addition, two known 
\hii\ absorbers were re-observed, to confirm earlier detections.
The new targets were selected from the literature, mostly from the CORALS sample of 
\citet{ellison01}, the UCSD sample of \citet{jorgenson06} and the SDSS. Five targets 
were observed with the GMRT, 22 with the GBT, and one with the WSRT, all between 
2003 and 2009.

\subsubsection{GBT observations and data analysis}
\label{sec:gbt}

The GBT observations were carried out with the PF1-800, PF1-450 and PF1-342 receivers, 
in projects AGBT03A-015, AGBT05B-018, AGBT06B-042, AGBT08A-076 and AGBT09A-025. Most 
observations used the Spectral Processor (SP) as the backend, with two linear 
polarizations and bandwidths of 1.25, 2.5 or 5 MHz, sub-divided into 1024 channels. 
The SP was not available for the observations of two sources, 1122$-$168 and 0454+039; 
these were hence observed with the Auto-Correlation Spectrometer (ACS), using two 
circular polarizations, a bandwidth of 50~MHz sub-divided into 16384 channels, 
and nine-level sampling. All observations used position-switching for bandpass 
calibration, with On and Off times of 5m each, and data recorded every 5 or 
10 seconds. A switching noise diode was used to measure system temperatures during 
all observations. Most runs also included observations of standard flux density 
calibrators like 3C286 or 3C48 to check the flux scale.

All GBT data were analysed in the {\sc aips}++ single-dish package \textsc{DISH}, using 
standard procedures. All data were first visually inspected, both before and after 
calibration, and individual records affected by either radio frequency interference 
(RFI) or backend failures were edited out. For each target, the data were calibrated 
and the quasar flux density was measured in the centre of the band from the averaged 
spectrum. Some of the observations were affected by severe out-of-band RFI (although the 
receiver did not saturate), and the 
flux densities were found to vary dramatically from one record to another, typically 
yielding average flux densities very different from those in the literature and often 
also differing values in the two polarizations. In such cases, we chose to use flux 
densities from the literature, interpolating between measurements to obtain values at 
the respective redshifted \hii\ line frequencies.  In general, the measured system 
temperatures ($T_{\rm sys}$) were found to match expected values (except in cases where 
the data were rendered unusable by extremely strong RFI) and the root-mean-square (RMS) 
noise values on the spectra were close to the expected values, based on the $T_{\rm sys}$ 
estimates. After the flux density measurements, a second-order baseline was fitted to 
line- and RFI-free channels in each record during the process of calibration and the 
residual spectra were averaged together to obtain the \hii\ spectrum for each 
source. In some cases, a further first- or second-order baseline was fit to the 
spectrum, again excluding line- and RFI channels, to obtain the final \hii\ 
spectrum for each DLA. Finally, data on eight targets were rendered unusable by 
strong RFI at or near the redshifted \hii\ line frequencies.

\begin{figure*}
\centering
\epsfig{file=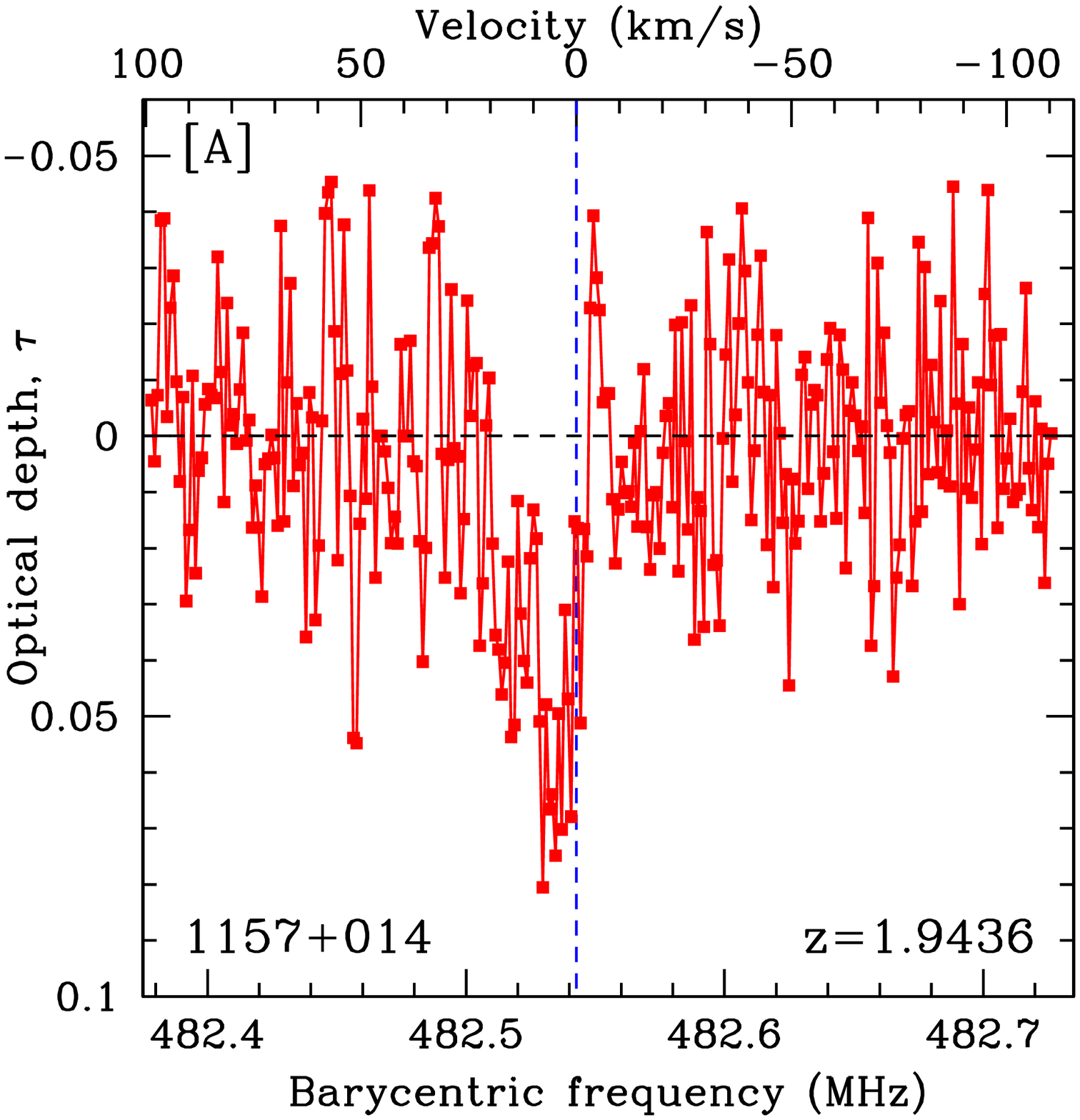,height=3.3truein,width=3.3truein}
\epsfig{file=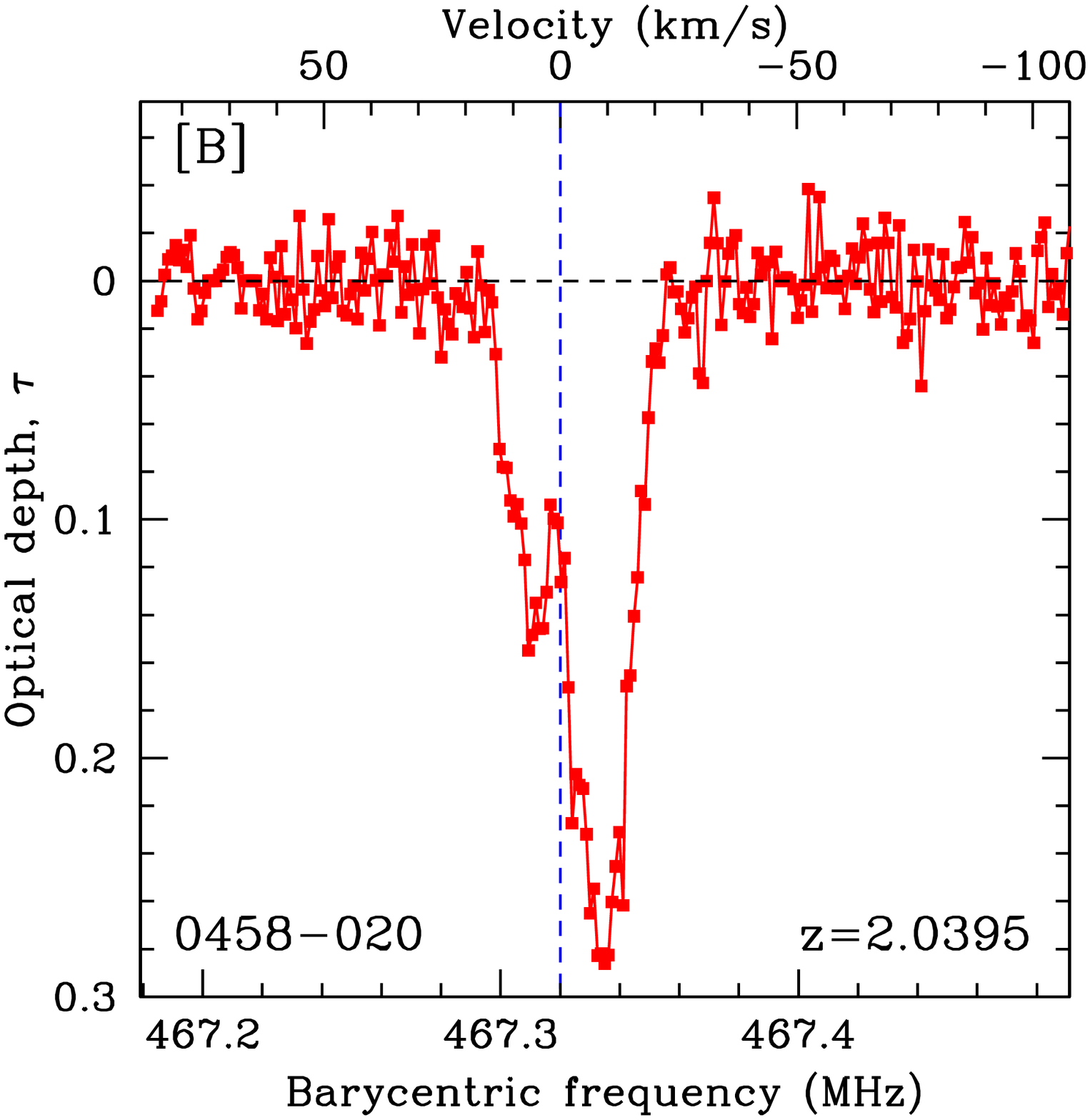,height=3.3truein,width=3.3truein}
\caption{Detections of \hii\ absorption in two known \hii\ absorbers, [A]~at $z = 1.9436$ 
towards 1157+014 and [B]~$z = 2.0395$ towards 0458$-$020, with \hii\ optical depth 
plotted against barycentric frequency, in MHz. The top axis in each panel shows 
velocity (in \kms), relative to the DLA redshift. See main text for discussion.}
\label{fig:det2}
\end{figure*}

\begin{figure*}
\centering
\epsfig{file=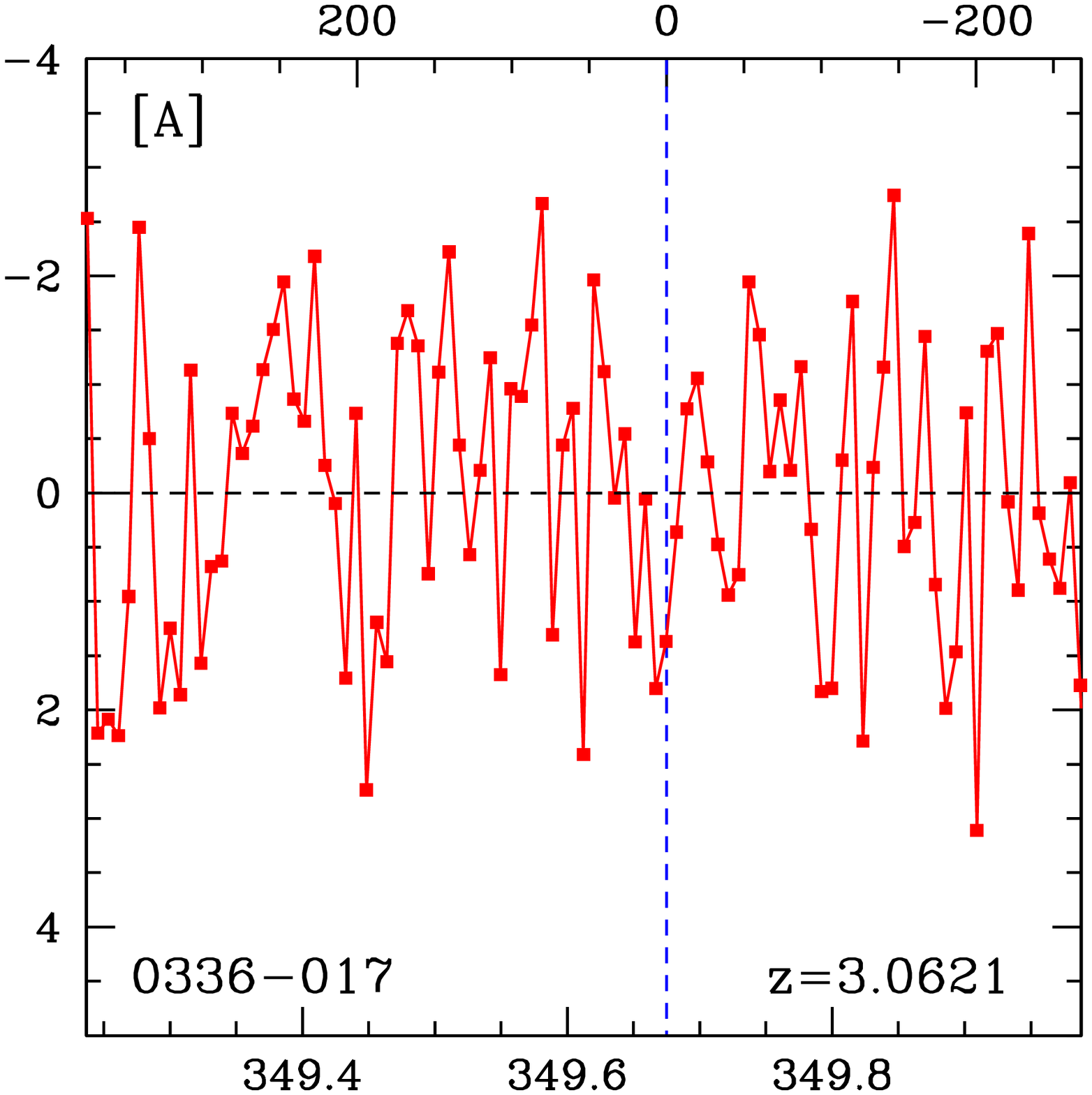,height=2.3truein,width=2.3truein}
\epsfig{file=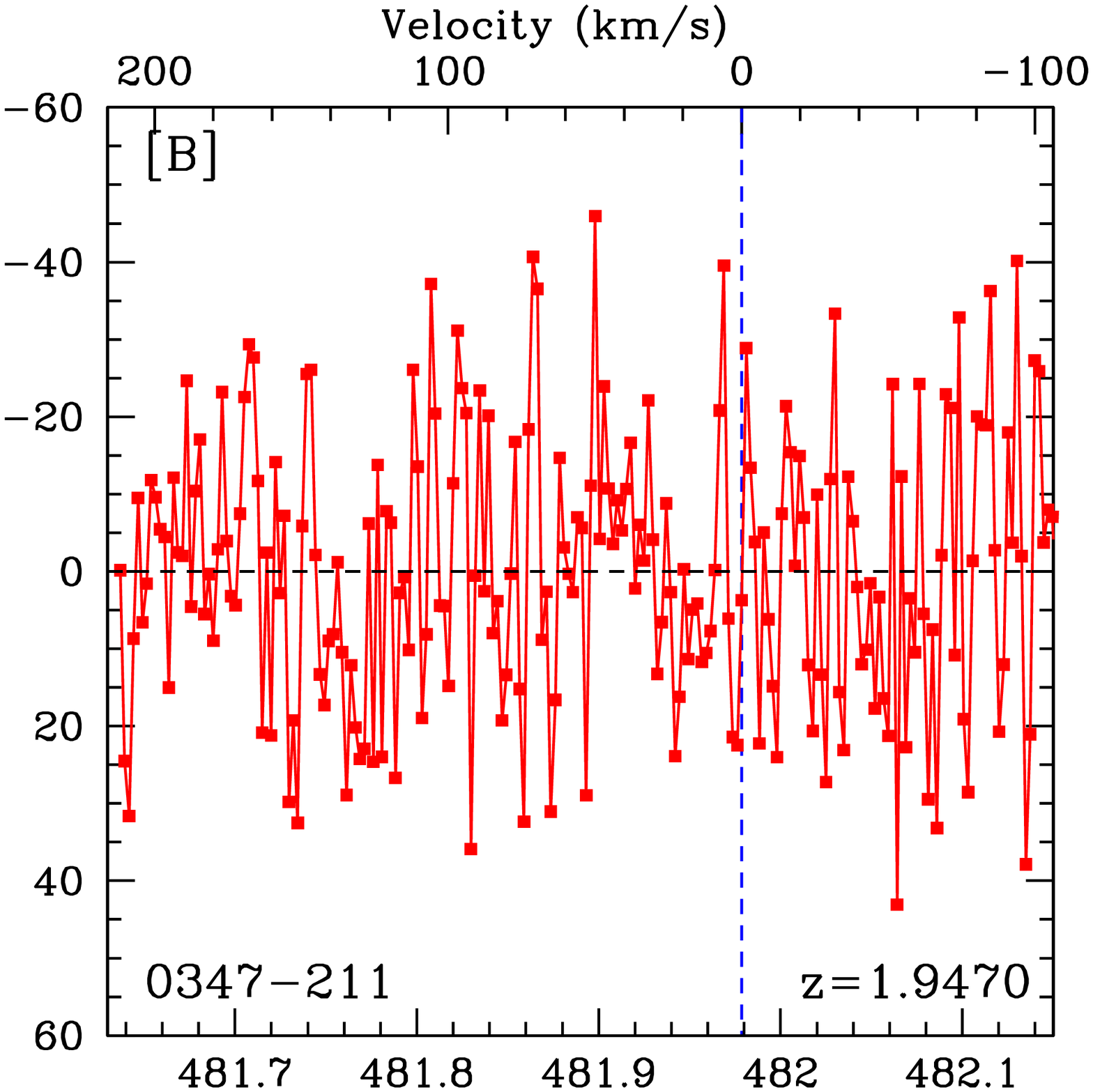,height=2.3truein,width=2.3truein}
\epsfig{file=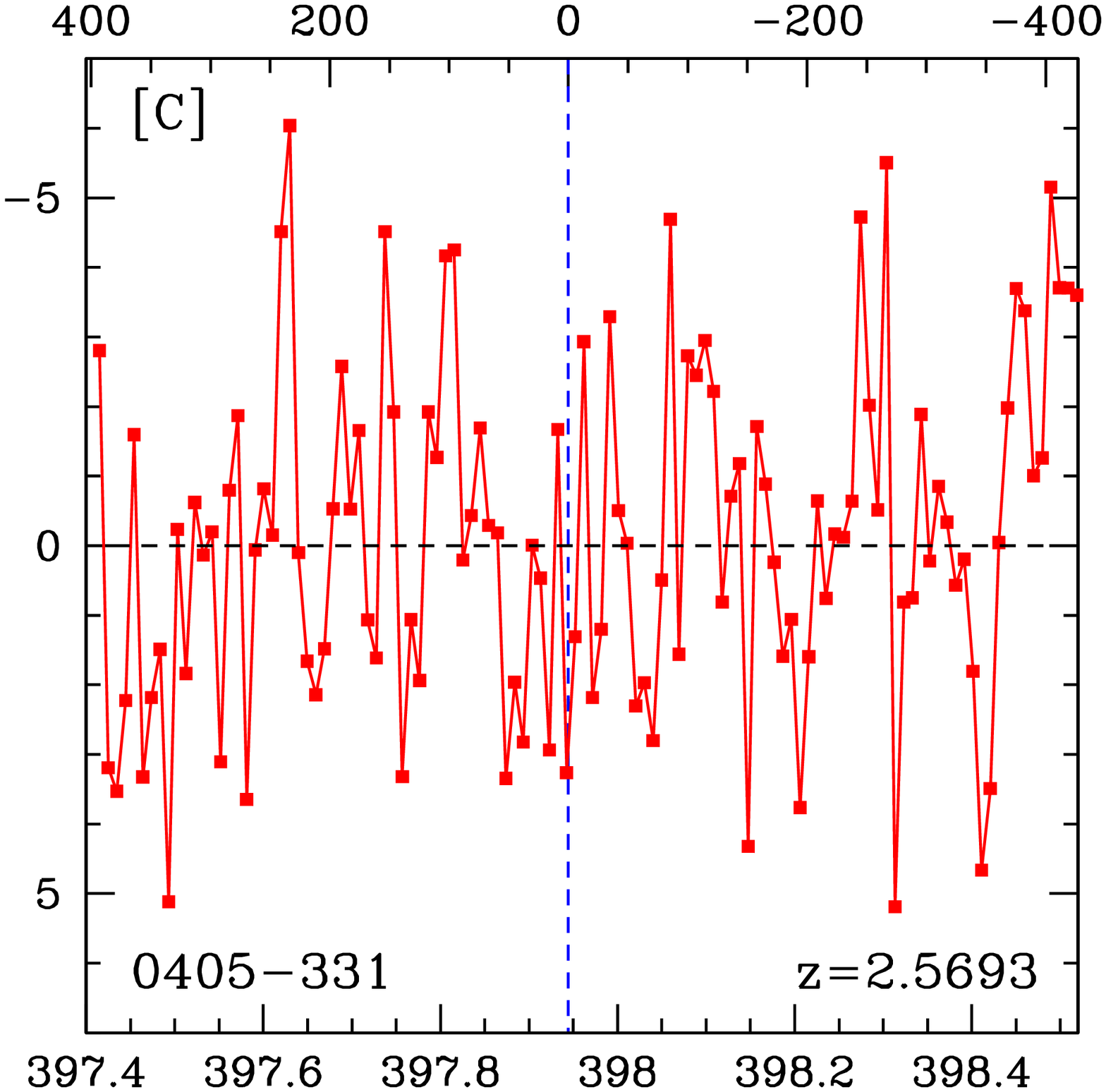,height=2.3truein,width=2.3truein}
\epsfig{file=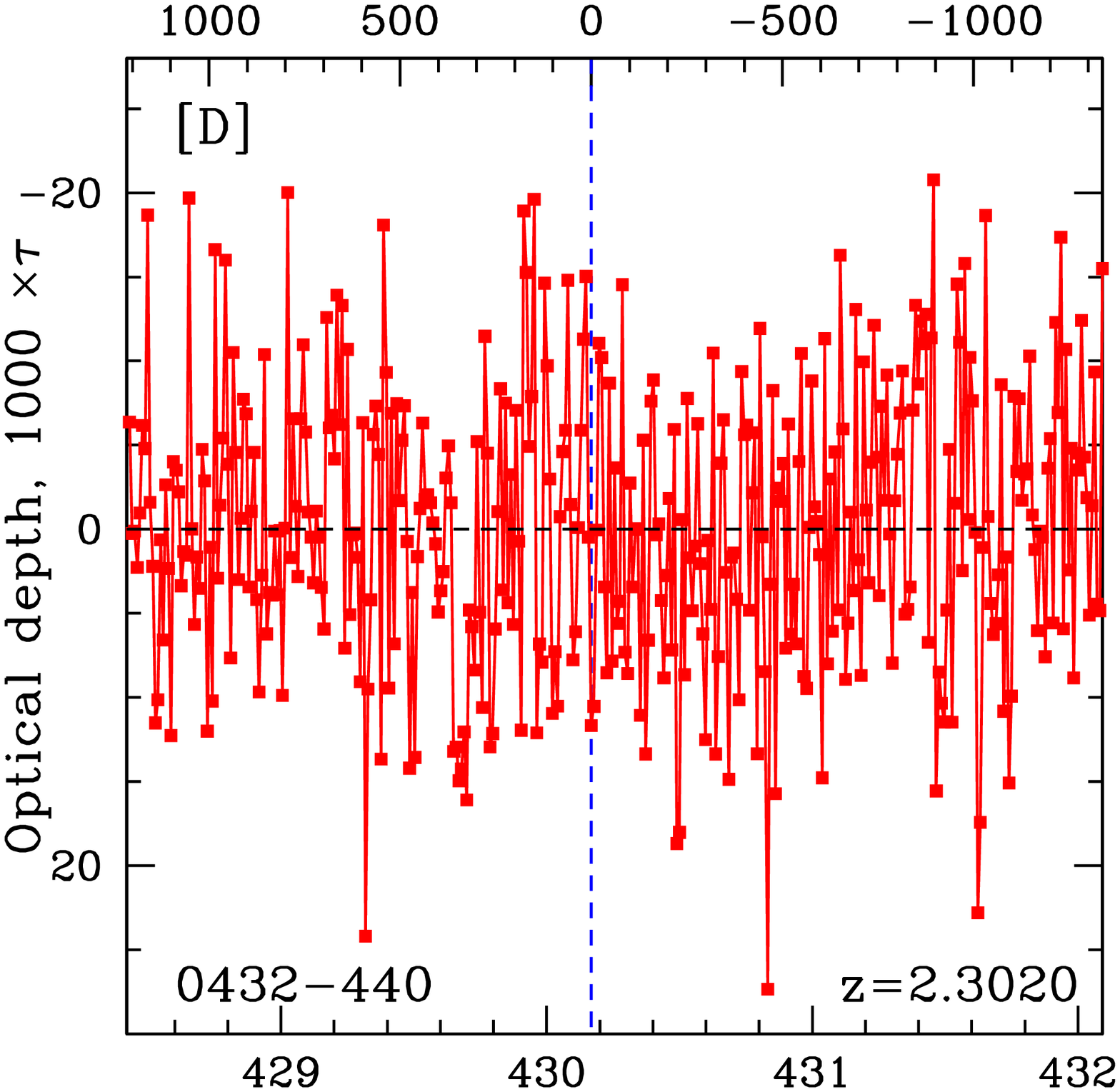,height=2.3truein,width=2.3truein}
\epsfig{file=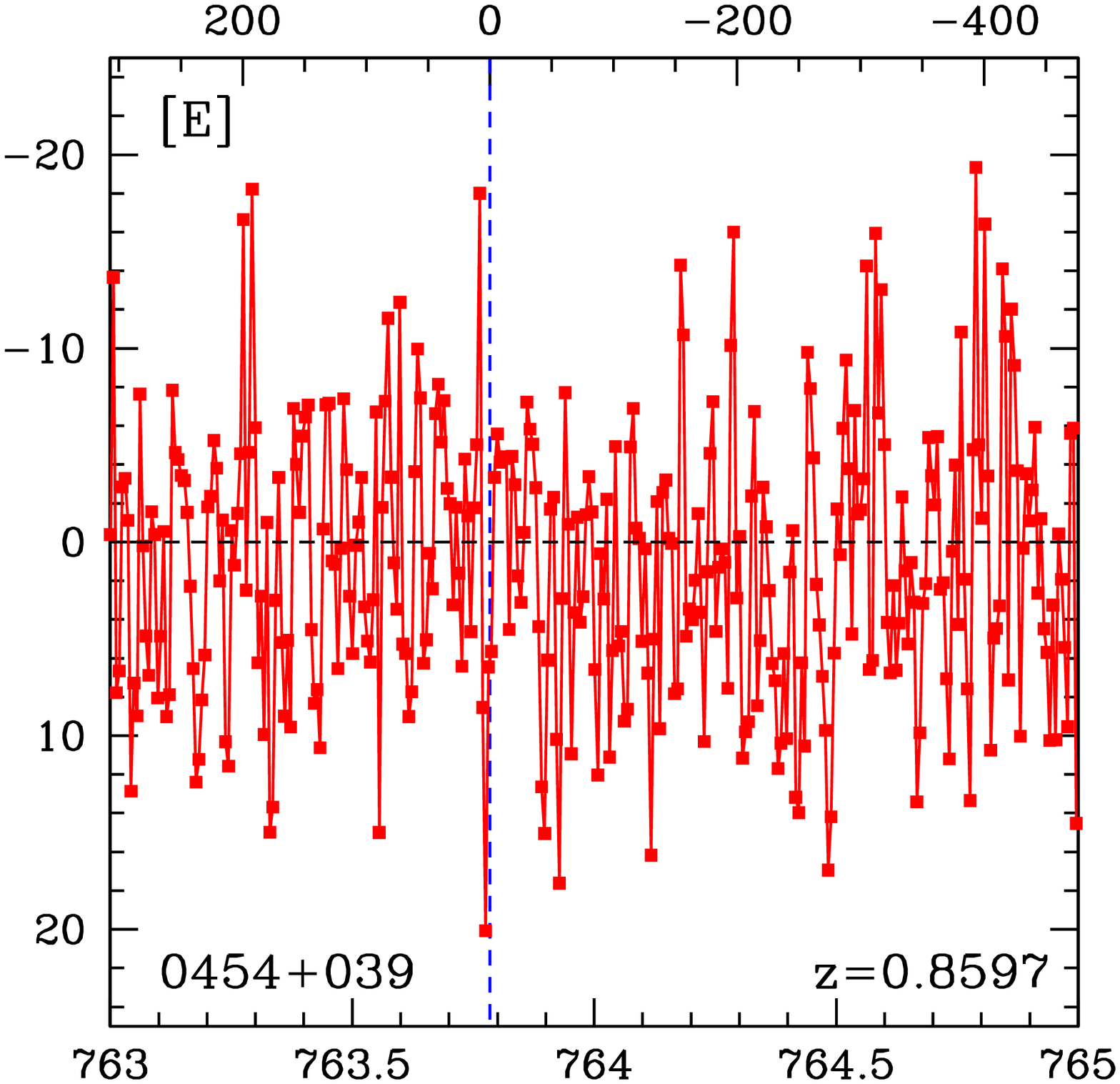,height=2.3truein,width=2.3truein}
\epsfig{file=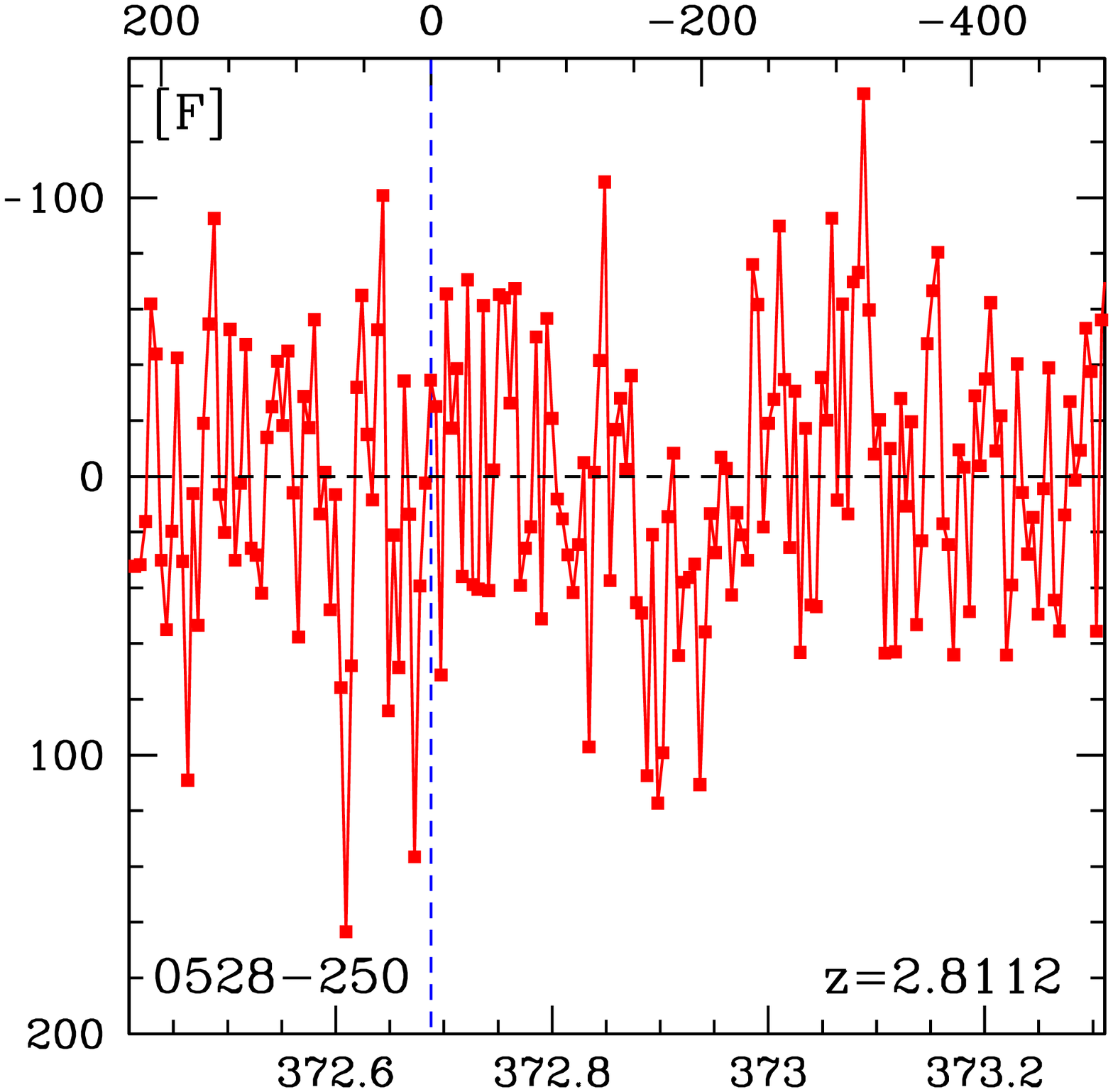,height=2.3truein,width=2.3truein}
\epsfig{file=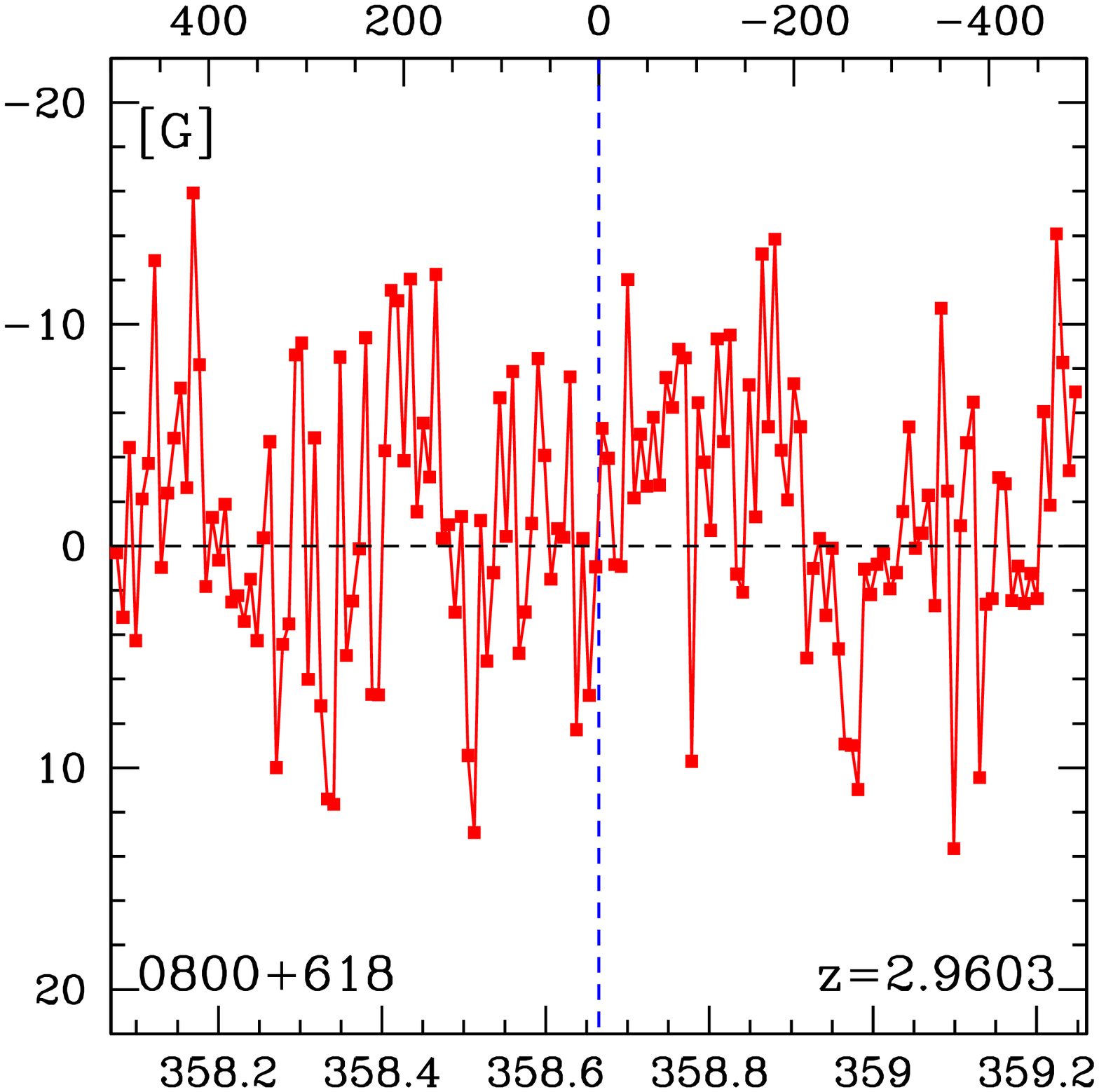,height=2.3truein,width=2.3truein}
\epsfig{file=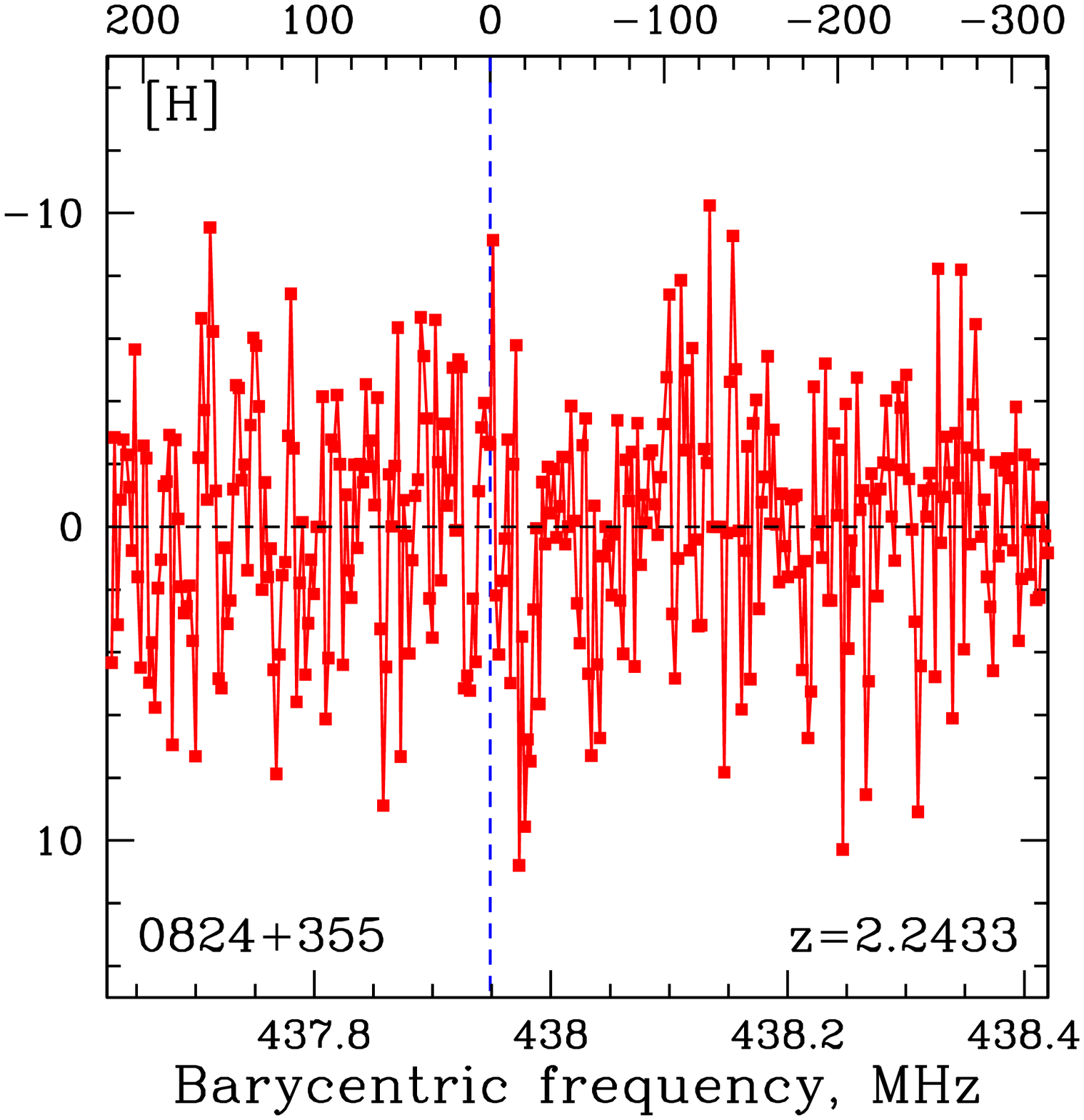,height=2.3truein,width=2.3truein}
\epsfig{file=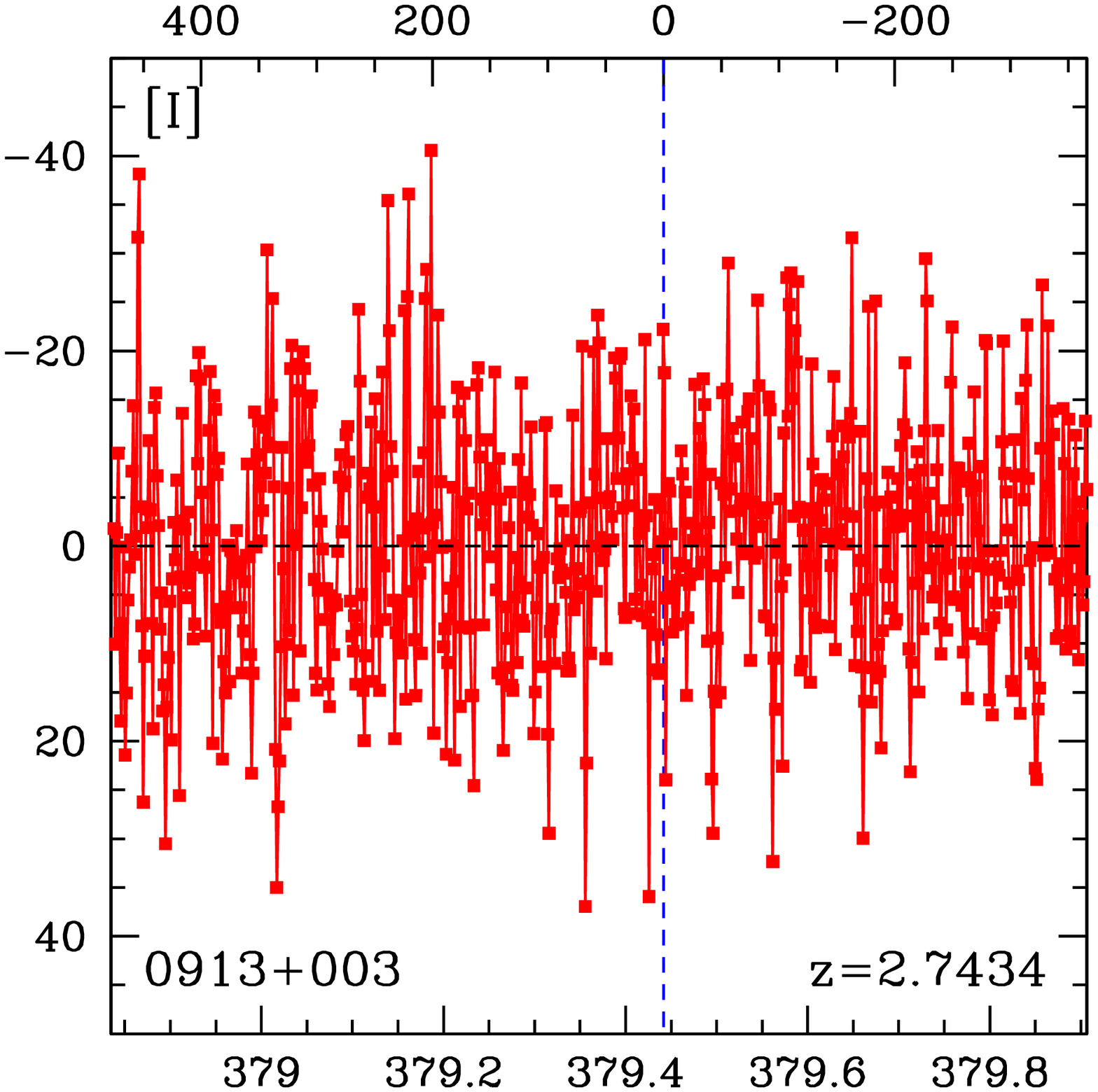,height=2.3truein,width=2.3truein}
\caption{The sixteen non-detections of \hii\ absorption, in order of increasing right ascension, 
with optical depth ($10^3 \times \tau$) plotted against barycentric frequency, in MHz. The top axis 
in each panel shows velocity (in \kms), relative to the DLA redshift, while the expected \hii\ line 
frequency, from the optically-determined redshift, is indicated by the dashed line in each panel. 
The panels contain spectra for
[A]~the $z=3.0621$ DLA towards 0336$-$017, 
[B]~the $z=1.9470$ DLA towards 0347$-$211, 
[C]~the $z = 2.5693$ DLA towards 0405$-$311, 
[D]~the $z= 2.3020$ DLA towards 0432$-$440,
[E]~the $z=0.8597$ DLA towards 0454+039, 
[F]~the $z = 2.8112$ DLA towards 0528$-$250,
[G]~the $z=2.9603$ DLA towards 0800+618,
[H]~the $z=2.2433$ DLA towards 0824+355,
[I]~the $z=2.7434$ DLA towards 0913$+$003,
[J]~the $z=2.7670$ DLA towards 1013+615,
[K]~the $z=0.6819$ DLA towards 1122$-$168, and 
[L]~the $z=2.7799$ DLA towards 1354$-$170.
[M]~The $z=2.7076$ DLA towards 1402+044,
[N]~the $z=3.4483$ DLA towards 1418$-$064, 
[O]~the $z=2.5200$ DLA towards 1614+051, and
[P]~the $z=2.9082$ DLA towards 2342+342.
The dashed vertical line marks the expected redshifted \hii\ frequency. The shaded 
vertical region in panel~[L] (for 1354$-$170) indicates a frequency range affected by RFI.}
\label{fig:nondetect}
\end{figure*}

\begin{figure*}
\setcounter{figure}{1}
\centering
\epsfig{file=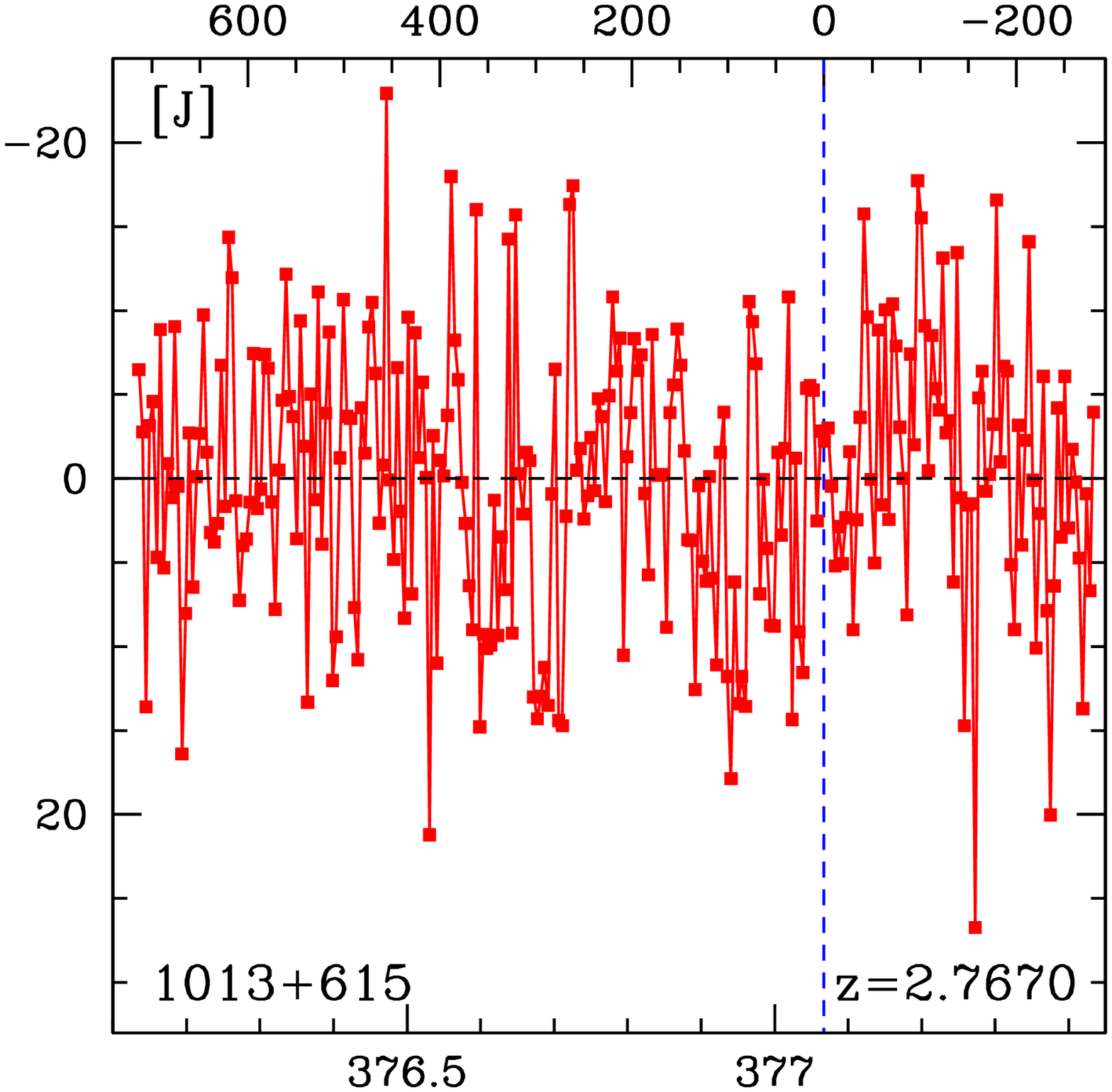,height=2.3truein,width=2.3truein}
\epsfig{file=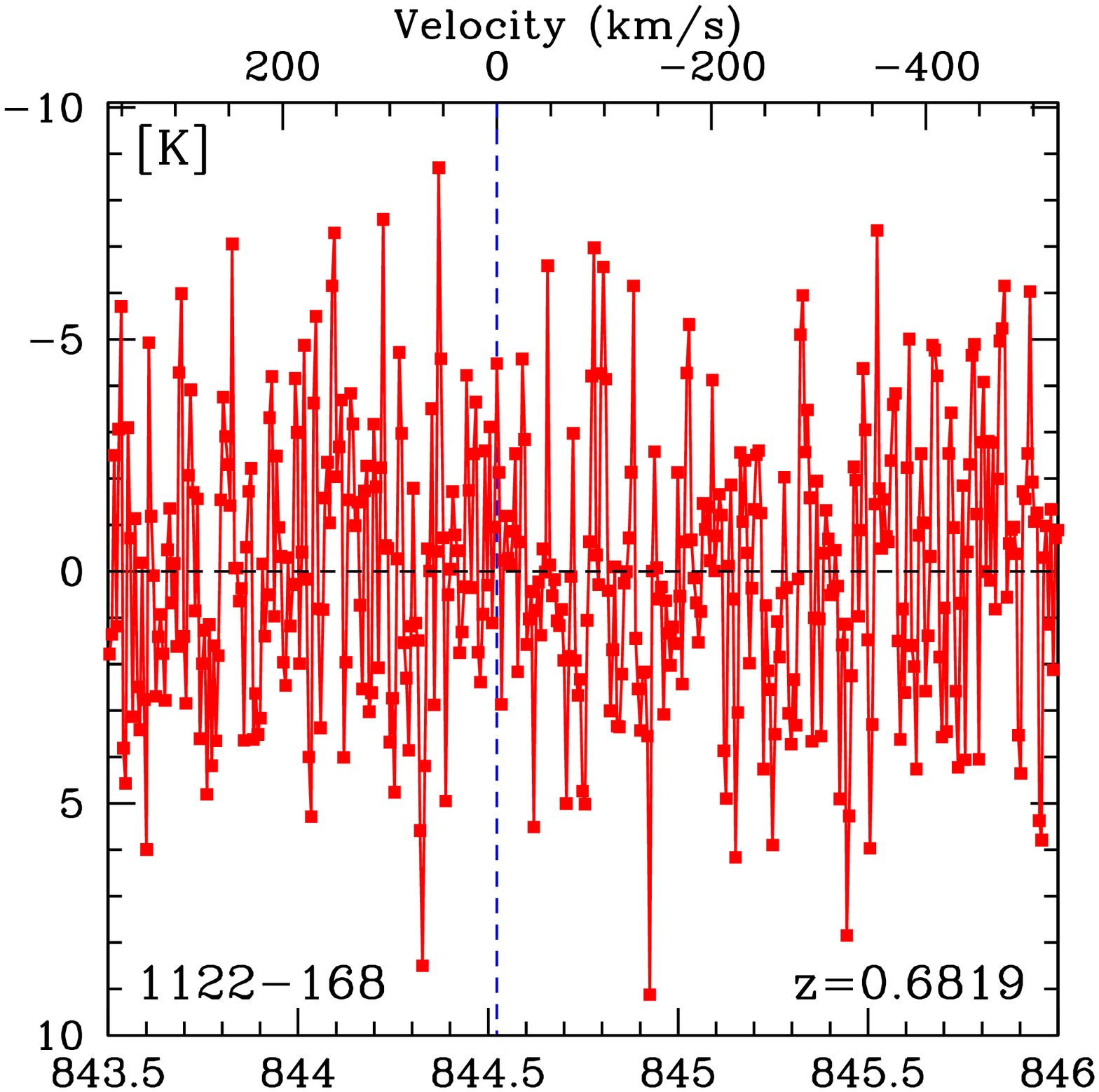,height=2.3truein,width=2.3truein}
\epsfig{file=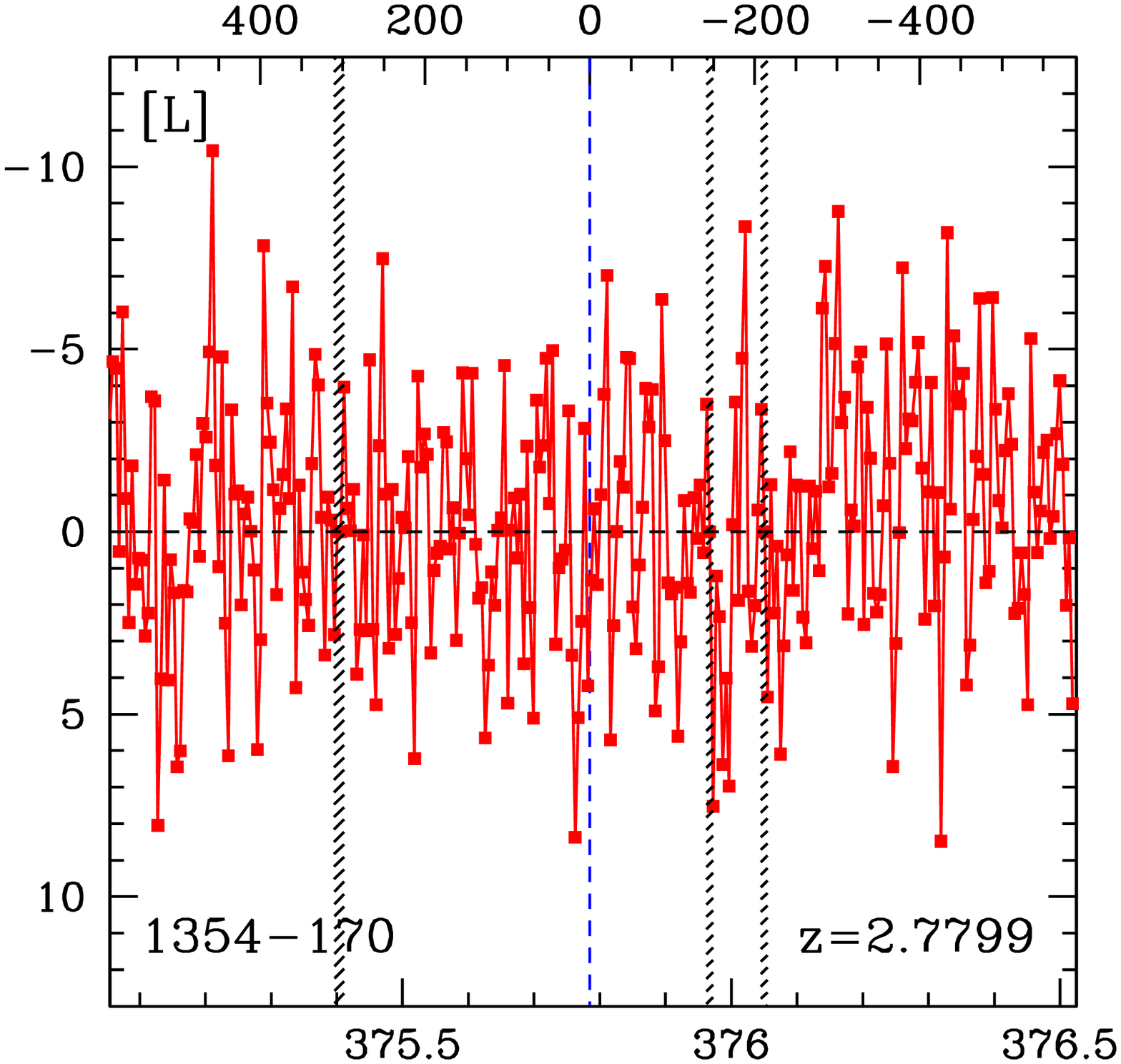,height=2.3truein,width=2.3truein}
\epsfig{file=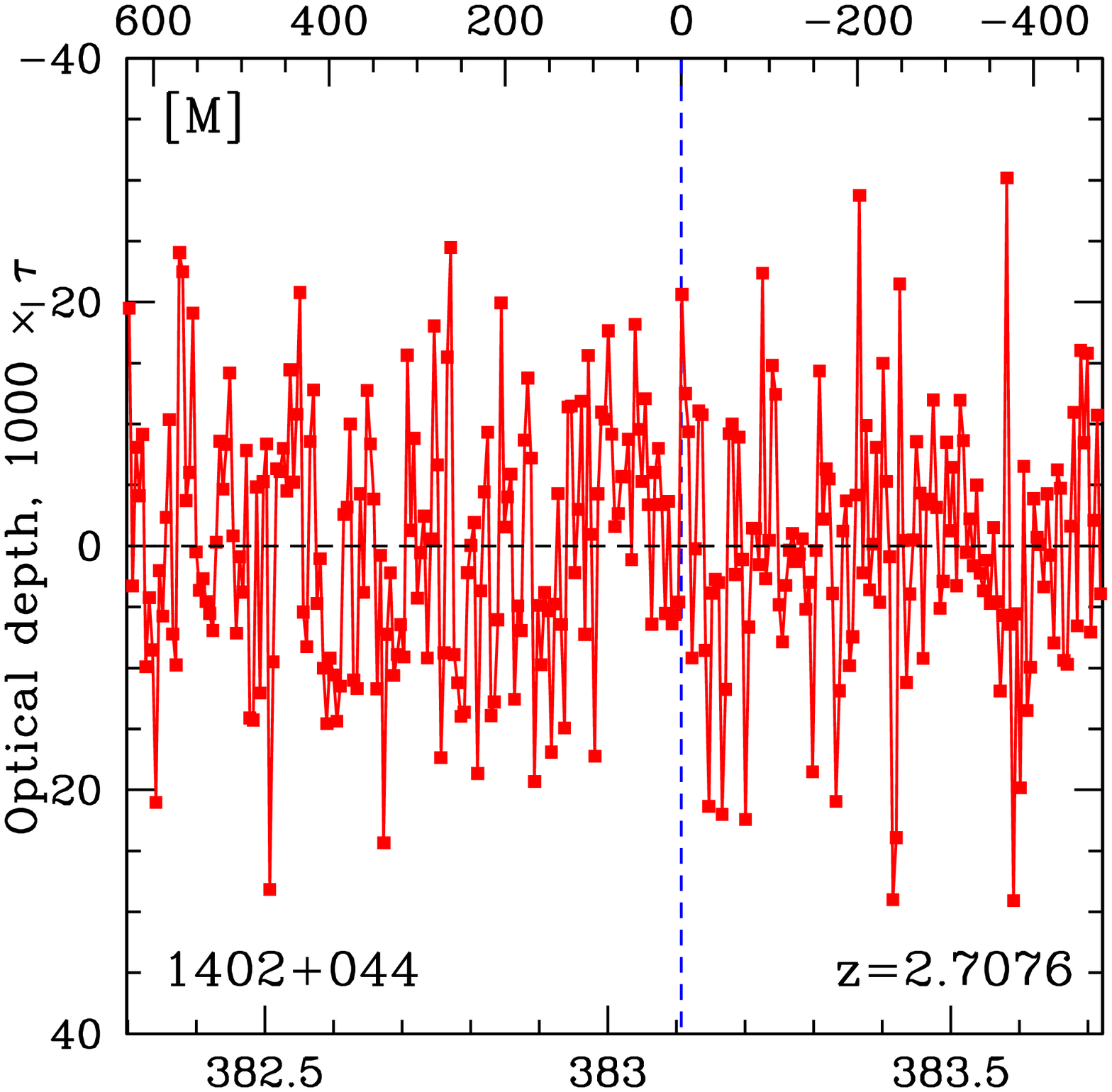,height=2.3truein,width=2.3truein}
\epsfig{file=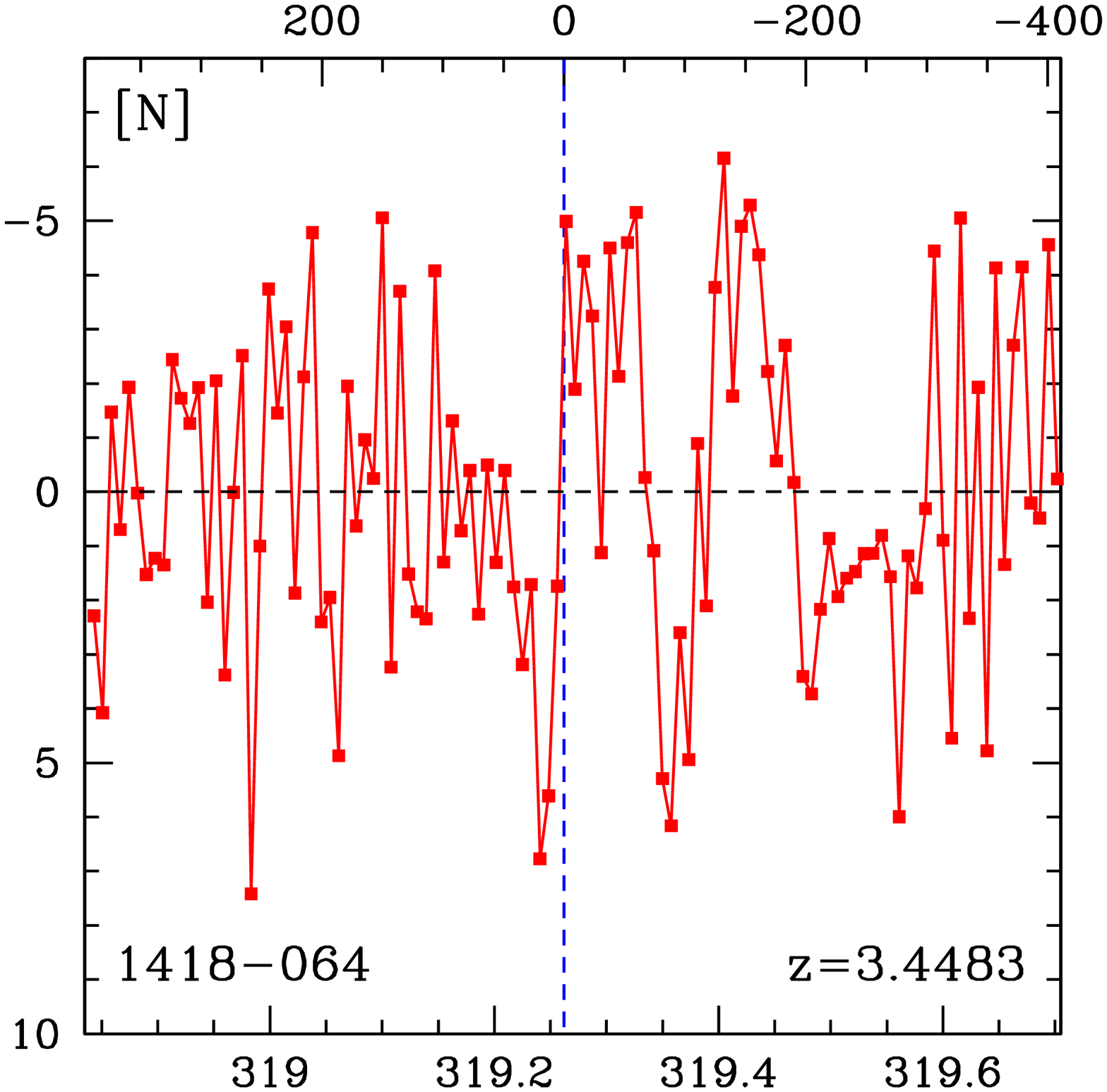,height=2.3truein,width=2.3truein}
\epsfig{file=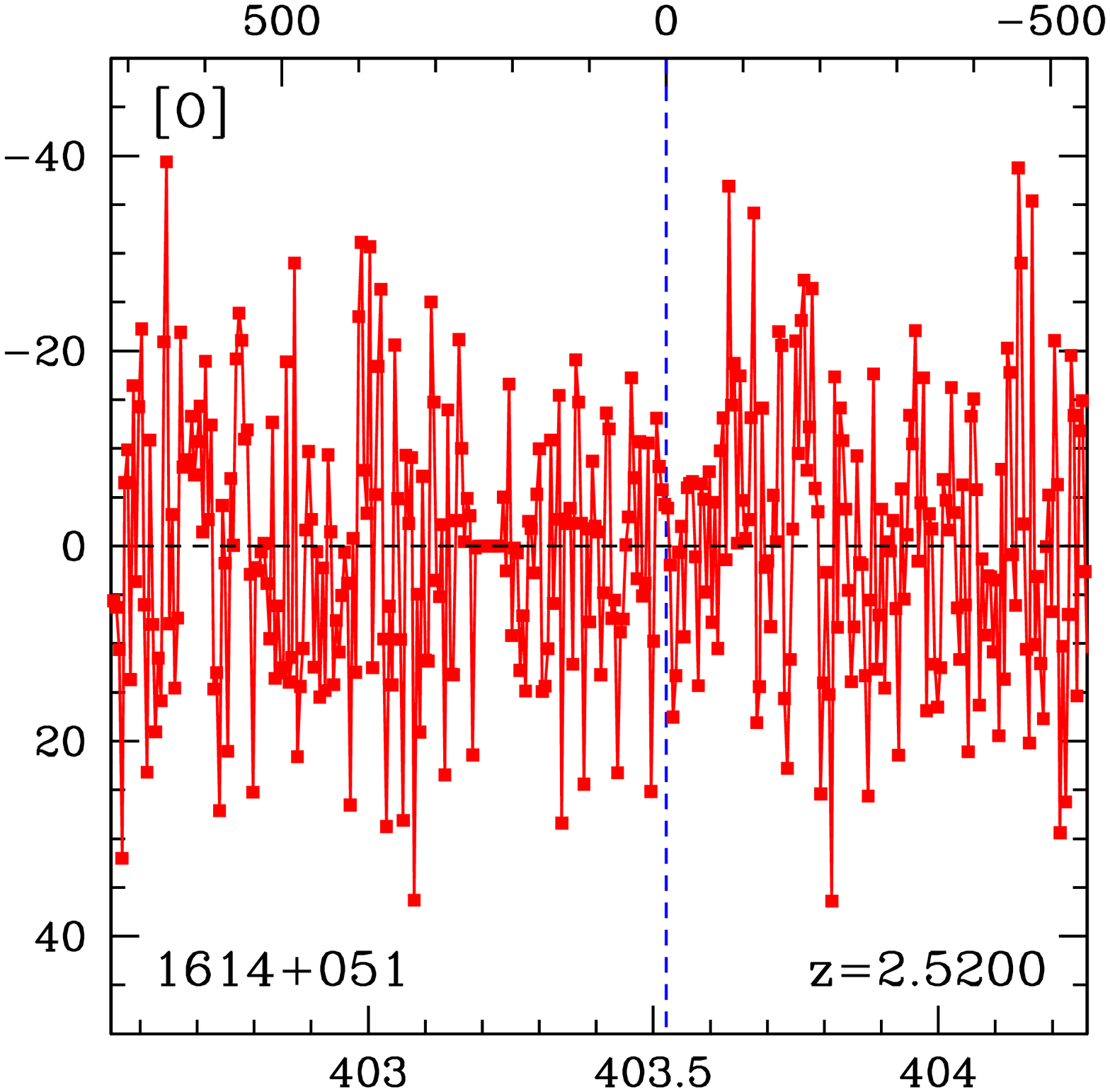,height=2.3truein,width=2.3truein}
\epsfig{file=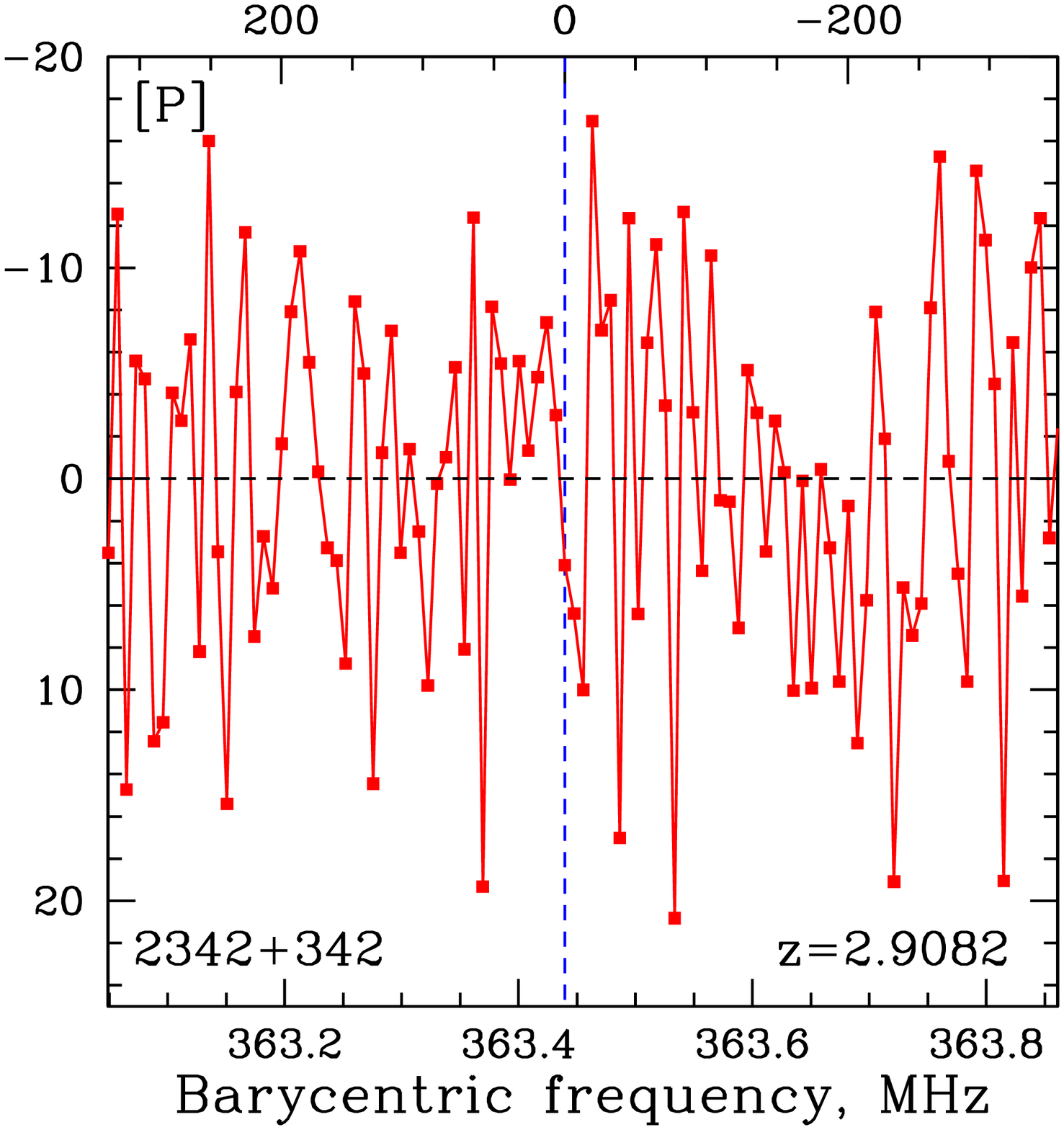,height=2.3truein,width=2.3truein}
\caption{(contd.)
}
\label{fig:nondetect1}
\end{figure*}

\subsubsection{GMRT and WSRT observations, and data analysis}
\label{sec:gmrtwsrt}

\setcounter{table}{0}
\begin{table*}
\caption{\label{table:21cm} Summary of \hii\ observations.}
\begin{center}
\begin{tabular}{|c|c|c|c|c|c|c|c|c|c|c|}
\hline 
QSO  & $z_{\rm QSO}$ & $z_{\rm DLA}$ & $\nu_{\rm 21cm}$ &  Telescope & Time   & BW  & Resn. & RMS & $S_{\rm 21cm}^d$ & $\int \tau_{\rm 21} {\rm d}V$ \\
     &               &               & MHz              &            & hours  & MHz & \kms\ & mJy &   Jy             & \kms\         		  \\
\hline 
0149+335   & 2.431 & 2.1408 & 452.24 & GBT  & 2   & 2.5        & 3.2      & RFI     & $-$       & $-$  	     	\\
0201+113   & 3.610 & 3.3869 & 323.78 & GMRT & 20  & 1.0        & 7.2      & 1.0     & 0.42      & $0.714 \pm 0.017$ 	\\
0336$-$017 & 3.200 & 3.0621 & 349.67 & GMRT & 11  & 1.0        & 6.7      & 1.1     & 0.83      & $< 0.07$          	\\
0347$-$211 & 2.944 & 1.9470 & 481.98 & GBT  & 2   & 1.25       & 1.5      & 16.8    & 0.94      & $< 0.29$          	\\
0405$-$331 & 2.570 & 2.5693 & 397.95 & GBT  & 6   & 5,1.25     & 1.8,7.4  & 2.5$^a$ & 1.0       & $< 0.12$     	\\
0432$-$440 & 2.649 & 2.3020 & 430.17 & GBT  & 0.5 & 5.0        & 6.8      & 5.6     & 0.61      & $< 0.22$ 	     	\\
0438$-$436 & 2.863 & 2.3474 & 424.33 & GBT  & 1.5 & 1.25       & 1.7      & 10.4    & 7.1       & $0.216 \pm 0.027$ 	\\
0405$-$443 & 3.020 & 1.9130 & 487.61 & GBT  & 0.2 & 1.25       & 1.5      & RFI     & $-$       & $-$ 	     	\\
0454+039   & 1.350 & 0.8597 & 763.78 & GBT  & 0.5 & 50         & 2.4      & 3.2     & 0.46      & $< 0.15$        	\\
0458$-$020 & 2.229 & 2.0395 & 467.32 & GBT  & 1   & 0.625,1.25 & 0.78,1.6 & 16.2    & 1.5       & $5.866 \pm 0.059^b$  \\
0528$-$250 & 2.779 & 2.8112 & 372.69 & WSRT & 50  & 1.25       & 3.9      & 7.0$^c$ & 0.140$^c$ & $< 1.3$   		\\
0800+618   & 3.033 & 2.9603 & 358.66 & GMRT & 5   & 2.0        & 6.5      & 4.0     & 0.70      & $< 0.19$  		\\
0824+355   & 2.249 & 2.2433 & 437.95 & GBT  & 2   & 1.25       & 1.7      & 9.6     & 2.6       & $< 0.068$   	\\
0913+003   & 3.074 & 2.7434 & 379.44 & GBT  & 1   & 12.5       & 1.2      & 10.2    & 0.84      & $< 0.18$ 		\\
0944+636   & 2.617 & 2.4960 & 406.29 & GBT  & 2   & 2.5        & 3.6      & RFI     & $-$       & $-$			\\
1013+615   & 2.800 & 2.7670 & 377.07 & GBT  & 4   & 2.5        & 3.9      & 8.1     & 1.00      & $< 0.20$          	\\
1122$-$168 & 2.397 & 0.6819 & 844.52 & GBT  & 1.5 & 50.0       & 2.2      & 1.7     & 0.58      & $< 0.059$ 	     	\\
1157+014   & 2.000 & 1.9436 & 482.47 & GBT  & 2   & 0.625      & 0.76     & 14.2    & 0.70      & $1.020 \pm 0.091$ 	\\
1215+333   & 2.610 & 1.9991 & 473.61 & GBT  & 0.2 & 1.25       & 1.6      & RFI     & $-$       & $-$            	\\
1230$-$101 & 2.394 & 1.9314 & 484.55 & GBT  & 1   & 1.25       & 1.5      & RFI     & $-$       & $-$            	\\
1354$-$170 & 3.150 & 2.7799 & 375.78 & GBT  & 2.5 & 2.5        & 3.9      & 8.7     & 2.6       & $< 0.10$       	\\
1402+044   & 3.215 & 2.7076 & 383.11 & GBT  & 1.5 & 2.5        & 3.8      & 11.5    & 1.1       & $< 0.21$       	\\
1418$-$064 & 3.689 & 3.4483 & 319.31 & GMRT & 11  & 1.0        & 7.3      & 1.3     & 0.44      & $< 0.10$      	\\
1614+051   & 3.215 & 2.5200 & 403.52 & GBT  & 3   & 2.5        & 3.6      & 16      & 1.1       & $< 0.30$       	\\
1645+635   & 2.380 & 2.1253 & 454.49 & GBT  & 0.5 & 1.25       & 1.6      & RFI     & $-$       & $-$            	\\
1755+578   & 2.110 & 1.9698 & 478.28 & GBT  & 1   & 1.25       & 1.5      & RFI     & $-$       & $-$            	\\
1850+402   & 2.120 & 1.9888 & 475.24 & GBT  & 1   & 1.25       & 1.5      & RFI     & $-$       & $-$            	\\
2342+342   & 3.053 & 2.9082 & 363.44 & GMRT & 10  & 1.0        & 6.4      & 1.4     & 0.29      & $< 0.22$       	\\
\hline
\end{tabular}
\end{center}
\tablenotetext{a}{The quoted RMS noise is on the final \hii\ spectrum, produced by smoothing the two spectra to the 
same velocity resolution and then averaging with weights based on the RMS noise values.}
\tablenotetext{b}{The quoted integrated \hii\ optical depth for the $z = 2.0395$ DLA towards 0458$-$020 is 
the average of the values obtained in the two observing runs in March and May 2008, while the RMS 
noise is at a resolution of 2.4~kHz, averaging over the two runs. See Section~\ref{sec:notes} for details.}
\tablenotetext{c}{The low elevation of 0528$-$250 during the WSRT observations resulted in a multiplicative error 
by a factor of 1.71 in the flux density scale. The measured flux density and RMS noise were  82~mJy and 4.1~mJy, 
respectively; however, the quoted values in the table are on the flux scale of \citet{carilli96}. 
Note that the flux scale has no effect on the inferred optical depth.}
\tablenotetext{d}{$S_{\rm 21cm}$ is the source flux density at the redshifted \hii\ line frequency.}
\end{table*}

The GMRT observations used the 327-MHz receivers and the 30-antenna hardware correlator, 
with two polarizations and a bandwidth of 1~MHz sub-divided into 128 channels, or 
2~MHz sub-divided into 256 channels. The number of working antennas varied between 22 and 
30 on the different runs, due to maintenance 
activities and technical problems. Observations of the standard calibrators 3C48, 3C147 
and 3C286 were used to test the system performance at the beginning of each observing run,
and then to calibrate the flux density scale. An observing run typically consisted of 
a full-synthesis track on the target source, interleaving 40-minute scans on the target 
with 7-minute scans on compact phase calibrators to calibrate the antenna gains. The 
system passband was calibrated with observations of bright radio sources, usually
the flux density calibrators themselves, approximately every $3$~hours.

The WSRT observations of 0528$-$250 used the 92cm receivers, with two 
DZB/IVC sub-bands of bandwidth 1.25~MHz each sub-divided into 512 channels, 
with two polarizations. The two intermediate-frequency DZB/IVC sub-bands were 
used in order to reduce the digital noise. The standard calibrators 3C48 and 
3C286 were observed at the start and end of each run to calibrate the flux 
density scale and the system bandpass. A total of 50~hours of observing 
time were obtained on the target source.

All GMRT and WSRT data were analysed in ``classic'' {\sc aips}, following standard 
procedures. After initial editing of bad data (e.g. due to malfunctioning antennas, 
shadowing, correlator problems, RFI, etc), the antenna gains, passband shapes and 
amplitude scales were determined, using the data on the various calibrators. 
After applying the initial calibration, a number of channels were averaged together 
to obtain a ``channel-0'' continuum visibility dataset and a series of 
self-calibration and imaging cycles were then used to accurately determine the 
antenna gains. This was accompanied with intermediate flagging steps to further edit 
out bad data. For the GMRT, 3-D imaging procedures were used for all 327~MHz data, 
with the field sub-divided into 37~facets during imaging. In the case of the WSRT, 
the low source elevation meant that additional care had to be taken to remove data 
affected by shadowing by foreground trees.  The self-calibration and imaging procedure 
consisted of 3-4 phase-only calibration and imaging cycles, followed by a single 
amplitude-and-phase self-calibration and imaging step. This was followed by 
subtraction of the image from the calibrated visibilities, with the residual U-V 
data then examined for any systematic behaviour and bad data edited out, after which 
the self-calibration and imaging cycles were repeated. This procedure was carried out 
for each source until no further improvement was obtained in the image and U-V 
residuals, typically after 2-3 such cycles. The final image was then subtracted out 
from the calibrated multi-channel U-V visibility dataset, using the task {\sc uvsub}. 
Following this, the task {\sc uvlin} was used to subtract out any residual emission, 
by a first-order polynomial fit to the spectrum on each visibility baseline. The 
residual U-V data were then shifted to the heliocentric frame using the task 
{\sc cvel}, and were imaged to produce the final spectral cube. A spectrum was then 
obtained by taking a cut through the location of the target source. In some cases, 
a second-order baseline was fit to this spectrum to obtain the final \hii\ spectrum 
of  the source.  Finally, the source flux density was measured by using the task 
{\sc jmfit} to fit a gaussian model to the final continuum image; all target sources 
were unresolved by the WSRT and GMRT synthesized beams and a single gaussian was 
hence used as the model in all cases.

In the case of the WSRT data, the extremely low elevation of 0528$-$250 meant 
that the flux density scale could not be determined accurately. We found that both the 
source flux density and the RMS noise on the spectrum were lower by a factor of 
$\approx 1.71$ than the expected values, indicating a multiplicative problem with 
the flux scale. Fortunately, this scaling issue does not affect the \hii\ optical 
depth spectrum. Note that the quoted flux density and RMS noise values in 
Table~\ref{table:21cm} have been corrected by this factor of 1.71, for consistency 
with the literature.

\begin{figure*}
\centering
\epsfig{file=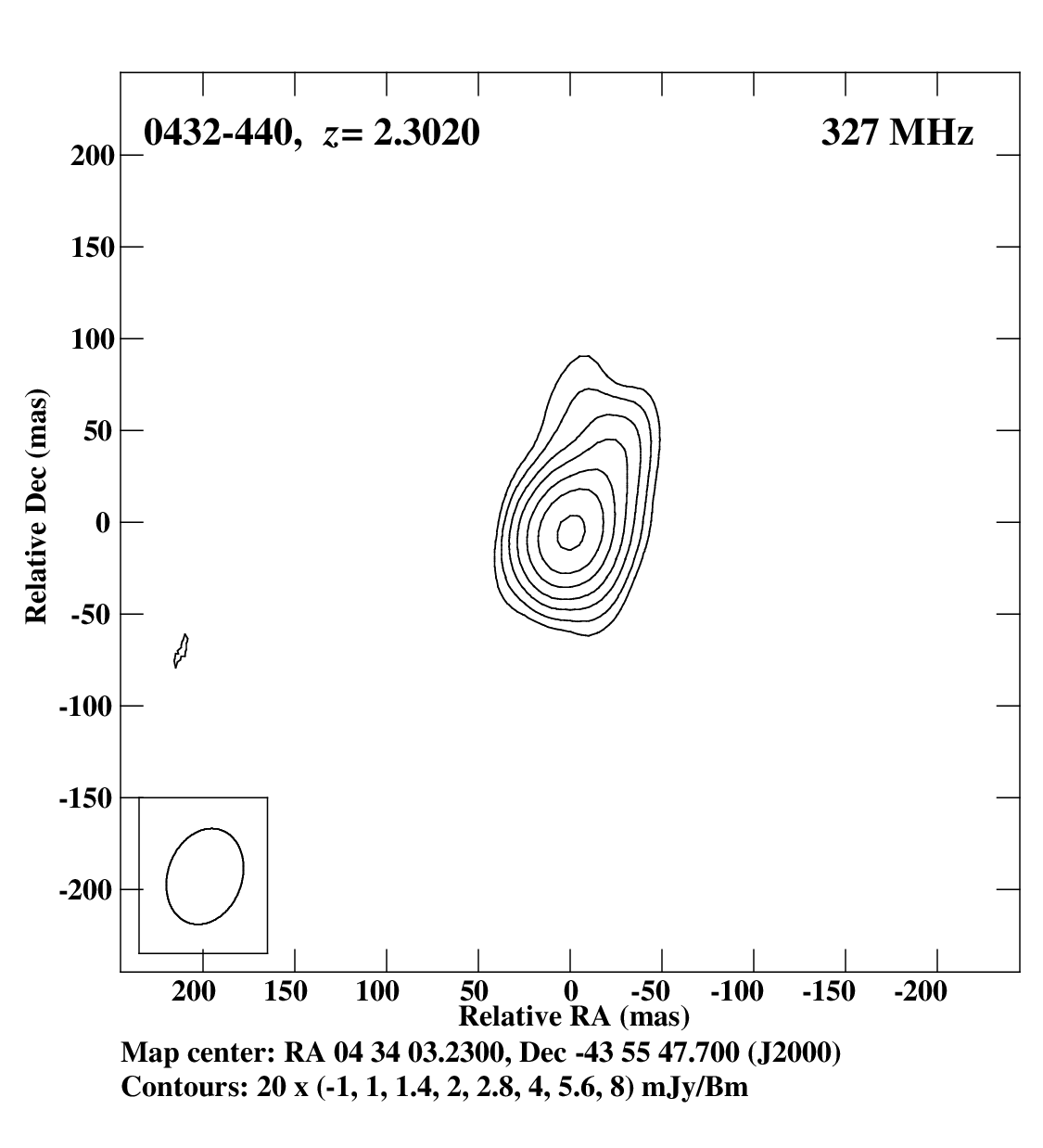,height=2.2truein,width=2.2truein}
\epsfig{file=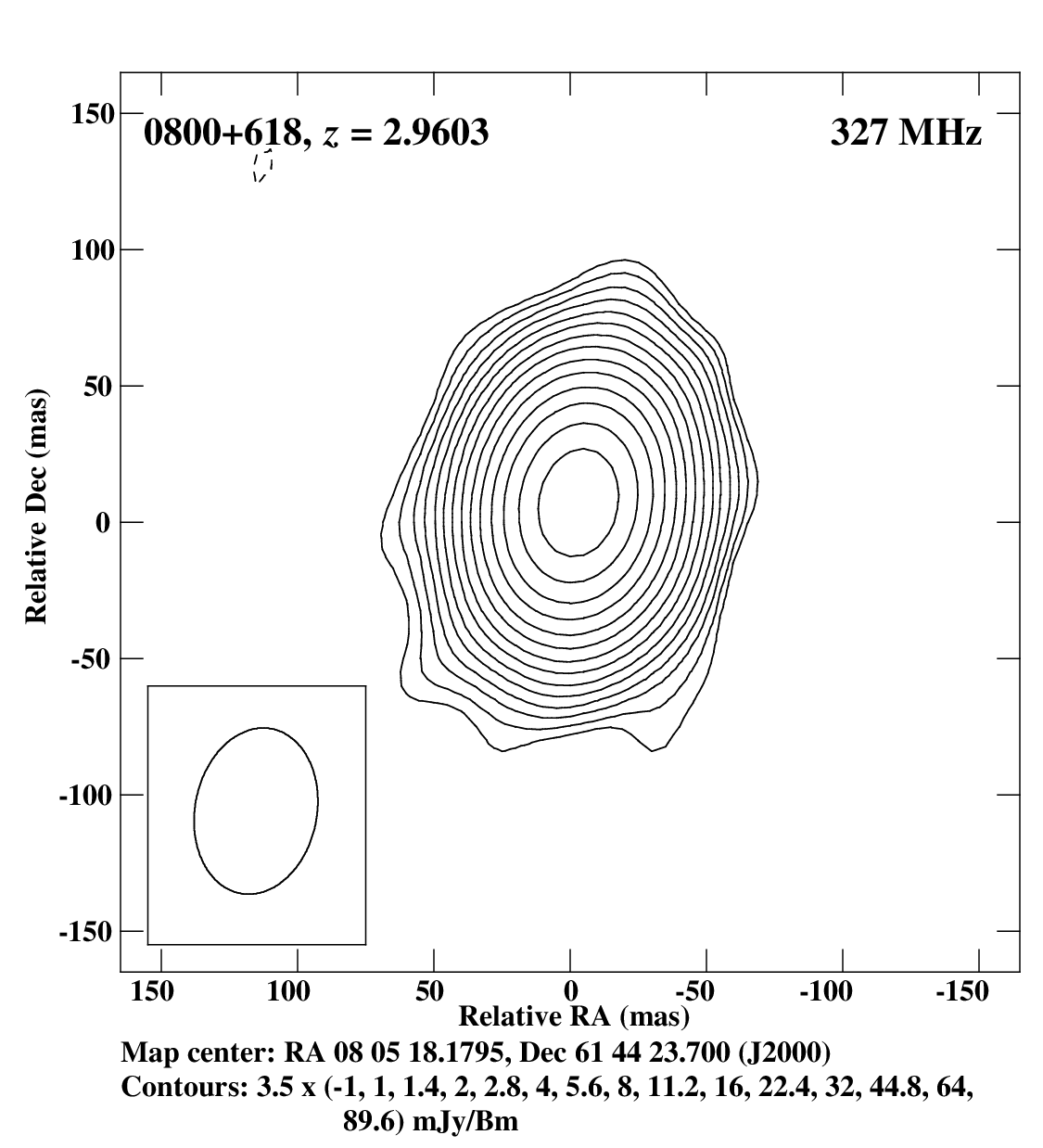,height=2.2truein,width=2.2truein}
\epsfig{file=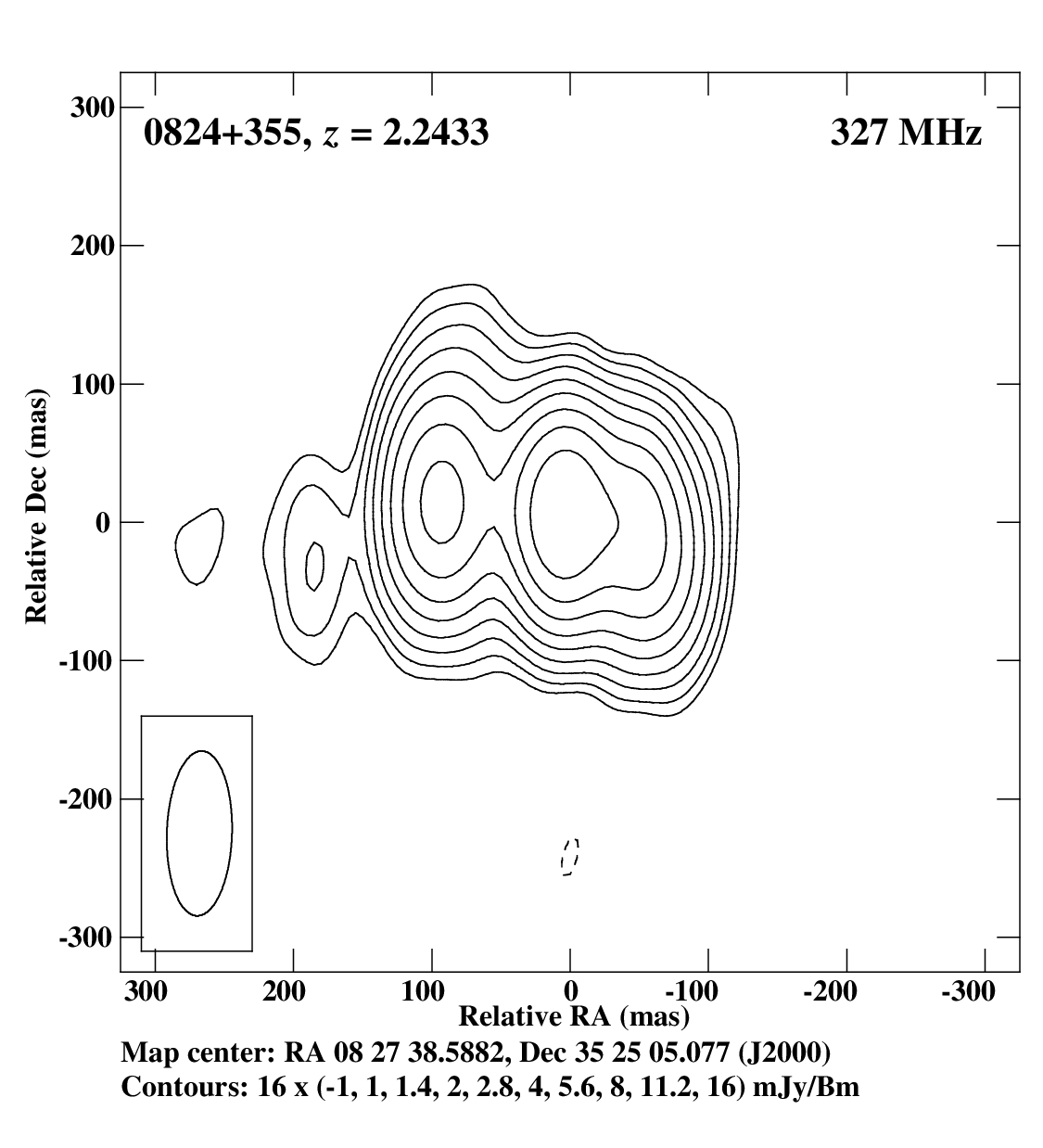,height=2.2truein,width=2.2truein}
\epsfig{file=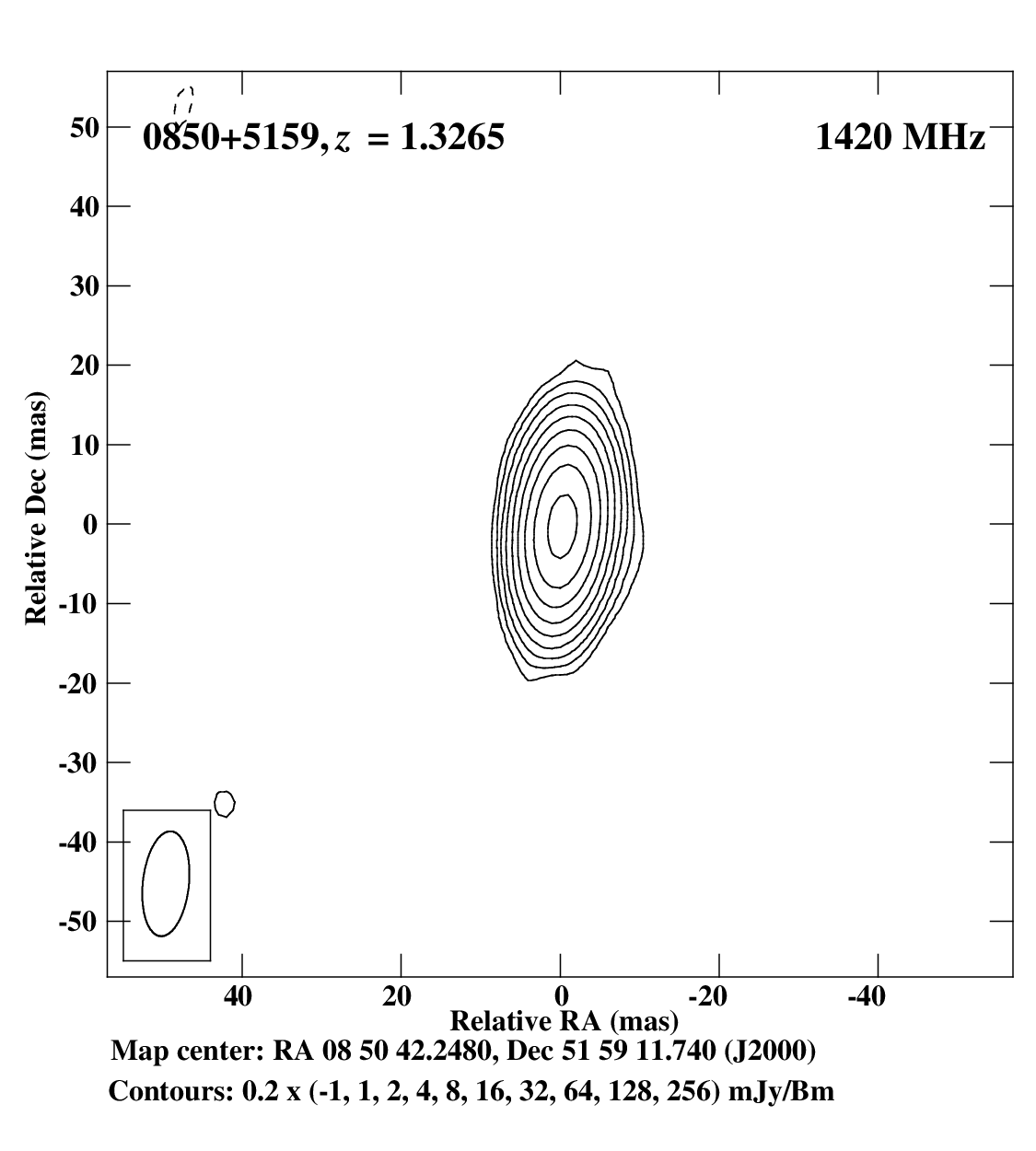,height=2.2truein,width=2.2truein}
\epsfig{file=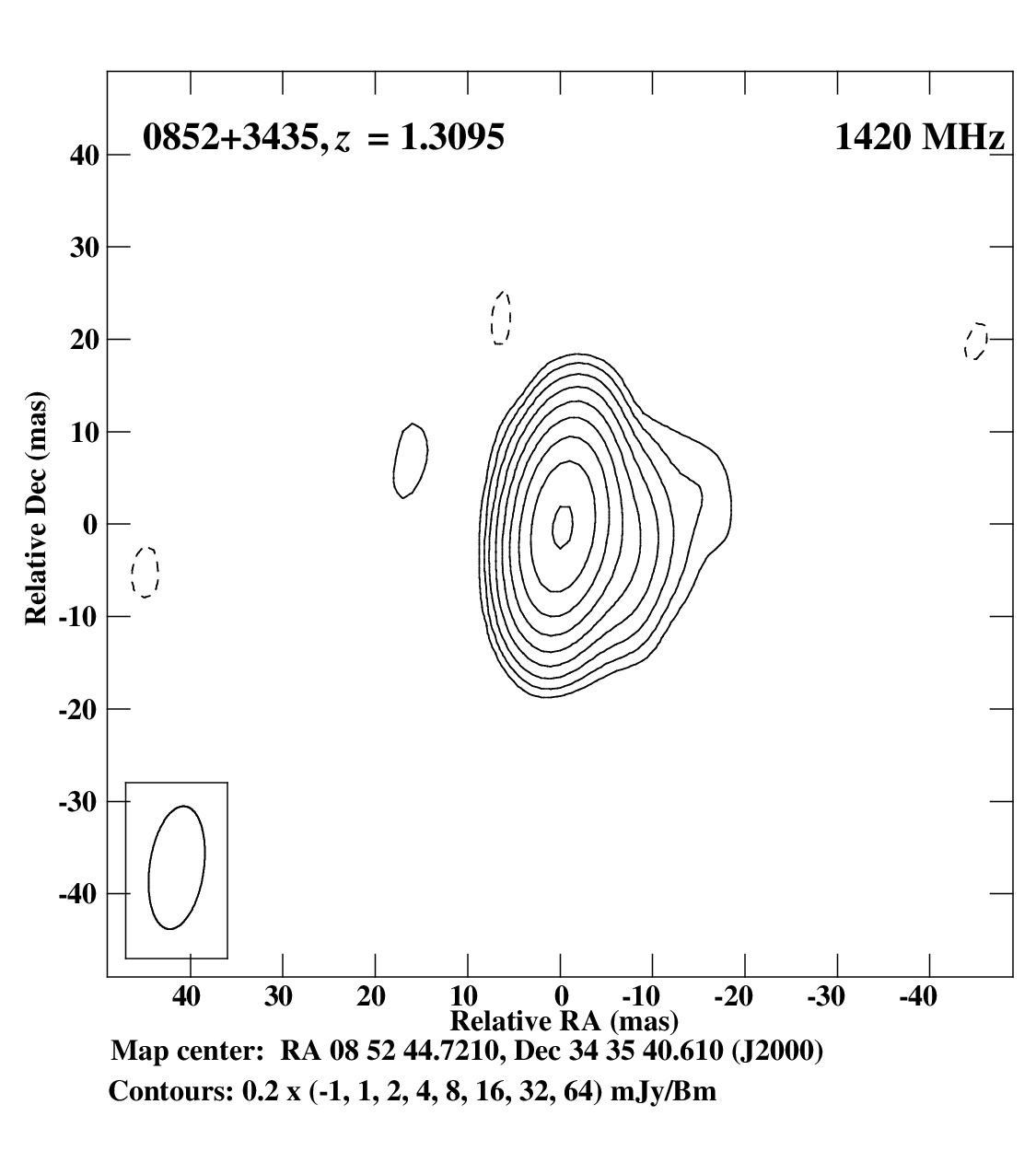,height=2.2truein,width=2.2truein}
\epsfig{file=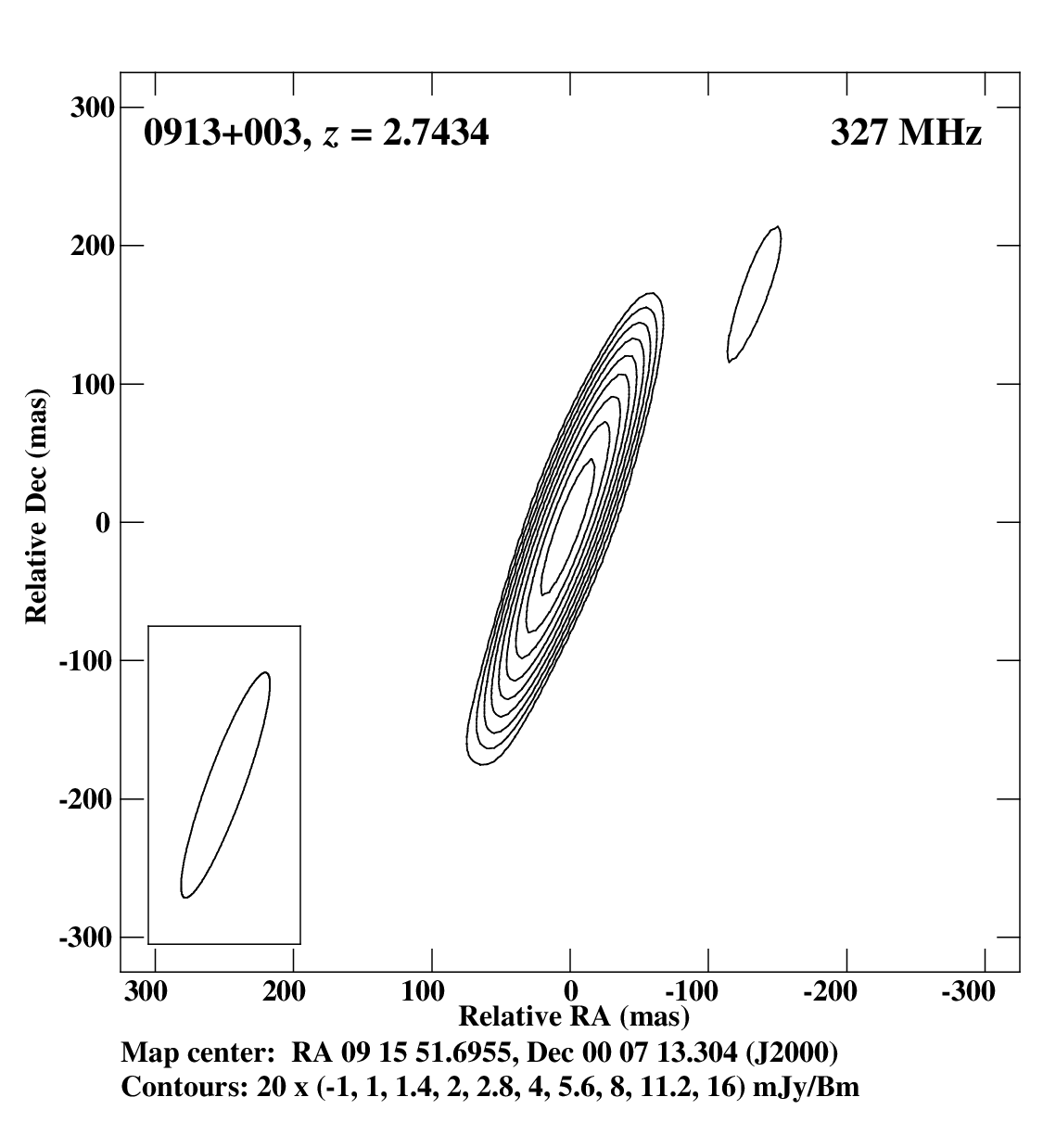,height=2.2truein,width=2.2truein}
\epsfig{file=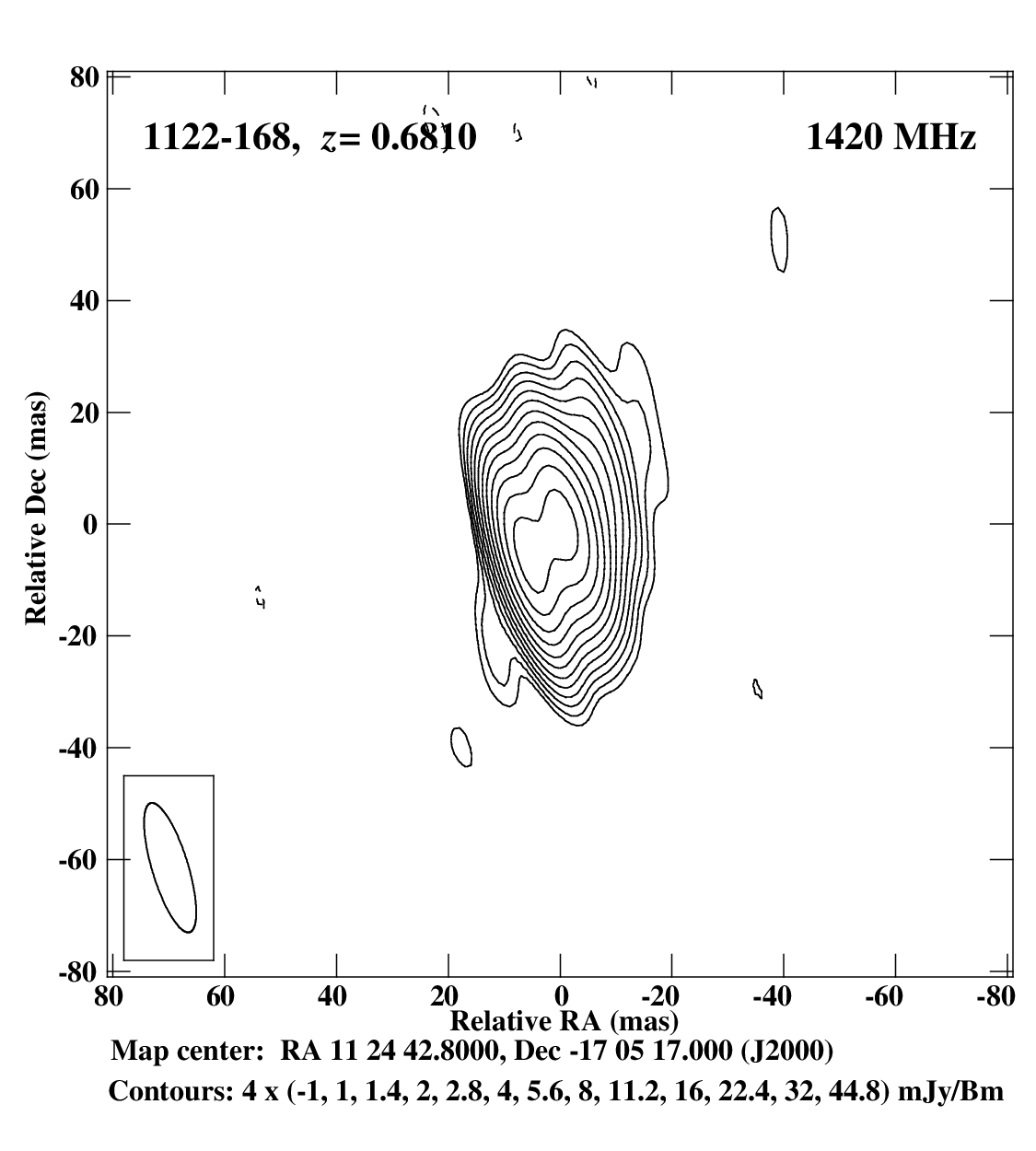,height=2.2truein,width=2.2truein}
\epsfig{file=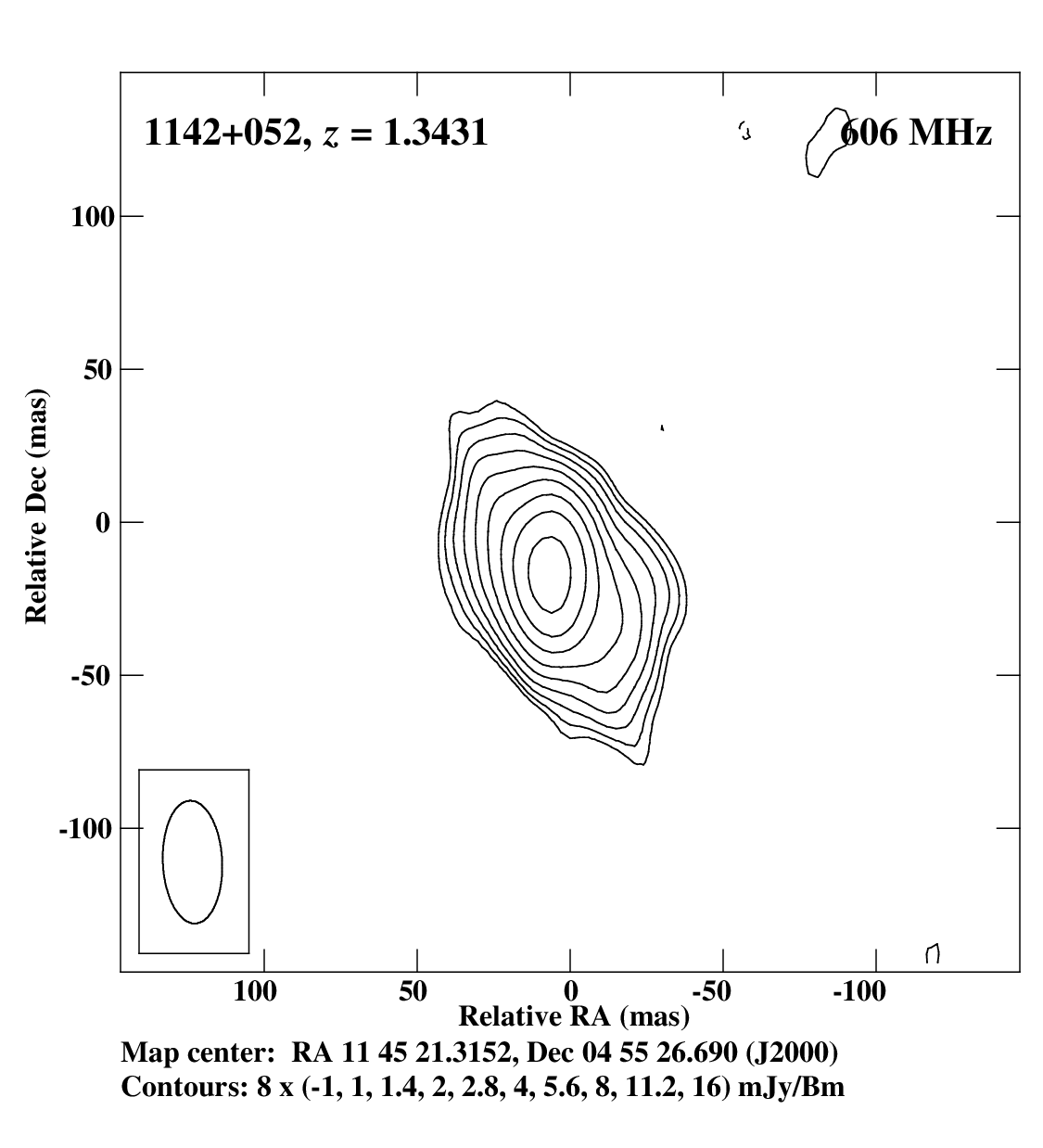,height=2.2truein,width=2.2truein}
\epsfig{file=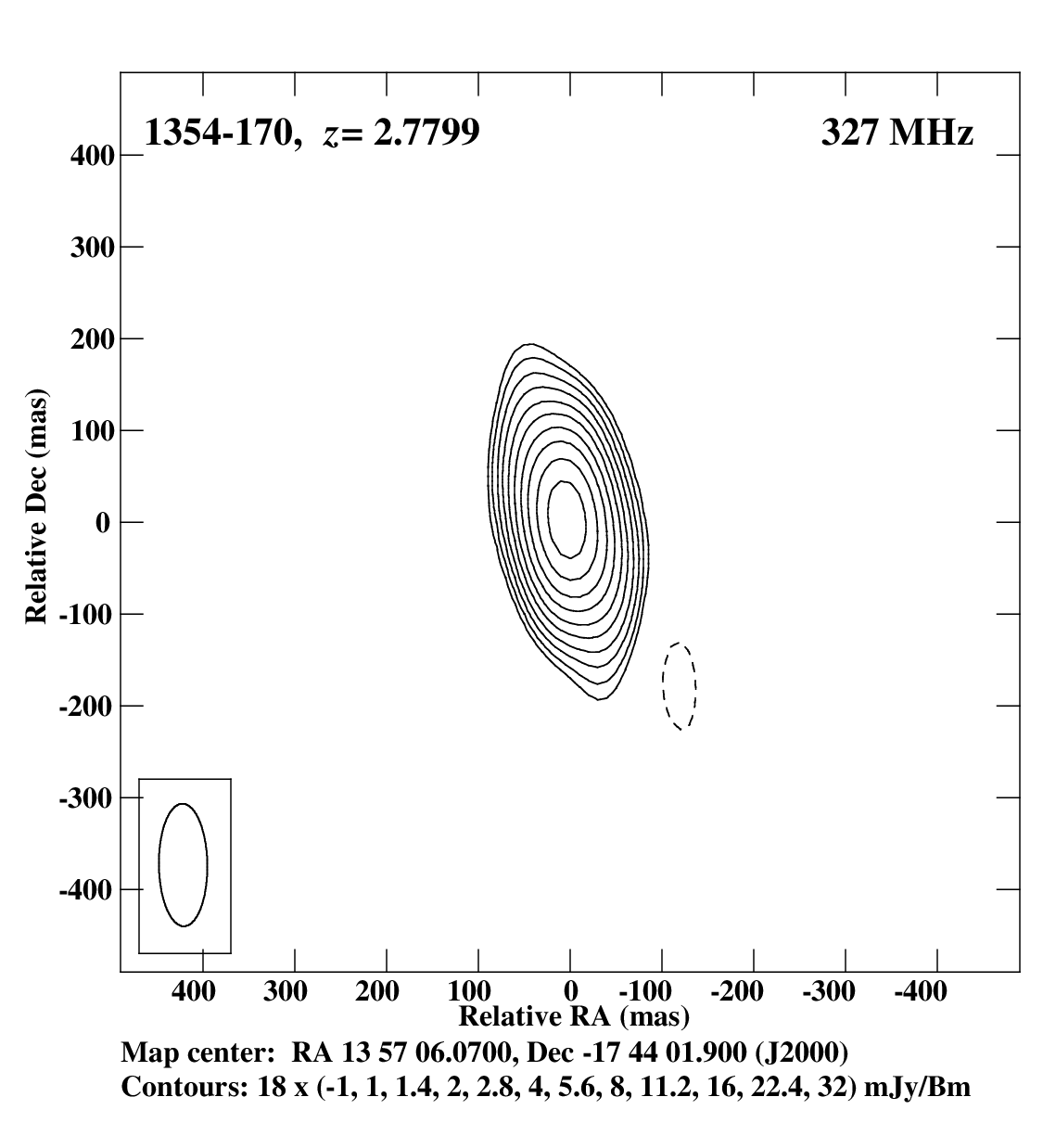,height=2.2truein,width=2.2truein}
\epsfig{file=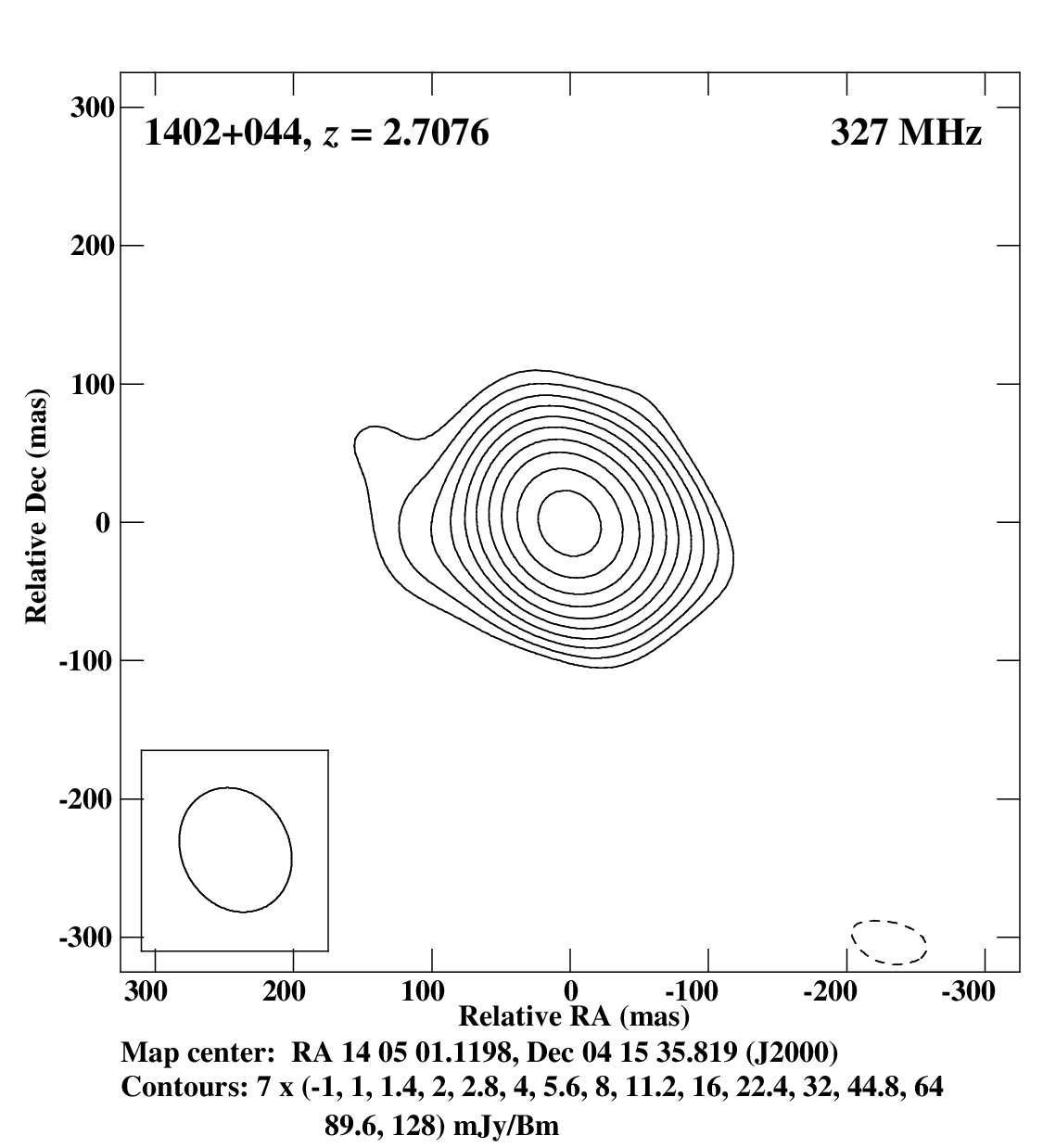,height=2.2truein,width=2.2truein}
\epsfig{file=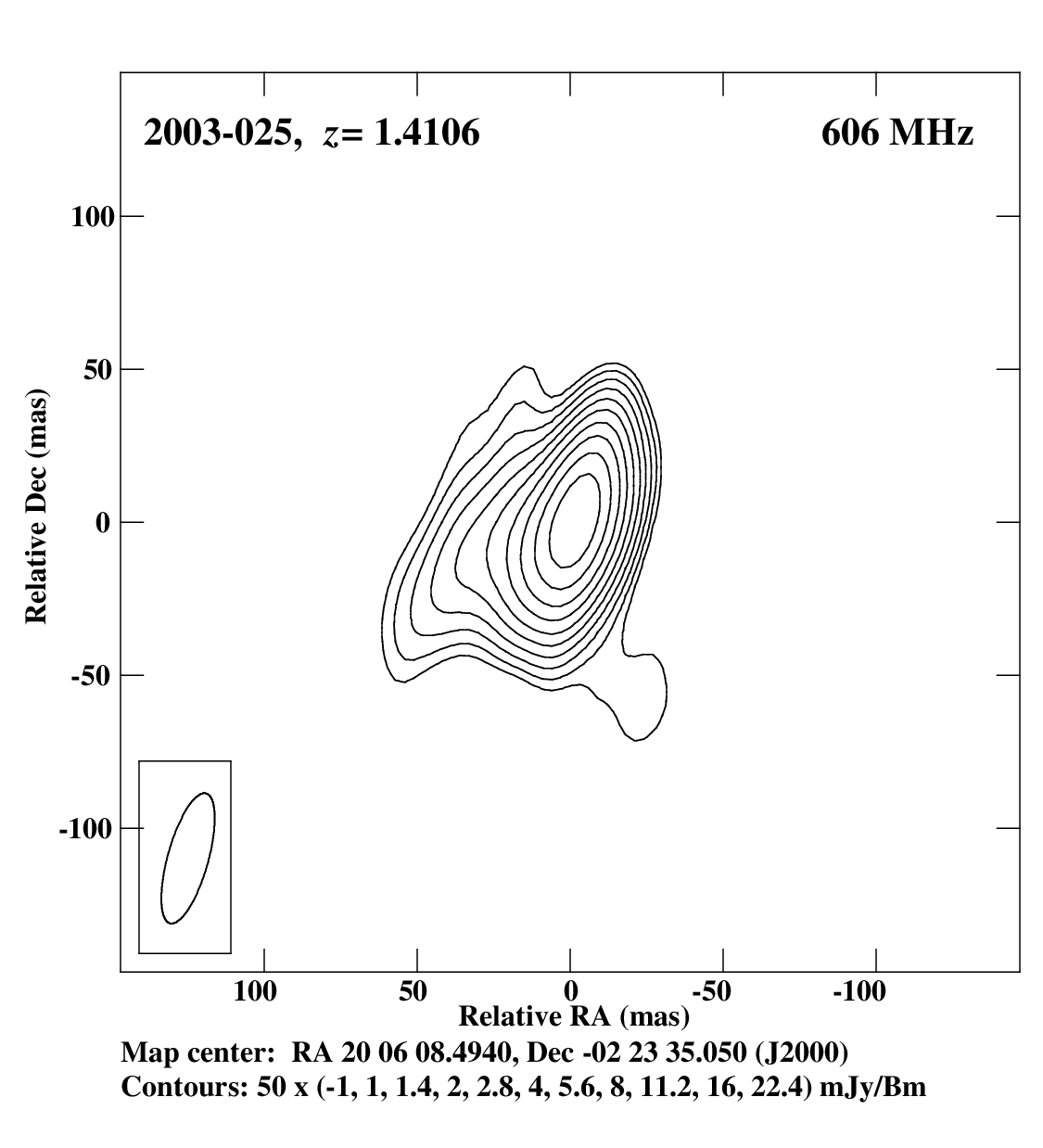,height=2.2truein,width=2.2truein}
\epsfig{file=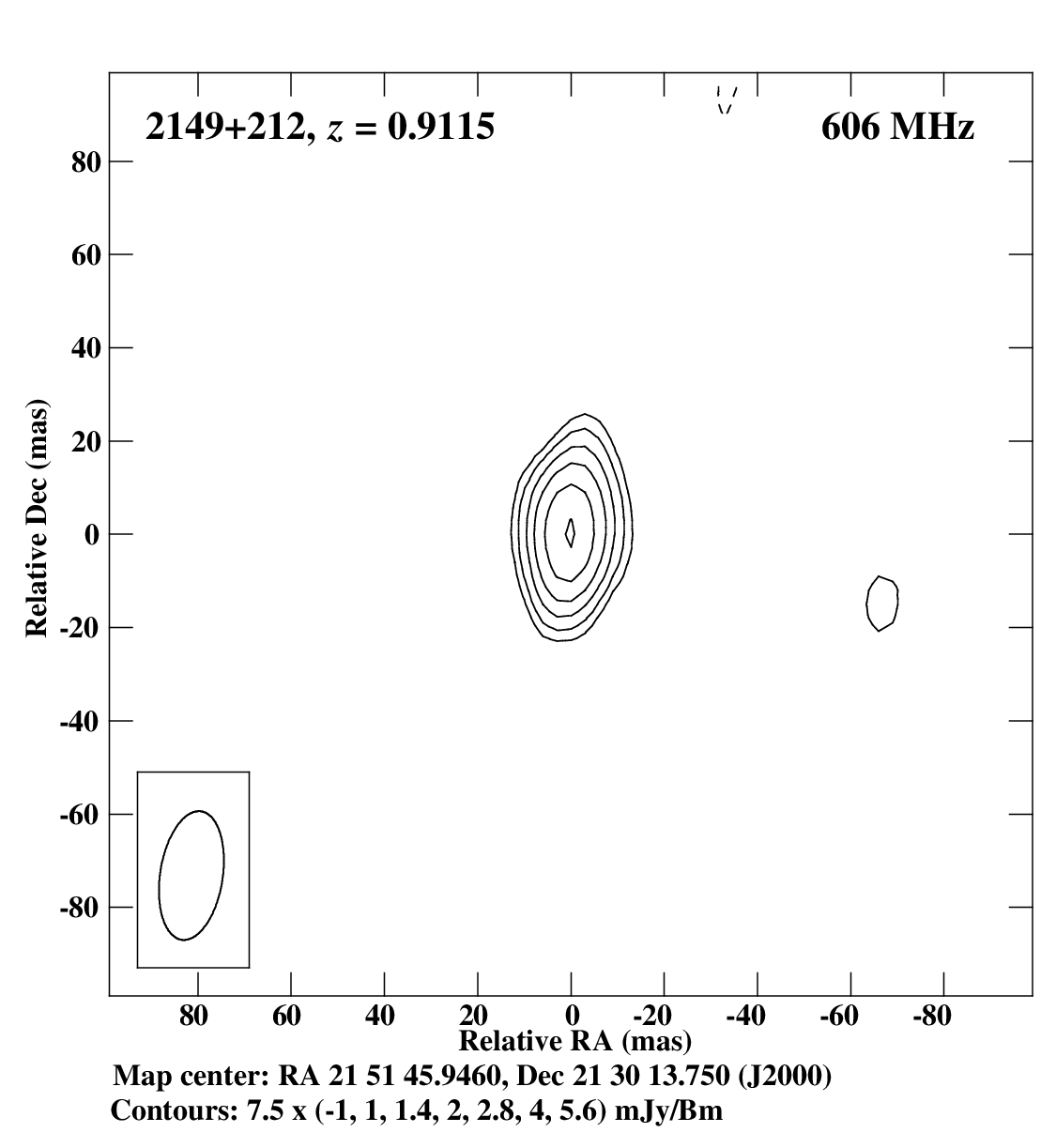,height=2.2truein,width=2.2truein}
\caption{VLBA images of the compact radio structure of the 13 background quasars; 
the quasar name, DLA redshift and observing frequency are indicated at the top of 
each panel. The dashed contour is the largest negative value in each image, with 
the outermost positive contour at the same level; the penultimate solid contour 
thus indicates the lowest believable structure in each image. The synthesized
beam is shown at the bottom left corner of each panel.}
\label{fig:vlba}
\end{figure*}

\begin{figure*}
\centering
\setcounter{figure}{2}
\epsfig{file=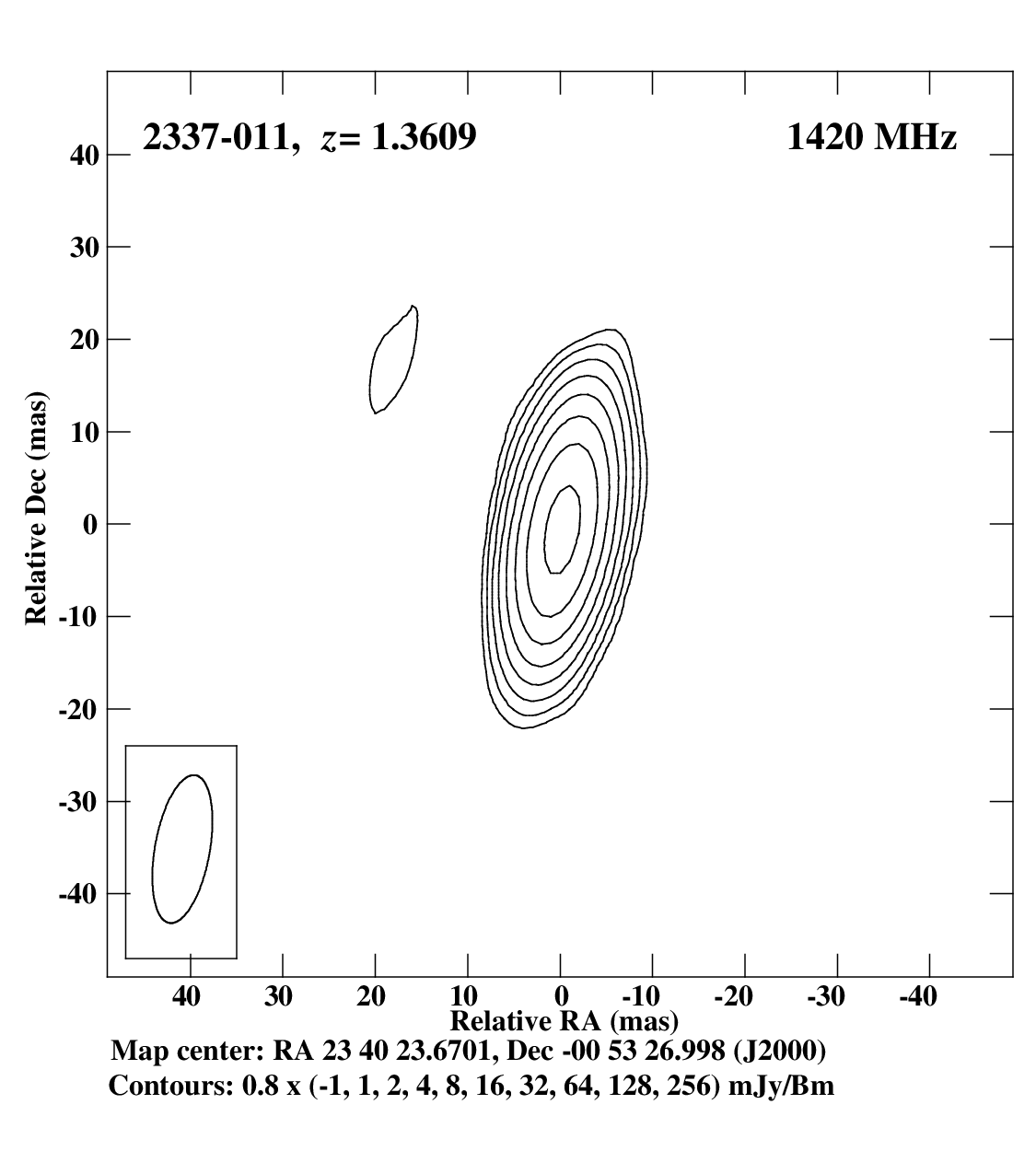,height=2.2truein,width=2.2truein}
\caption{(contd.)}
\label{fig:vlba2}
\end{figure*}

\subsubsection{Spectra and results}
\label{sec:hi21cm-results}

New detections of \hii\ absorption were obtained in two DLAs, at $z \sim 3.387$ towards 
0201+113 and $z \sim 2.347$ towards 0438$-$436. These results were discussed in 
detail by \citet{kanekar06} and \citet{kanekar07}, and the spectra are not shown here. 
\hii\ absorption was also detected in two known \hii\ absorbers, at $z \sim 1.9436$ towards 
1157+014 \citep{wolfe81} and $z = 2.0395$ towards 0458$-$020 \citep{wolfe85}; these spectra 
are shown in Fig.~\ref{fig:det2}. For the non-detections, the final \hii\ spectra are shown 
in Figure~\ref{fig:nondetect}, in order of increasing right ascension. 

The observational details and results of the GBT, GMRT and WSRT spectroscopy are summarized 
in Table~\ref{table:21cm}. The columns of this table are: (1)~the quasar name, (2)~the quasar 
emission redshift, (3)~the DLA redshift, (4)~the redshifted \hii\ line frequency, (5)~the 
telescope used for the \hii\ search, (6)~the on-source integration time, in hours, (7)~the 
bandwidth, in MHz, (8)~the velocity resolution, in \kms, after Hanning-smoothing and re-sampling, 
(9)~the RMS noise at this velocity resolution, in mJy, (10)~the source flux density, in Jy, 
(11)~for detections, the integrated \hii\ optical depth $\int \tau_{\rm 21cm} {\rm d}V$, or, 
for non-detections, the $3\sigma$ limit on $\int \tau_{\rm 21cm} {\rm d}V$, assuming a gaussian 
line profile with a line FWHM of $\Delta V=15$~\kms, with the RMS noise computed at a similar 
velocity resolution. For the eight GBT targets where the final \hii\ spectrum was unusable 
due to RFI, column (10) contains ``RFI'', and the remaining column entries have been left 
blank. We note, in passing, that the assumed line FWHM of 15~\kms\ is comparable to the widths 
of spectral components in known redshifted \hii\ absorbers; assuming a narrower line FWHM 
would yield a more stringent lower limit on the spin temperature (see 
Section~\ref{sec:tspin} for discussion).

\subsection{VLBA imaging studies}
\label{sec:vlba}

\setcounter{table}{1}
\begin{table*}
\caption{\label{table:vlba} Results from VLBA low-frequency imaging of quasars behind high-$z$ DLAs. 
}
\begin{center}
\begin{tabular}{|c|c|c|c|c|c|c|c|c|c|c|c|}
\hline
                    &     &  &  & & & & & & & \\
QSO  & $z_{\rm QSO}$ & $z_{\rm abs}$ & $\nu_{\rm 21cm}$ & $\nu_{\rm VLBA}$ & $S_{\rm tot}^a$ & Beam             & S$_{\rm fit}$ & Angular size & Spatial extent  & $f^c$\\
	&       &        & MHz    & MHz  & Jy   & mas $\times$ mas  &   Jy    &  mas $\times$ mas  & pc $\times$ pc    & \\
\hline
	&       &        & & & & &   &  & & \\
0432$-$440   & 2.649 & 2.3020 & 430.17 & 327  & 0.69 & $54 \times 40$    & $0.26$  & $58.2^{+2.5}_{-2.5} \times 24.4^{+2.1}_{-2.3}$ & $490^{+21}_{-21} \times 205^{+18}_{-19}$  & 0.38 \\
	&       &        & & & & &   &  & & \\
0800+618     & 3.033 & 2.9603 & 358.66 & 327  & 0.70 & $61 \times 45$    & $0.44$  & $13.5^{+0.8}_{-0.8} \times 10.3^{+0.6}_{-0.8}$ & $107^{+6}_{-6} \times 82^{+5}_{-6}$  & 0.63 \\
	&       &        & & & & &   &  & & \\
0824+355     & 2.249 & 2.2433 & 437.95 & 327  & 2.00 & $119 \times 47$   & $0.40$  & $27.8^{+6.2}_{-27.8} \times 2.4^{+0.8}_{-2.4}$  & $235^{+52}_{-235} \times 20^{+7}_{-20}$   & 0.20 \\
	        &       &        &        &      &      &                & $0.37$  & $66.2^{+9.2}_{-9.6} \times 41.1^{+9.9}_{-15.6}$ & $560^{+78}_{-81} \times 348^{+84}_{-132}$ & -- \\
	        &       &        &        &      &      &                & $0.36$  & $79.2^{+9.4}_{-9.7} \times 51.7^{+3.4}_{-5.1}$  & $670^{+79}_{-82} \times 437^{+29}_{-43}$  & -- \\
	&       &        & & & & &   &  & & \\
J0850+5159  & 1.894 & 1.3265 & 610.53 & 1420 & 0.061 & $13.3 \times 5.8$ & $0.069$ & $1.49^{+0.04}_{-0.03} \times 0.95^{+0.17}_{-0.21}$   & $12.9^{+0.3}_{-0.3} \times 8.0^{+1.5}_{-1.8}$ & 1.0  \\
	&       &        & & & & &   &  & & \\
J0852+3435  & 1.655 & 1.3095 & 615.03 & 1420 & 0.066 & $13.4 \times 5.9$ & $0.062$ & $2.9^{+0.03}_{-0.03} \times 0.89^{+0.25}_{-0.35}$   & $25.0^{+0.3}_{-0.3} \times 7.7^{+2.2}_{-3.0}$ & 0.93 \\
	&       &        & & & & &   &  & & \\
0913+003     & 3.074 & 2.7434 & 379.44 & 327  & 0.48 & $173 \times 27$   & $0.26$  & $16.6^{+17.4}_{-9.3} \times 7.6^{+2.2}_{-7.6}$  & $134^{+141}_{-75} \times 62^{+18}_{-62}$   & 0.54 \\
	&       &        & & & & &   &  & & \\
1122$-$168   & 2.397 & 0.6819 & 844.52 & 1420 & 0.29 & $24.1 \times 6.5$ & $0.012$ & $4.2^{+1.5}_{-2.4} \times 0.0^{+0.0}_{-0.0}$~$^b$  & $30.7^{+11.0}_{-17.5} \times 0.0^{+0.0}_{-0.0}$~$^b$ & 0.04 \\
	        &       &        &        &      &      &                   & $0.043$ & $12.5^{+0.4}_{-0.3} \times 5.7^{+0.9}_{-1.0}$ & $91.3^{+2.9}_{-2.2} \times 41.6^{+6.6}_{-7.3}$ & -- \\
	&       &        & & & & &   &  & & \\
1142+052     & 1.345 & 1.3431 & 606.21 & 606  & 1.01 & $40 \times 19$    & $0.30$  & $36.6^{+1.6}_{-1.6} \times 19.1^{+1.8}_{-2.1}$ & $317^{+14}_{-14} \times 165^{+16}_{-18}$  & 0.30 \\
	&       &        & & & & &   &  & & \\
1354$-$170   & 3.150 & 2.7799 & 375.78 & 327  & 0.83 & $134 \times 53$   & $0.81$  & $88.3^{+1.8}_{-1.8} \times 38.6^{+2.0}_{-2.1}$ & $712^{+15}_{-15} \times 311^{+16}_{-17}$  & 0.97 \\
	&       &        & & & & &   &  & & \\
1402+044     & 3.215 & 2.7076 & 383.11 & 327  & 1.26 & $93 \times 78$    & $0.43$  & $44.5^{+6.5}_{-7.3} \times 23.4^{+10.6}_{-23.4}$ & $362^{+53}_{-59} \times 190^{+86}_{-190}$  & 0.34 \\
	&       &        & & & & &   &  & & \\
2003$-$025   & 1.457 & 1.4106 & 589.23 & 606  & 3.7  & $44 \times 13$    & $1.97$  & $14.9^{+0.6}_{-0.5} \times 11.0^{+1.1}_{-1.3}$ & $129^{+5}_{-4} \times 95^{+10}_{-11}$  & 0.53 \\
	        &       &        &        &      &      &                   & $1.48$  & $48.5^{+2.1}_{-2.1} \times 31.2^{+3.2}_{-3.4}$ & $421^{+18}_{-18} \times 271^{+28}_{-30}$  & -- \\ 
	&       &        & & & & &   &  & & \\
2149+212     & 1.538 & 0.9115 & 743.08 & 606  & 1.92 & $28 \times 13$    & $0.05$  & $13.3^{+4.2}_{-13.3} \times 7.1^{+5.5}_{-7.1}$  & $107^{+34}_{-107} \times 57^{+44}_{-57}$   & 0.03 \\
	&       &        & & & & &   &  & & \\
2337$-$011   & 2.085 & 1.3606 & 601.71 & 1420 & 0.12 & $16.2 \times 5.9$ & $0.13$  & $1.5^{+0.2}_{-0.3} \times 0.60^{+0.19}_{-0.60}$   & $13.0^{+1.7}_{-2.6} \times 5.2^{+1.6}_{-5.2}$ & 1.0  \\
	&       &        & & & & &   &  & & \\
\hline
\end{tabular}
\end{center}
\tablenotetext{a}{The total flux density $S_{\rm tot}$ at the VLBA observing frequency was estimated
by extrapolating from measured values in the literature, usually from the 365~MHz Texas survey 
\citep{douglas96} and the 1.4~GHz Very Large Array FIRST and NVSS surveys \citep{becker95,condon98}.}
\tablenotetext{b}{For 1122$-$168, the deconvolved angular size and spatial extent are both $0.0$ along
one axis. This is because the angular extent of the core along this axis is significantly smaller 
than the VLBA synthesized beam.}
\tablenotetext{c}{We assume that the DLA covering factor $f$ is equal to the quasar core fraction, i.e. 
the ratio of the core flux density to the total integrated flux density, $S_{\rm tot}$. For the three 
sources, 1122$-$168, 0824+355 and 2003$-$025, where the model contained multiple Gaussian components, the 
more compact component has been identified with the core. In such cases, the covering factor estimates 
assume that only the core has been covered by the foreground DLA.}
\end{table*}

We used the VLBA 327~MHz, 606~MHz and 1.4~GHz receivers in proposals BK89, BK131, BK153, BK159 
and BK175 to obtain high spatial resolution images of the compact radio structure 
of the quasars behind a sample of DLAs and \hii\ absorbers, to measure the quasar core fractions,
and thus estimate the DLA covering factors. For each target, the VLBA observing frequency was chosen 
to be within a factor of 
$\approx 2$ of the redshifted \hii\ line frequency, to ensure that the spatial extent of 
the background radio continuum was determined close to the line frequency. In practice, the 
VLBA observing frequency was within a factor of $1.5$ of the redshifted \hii\ line frequency 
for all targets except for three \hii\ absorbers at $z \sim 1.3$ lying towards faint background 
quasars (J0850+5159, J0852+3435 and 2337$-$011, all with flux densities 
$\lesssim 100$~mJy at the redshifted \hii\ line frequency). These three sources were observed 
with the higher-sensitivity L-band receivers of the VLBA.

The results from the initial VLBA data of proposals BK89 and BK131 were presented by 
\citet{kanekar09a}. We will not describe these data again here, but will instead 
simply use the core flux density estimates from that paper. 13 DLAs and \hii\ 
absorbers were observed with the VLBA in proposals BK153, BK159 and BK175, six with 
the 327~MHz receivers, three with the 606~MHz receivers and four with the 1.4~GHz
receivers. Bandwidths of 12, 4 and 32~MHz were used at 327~MHz, 606~MHz and 1.4~GHz, 
respectively, sub-divided into 32 spectral channels and with two polarizations 
and two-bit sampling. The on-source time was $\approx 2$~hours for each source.
One or more of the strong fringe finders 0438$-$436, 3C454.3, 3C84, 3C147, 
3C286 and 3C345 were observed during each run for bandpass calibration; phase 
referencing was not used. 

The VLBA data were also reduced in ``classic'' {\sc aips}, following 
procedures similar to those described in section~\ref{sec:gmrtwsrt}. 
However, for the VLBA data, the flux density scale was calibrated 
by using online measurements of the antenna gains and system temperatures, 
and the calibration steps included ionospheric corrections and fringe-fitting 
to determine the delay rates. Unfortunately, the ionospheric phase stability
was relatively poor during the 327~MHz and 606~MHz observations and only 
5-6 antennas could be retained during the self-calibration procedure. Further,
the final VLBA images were produced with phase self-calibration alone, again 
after the usual iterative procedure involving a number of self-calibration 
and imaging cycles. The core flux density of each source was measured using the 
task {\sc jmfit}, by fitting an elliptical gaussian model to the core radio emission.
The core emission was found to be fairly compact in most cases, with little 
extended emission; a single gaussian was hence used as the model for most sources. 
For three sources, adding more source components yielded a clear decrease in the 
$\chi^2$ and thus an appreciably better fit: a two-component model was hence used 
for two of these, 2003$-$025 and 1122$-$168, and a three-component model for 
0824+355.

The VLBA images are presented in Figure~\ref{fig:vlba}, in order of 
increasing right ascension, with the results summarized in Table~\ref{table:vlba}.
The columns of this table contain: (1)~the quasar name, (2)~the quasar emission redshift, 
(3)~the DLA redshift, (4)~the redshifted \hii\ line frequency, in MHz, (5)~the VLBA observing frequency 
$\nu_{\rm VLBA}$, in MHz, (6)~the total source flux density $S_{\rm tot}$ at the 
VLBA observing frequency, obtained from single-dish or low-resolution interferometer 
studies, (7)~the VLBA beam, 
in mas~$\times$~mas, (8)~the core flux density $S_{\rm VLBA}$, in Jy, measured 
from the VLBA image (for the three sources with multiple gaussian components, the most 
compact component was identified as the core), (9)~the deconvolved size of the 
radio emission, in mas~$\times$~mas, (10)~the spatial extent of the core radio emission 
at the DLA redshift, and (11)~the DLA covering factor $f$, assumed to be equal to the quasar 
core fraction, i.e. the ratio of the core flux density measured in the VLBA image $S_{\rm VLBA}$ 
to the total source flux density $S_{\rm tot}$ at the VLBA frequency (see Section~\ref{sec:tspin}). 
For systems without measurements of the total flux density at the VLBA observing frequency, 
$S_{\rm tot}$ was estimated by extrapolating from measurements at other frequencies. It should 
be emphasized that the estimates of the deconvolved size of the core emission and its spatial 
extent are upper limits, because any residual phase errors would increase the apparent size of 
the core emission. Finally, we note that error bars have not been included on the covering factor 
estimates. While we estimate that the errors on the VLBA flux density calibration are 
$\approx 10$\%, the VLBA and total flux density measurements were not carried out simultaneously, 
implying that individual covering factor estimates could have larger errors due to 
source variability.

\subsection{Optical spectroscopy}
\label{sec:optical}

\setcounter{table}{2}
\begin{table}
\tablewidth{0pc}
\begin{centering}
\caption{\label{table:abundances} Ionic column densities}
\begin{tabular}{lcccc}
\hline
\hline
Ion & $\lambda$ & $\log f$ & $\log [N_0/{\rm cm}^{-2}]^b$ & $\log [N/{\rm cm}^{-2}]^b$  \\
& ($\AA$) & &  &  \\ 
\hline
\multicolumn{5}{c}{(1) 0738+313, $z = 0.0912$, HIRES, $\log[N_{\rm HI}/{\rm cm}^{-2}] = 21.18 \pm 0.06$ }\\
\hline
Mg{\sc i}\\
& 2852.964 & $0.2577$ & $> 12.98$ & $> 12.98$  \\
Ca{\sc ii}\\
&3934.777 &$ -0.1871$ & $  12.28 \pm 0.03$&$  12.26 \pm 0.03$ \\
&3969.591 &$ -0.4921$ & $  12.18 \pm 0.08$ \\
Ti{\sc ii}\\
&3230.131 &$ -1.1630$ & $< 12.74$&$  12.51 \pm 0.04$	\\
&3384.740 &$ -0.4461$ & $  12.51 \pm 0.04$& \\
\hline
\hline
\multicolumn{5}{c}{(2) 0738+313, $z = 0.2212$, HIRES, $\log [N_{\rm HI}/{\rm cm}^{-2}] = 20.90 \pm 0.07$}\\
\hline
Mg{\sc i}\\
&2852.964 &$  0.2577$ & $> 12.46$&$> 12.46$ \\
Mg{\sc ii}\\
&2796.352 &$ -0.2130$ & $> 13.49$&$> 13.75$ \\
&2803.531 &$ -0.5151$ & $> 13.75$& \\
Ca{\sc ii}\\
&3934.777 &$ -0.1871$ & $  11.86 \pm 0.07$&$  11.92 \pm 0.06$ \\
&3969.591 &$ -0.4921$ & $  12.12 \pm 0.09$ \\
Ti{\sc ii}\\
&3230.131 &$ -1.1630$ & $< 12.58$ & $< 11.91$ \\
&3384.740 &$ -0.4461$ & $< 11.91$ & \\
Fe{\sc ii}\\
&2586.650 &$ -1.1605$ & $> 14.30$&$> 14.30$ \\
&2600.173 &$ -0.6216$ & $> 13.82$& \\
\hline
\hline
\multicolumn{5}{c}{(3) 0952+179, $z = 0.2378$, UVES, $\log [N_{\rm HI}/{\rm cm}^{-2}] = 21.32 \pm 0.05$}\\
\hline
Mg{\sc i}\\
&2852.964 &$  0.2577$ & $  12.68 \pm 0.03$&$  12.68 \pm 0.03$ \\
Mg{\sc ii}\\
&2796.352 &$ -0.2130$ & $> 13.87$&$> 14.13$ \\
&2803.531 &$ -0.5151$ & $> 14.13$& \\
Ti{\sc ii}\\
&3073.877 &$ -0.9622$ & $  12.62 \pm 0.03$&$  12.62 \pm 0.03$ \\
Mn{\sc ii}\\
&2576.877 &$ -0.4549$ & $  12.50 \pm 0.03$&$  12.58 \pm 0.03$ \\
&2594.499 &$ -0.5670$ & $  12.67 \pm 0.03$& \\
Fe{\sc ii}\\
&2586.650 &$ -1.1605$ & $14.52 \pm 0.03$$^a$&$ 14.52 \pm 0.03$ \\
&2600.173 &$ -0.6216$ & $> 14.09$& \\
\hline
\hline
\multicolumn{5}{c}{(4) 1127$-$145, $z = 0.3127$, UVES+STIS, $\log [N_{\rm HI}/{\rm cm}^{-2}] = 21.70 \pm 0.08$}\\
\hline
Mg{\sc i}\\
&2852.964 &$  0.2577$ & $13.19 \pm 0.03$ & $13.19 \pm 0.03$$^a$ \\
Mg{\sc ii}\\
&2796.352 &$ -0.2130$ & $> 14.07$&$  > 14.35$ \\
&2803.531 &$ -0.5151$ & $> 14.35$ & \\
Ca{\sc ii}\\
&3934.777 &$ -0.1871$ & $ 12.69 \pm 0.03$&$  12.70 \pm 0.03$\\
&3969.591 &$ -0.4921$ & $ 12.73 \pm 0.03$& \\
Mn{\sc ii}\\
&2576.877 &$ -0.4549$ & $ 13.30 \pm 0.03$&$  13.26 \pm 0.03$\\
&2594.499 &$ -0.5670$ & $ 13.20 \pm 0.03$& \\
&2606.462 &$ -0.7151$ & $ 13.23 \pm 0.03$& \\
Fe{\sc ii}\\
&2344.214 &$ -0.9431$ & $> 14.71$&$> 15.16$\\
&2374.461 &$ -1.5045$ & $> 15.16$& \\
&2382.765 &$ -0.4949$ & $> 14.32$& \\
&2586.650 &$ -1.1605$ & $> 14.92$& \\
&2600.173 &$ -0.6216$ & $> 14.43$& \\
Zn{\sc ii}\\
&2026.136 &$ -0.3107$ & $  13.57 \pm 0.05$ & $  13.57 \pm 0.05$\\
\hline
\hline
\end{tabular}
\end{centering}
\end{table}
\setcounter{table}{2}
\begin{table}
\tablewidth{0pc}
\begin{centering}
\caption{\label{table:abundances2} (contd.) Ionic column densities}
\begin{tabular}{lcccc}
\hline
\hline
Ion & $\lambda$ & $\log f$ & $\log [N_0/{\rm cm}^{-2}]^b$ & $\log [N/{\rm cm}^{-2}]^b$  \\
& ($\AA$) & &  &  \\ 
\hline
\multicolumn{5}{c}{(5) 0827+243, $z = 0.5247$, UVES, $\log [N_{\rm HI}/{\rm cm}^{-2}] = 20.30 \pm 0.04$} \\
\hline
Ca{\sc ii}\\
&3934.777 &$ -0.1871$ & $  12.59 \pm 0.03$&$  12.59 \pm 0.03$ \\
&3969.591 &$ -0.4921$ & $  12.60 \pm 0.03$& \\
Ti{\sc ii}\\
&3073.877 &$ -0.9622$ & $< 11.86$ & $11.82 \pm 0.04$ \\
&3230.131 &$ -1.1630$ & $< 11.97$ & \\
&3242.929 &$ -0.6345$ & $  11.55 \pm 0.12$& \\
&3384.740 &$ -0.4461$ & $  11.92 \pm 0.04$& \\
Fe{\sc ii}\\
&2249.877 &$ -2.7397$ & $  14.79 \pm 0.05$ &$  14.84 \pm 0.03$ \\
&2260.781 &$ -2.6126$ & $  14.96 \pm 0.03$& \\
&2344.214 &$ -0.9431$ & $> 14.66$& \\
&2374.461 &$ -1.5045$ & $  14.84 \pm 0.03$& \\
&2382.765 &$ -0.4949$ & $> 14.43$& \\
\hline
\hline
\multicolumn{5}{c}{(6) 0311+430, $z = 2.2890$, GMOS, $\log [N_{\rm HI}/{\rm cm}^{-2}] = 20.30 \pm 0.11$}\\
\hline
Zn{\sc ii}\\
&2026.136 & $-0.3107$  & $< 12.5$ & $< 12.5$ \\
Cr{\sc ii}\\
& 2056.254 & $-0.9788$ & $< 13.05$ & $< 13.05$ \\
Mn{\sc ii}\\
& 2594.499 & $-0.5670$ & $12.37 \pm 0.11$ & $12.50 \pm 0.07$ \\
& 2606.462 & $-0.7151$ & $12.63 \pm 0.09$ \\
Fe{\sc ii}\\
&2249.877 &$ -2.7397$  & $14.80 \pm 0.10$ & $14.86 \pm 0.06$ \\
&2260.781 &$ -2.6126$  & $14.91 \pm 0.05$ \\
&2374.461 &$ -1.5045$  & $ > 14.32$ \\
\hline
\hline
\end{tabular}
\vskip -0.1in
\tablenotetext{a}{Note that the Fe{\sc ii}$\lambda$2586 line in the $z=0.2378$ DLA towards 0952+179 
and the Mg{\sc i} line in the $z=0.3127$ DLA towards 1127$-$145 are both close to saturation. If the 
lines are saturated, the quoted column densities would be lower limits.}
\tablenotetext{b}{The column densities measured from individual transitions are listed in the
penultimate column ($\log[N_0/{\rm cm}^{-2}]$), while the adopted column density for
the species is listed in the final column ($\log[N_0/{\rm cm}^{-2}]$). The adopted column
density of a given species was obtained by a weighted mean of the measured column densities 
from all unsaturated lines. All limits are at $3\sigma$ significance.}
\end{centering}
\end{table}

\subsubsection{VLT-UVES spectroscopy}
\label{sec:uves}

Very Large Telescope Ultraviolet Echelle Spectrograph (VLT-UVES) data were obtained as part 
of the programme 67.A-0567(A) (PI: Lane) for 1127$-$145 in May/June 2001. Five 3060-second 
exposures 
were obtained with both UVES arms in the 346+580~nm standard wavelength setting, with 
a 1.0~arcsec slit.  

Besides the above UVES observations, we also downloaded UVES data on a few DLAs of the sample from 
the VLT archive, along with the requisite calibration files obtained within a few nights of the 
science exposures.  The instrumental calibration plan dictates how frequently the calibration 
exposures are obtained, based on knowledge of the instrument stability.  The blue arm data downloaded 
by us on 0952+179 as part of program 69.A-0371(A) (PI: Savaglio) consist of three 5700~sec 
exposures with the 346~nm central wavelength setting. Data for 0827+243 were obtained as part 
of program 68.A-0170(A) (PI: Mallen-Ornelas) with both UVES arms in the 346+564 standard 
settings and four one-hour exposures.  In all cases, the data were obtained with a 
1.0~arcsec slit, with the detector binned by 2 in both the spatial and spectral directions.

The UVES data of 1127$-$145 were reduced with the UVES MIDAS pipeline version 2.1.0\footnote{UVES 
pipelines are available at http://www.eso.org/sci/data-processing/software/pipelines/}, while
the archival UVES data were reduced using a custom version of the UVES pipeline, based on 
Midas reduction routines. All data reduction followed the standard procedure for UVES data 
\citep[e.g.][]{ellison01c}, with wavelength calibration and normalization carried out for 
the individual spectra, before the spectra were optimally combined to produce the final 
one-dimensional spectrum on a heliocentric wavelength scale.

\subsubsection{HST-STIS spectroscopy of 1127$-$145}
\label{sec:1127}

The Space Telescope Imaging Spectrograph (STIS) on the Hubble Space Telescope (HST) was 
used in the E230M echelle mode in proposal 9173 (PI: J. Bechtold) to obtain an ultraviolet 
spectrum of 1127$-$145. These data were downloaded from the HST archive. The total 
exposure time on target amounted to 52,380~s with the $0.2''\times0.2''$ aperture. The HST-STIS 
data were reduced with the IDL version of CalSTIS v7.0 \citep{lindler99}, following standard 
procedures, combining data from different exposures with weights inversely proportional to
their variance.

\subsubsection{GMOS observations of 0311+430}
\label{sec:0311}

The $z = 2.2890$ DLA towards 0311+430 \citep{york07} was observed on 14 November 2007 
with the Gemini Multi-Object Spectrograph (GMOS) on the Gemini North telescope
(in proposal GN-2007B-Q-76).
The quasar was observed in nod and shuffle mode with the R600 grating, a
central wavelength of 800 nm and a $0.75''$slit.  The CCD was
binned by two pixels in both the spatial and spectral directions.  In
nod and shuffle mode, the target is observed at two spatial positions
on the slit in an ABBA configuration.  The exposure time at each
position was 30 seconds (120 seconds per cycle) with 19 cycles per
exposure.  Each exposure therefore has an effective integration time
of 2280 seconds, with 6 exposures taken in all.

The data were reduced using the Gemini IRAF package following standard
procedures, but with special modifications for the nod and shuffle
mode.  First, a bias was constructed, with an interactive fit to the
overscan region, using {\sc gbias}.  A dark calibration frame was
constructed using {\sc gnsdark} and subtracting the bias frame from the
previous step.  {\sc gsflat} was used to make the master flat field
from individual lamp exposures with interactive polynomial fits to
remove the underlying spectral shape of the illumination.  The
application of these calibration frames (bias, dark and flat) to the
science data was achieved using {\sc gsreduce}.  Sky subtraction is done
using {\sc gnsskysub} on the 2-dimensional (2D) spectra by subtracting
the shuffled image pixels from those in the unshuffled position.  This
step in the procedure is one of the great benefits of the nod and
shuffle mode, and results in superior subtraction with much lower sky
residuals than traditional long slit spectroscopy.  For our
reductions, with multiple exposures, {\sc gnsskysub} is called as part
of the {\sc gnscombine} routine which applies offsets to the various
dithers and yields a single 2D spectrum.  After the combination, the
2D frame has a positive and a negative spectrum at position A and B
respectively.  A first order wavelength correction (based on the image
header information) of the science image and mosaicing of the 3 CCDs
is achieved by running {\sc gsreduce} again.  {\sc gsreduce} is also
used to apply the master calibration files to the arc frames, before
the wavelength solution is determined and applied to the science frame
using {\sc gswavelength} and {\sc gstransform} respectively.  The 1-D
spectra (one positive and one negative) were then extracted using
{\sc gsextract} and averaged together after inversion of the negative
spectrum.  The final spectrum (obtained after binning by 2 in both spatial
and spectral directions) has a wavelength coverage of $6570-9450 \AA$,
a FWHM resolution of $\sim 3.1 \AA$ and a signal-to-noise ratio (S/N) 
per pixel of $30-50$.

\subsubsection{Keck- HIRES  spectroscopy}
\label{sec:keck}

The $z = 0.0912$ and $z = 0.2212$ DLAs towards 0738+313 were observed on 
26 December 2006 with the Keck/ HIRES  spectrometer \citep{vogt94}, configured with 
the blue cross-dispersor (i.e.\ HIRESb).  We employed the E3 decker which affords 
a spectral resolution of FWHM~$\approx 4$~\kms.  The grating was tilted to
XDANGL=1.039 giving a nearly continuous wavelength coverage from 
$\lambda \approx 3100-5900 \AA$, but with two gaps owing to the CCD mosaic. 
The science frames (two exposures at 900s each) and associated calibration files 
were reduced with the HIRedux data reduction 
pipeline\footnote{http://www.ucolick.org/$\sim$xavier/HIRedux/index.html} 
\citep[for details, see][]{prochaska07}.  The spectra were optimally extracted
and co-added (binning by 2 in the spatial direction), corrected to vacuum 
wavelengths and the heliocentric reference frame.  The S/N at $\lambda \approx 4000 \AA$ is
approximately 13 per 1.3~\kms\ pixel.

\subsubsection{Abundances and velocity widths}
\label{sec:v90}

The Keck-HIRES, VLT-UVES, GMOS-N and HST-STIS spectra were used to derive column densities and 
elemental abundances for six DLAs, at $z = 0.0912$ and $z = 0.2212$ towards 0738+313, $z = 0.2378$ 
towards 0952+179, $z = 0.3127$ towards 1127$-$145, $z = 0.5247$ towards 0827+243 and 
$z = 2.2890$ towards 0311+430. For all absorbers, the column densities for different species were 
estimated from the observed line profiles using the apparent optical depth method \citep[AODM;][]{savage91}. 

For the $z = 0.3127$ DLA towards 1127$-$145, the weakness of the Zn{\sc ii} lines in the HST-STIS 
spectrum prompted us to also try a different route. Here, the strong absorption lines in the VLT-UVES 
spectrum were first analysed with {\sc fitlyman} \citep{fontana95}, with Voigt profiles first fitted 
to the Ca{\sc ii} doublets alone, in order to determine the turbulent Doppler parameters and the 
redshifts of the individual components. These values were then fixed, together with the column density of 
the individual Ca{\sc ii} doublets, and the Zn{\sc ii} and Mn{\sc ii} lines included in the fit to 
determine the Zn{\sc ii} column density from the HST-STIS spectrum and the Mn{\sc ii} column density 
from the VLT-UVES spectrum. This approach yielded $\log[N_{\rm ZnII}/{\rm cm}^{-2}] = 13.45 \pm 0.08$, 
consistent with the AODM value of $\log[N_{\rm ZnII}/{\rm cm}^{-2}] = 13.57 \pm 0.05$; the AODM value will 
be used in the following discussion, as it involves fewer assumptions. We note, in passing, that the 
Voigt profile fit to the Mg{\sc i} line yielded a larger Mg{\sc i} column density than that obtained 
from the AODM, with $\log[N_{\rm MgI}/{\rm cm}^{-2}] = 13.95 \pm 0.07$. This may be because the 
Mg{\sc i} line is nearly saturated in this absorber; in such situations, the AODM estimate should 
be regarded as a lower limit

\begin{figure}
\centering
\epsfig{file=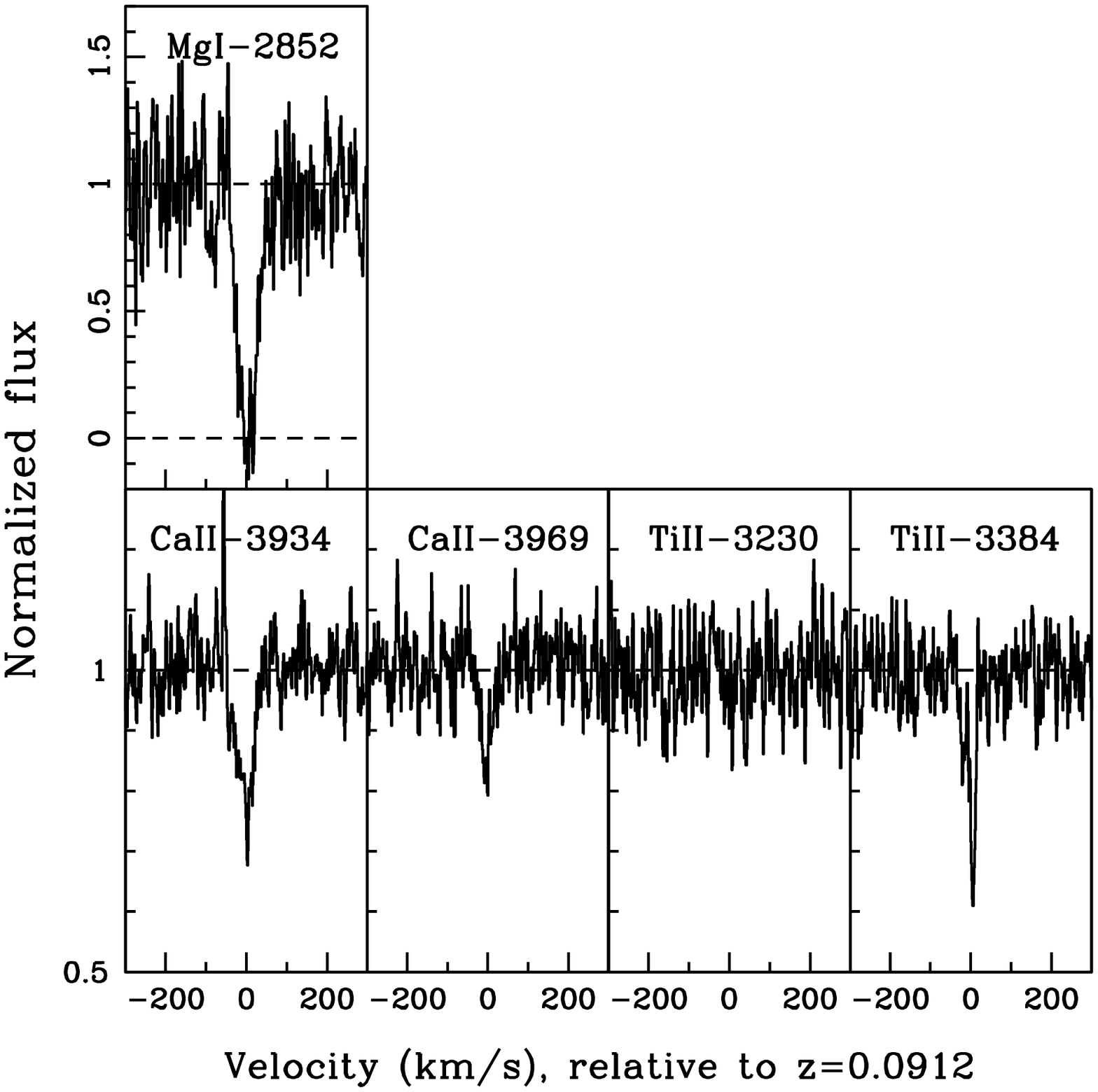,height=3.3in}
\caption{Low-ionization metal absorption profiles from the $z = 0.0912$ DLA 
towards 0738+313, from the Keck-HIRES spectrum. The x-axis represents velocity, 
in \kms, relative to the DLA redshift. }
\label{fig:0738lowz}
\end{figure}

\begin{figure}
\centering
\epsfig{file=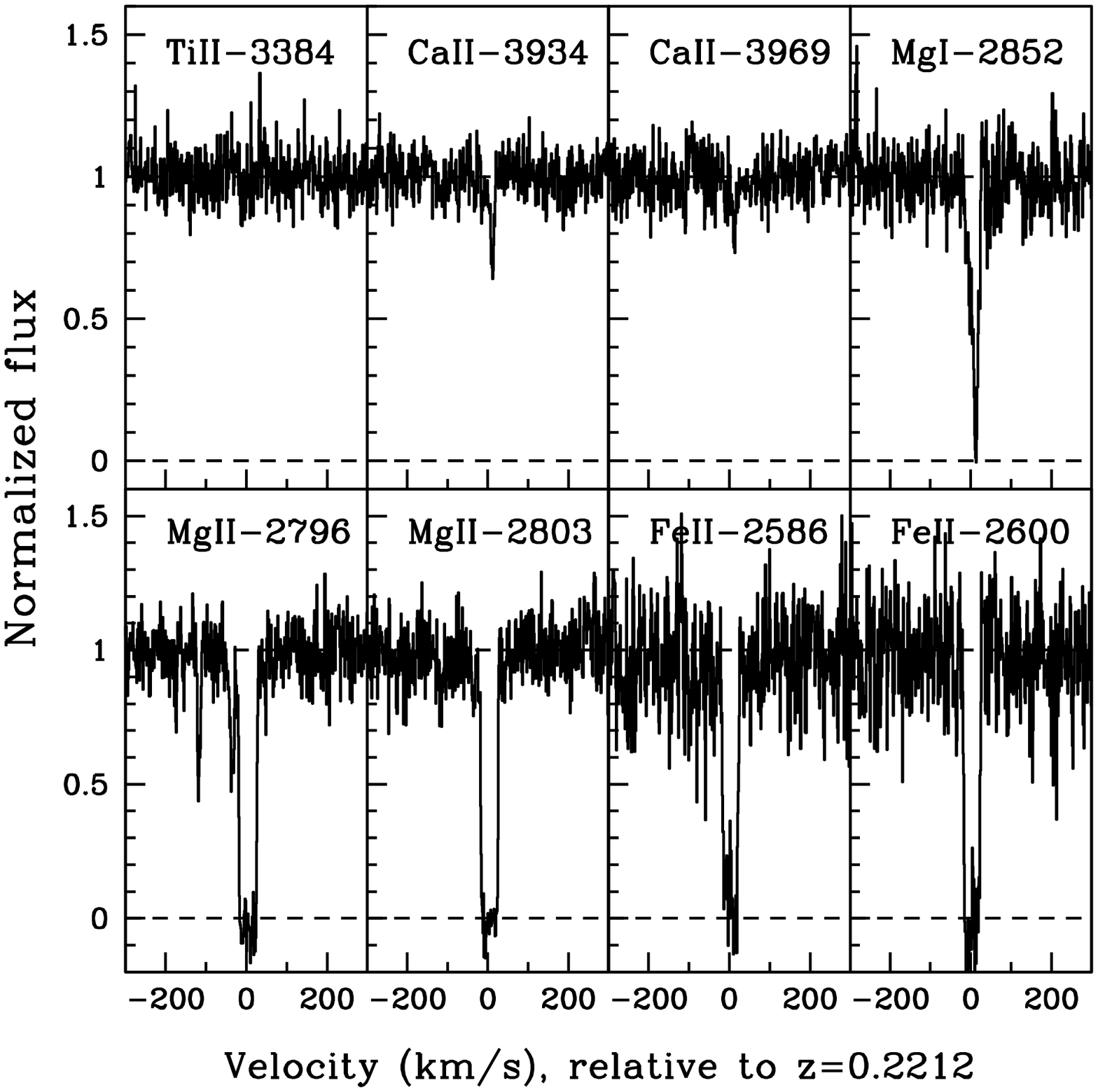,height=3.3in}
\caption{Low-ionization metal absorption profiles from the $z = 0.2212$ DLA 
towards 0738+313, from the Keck-HIRES spectrum. The x-axis contains velocity, 
in \kms, relative to the DLA redshift. }
\label{fig:0738highz}
\end{figure}

\begin{figure}
\centering
\epsfig{file=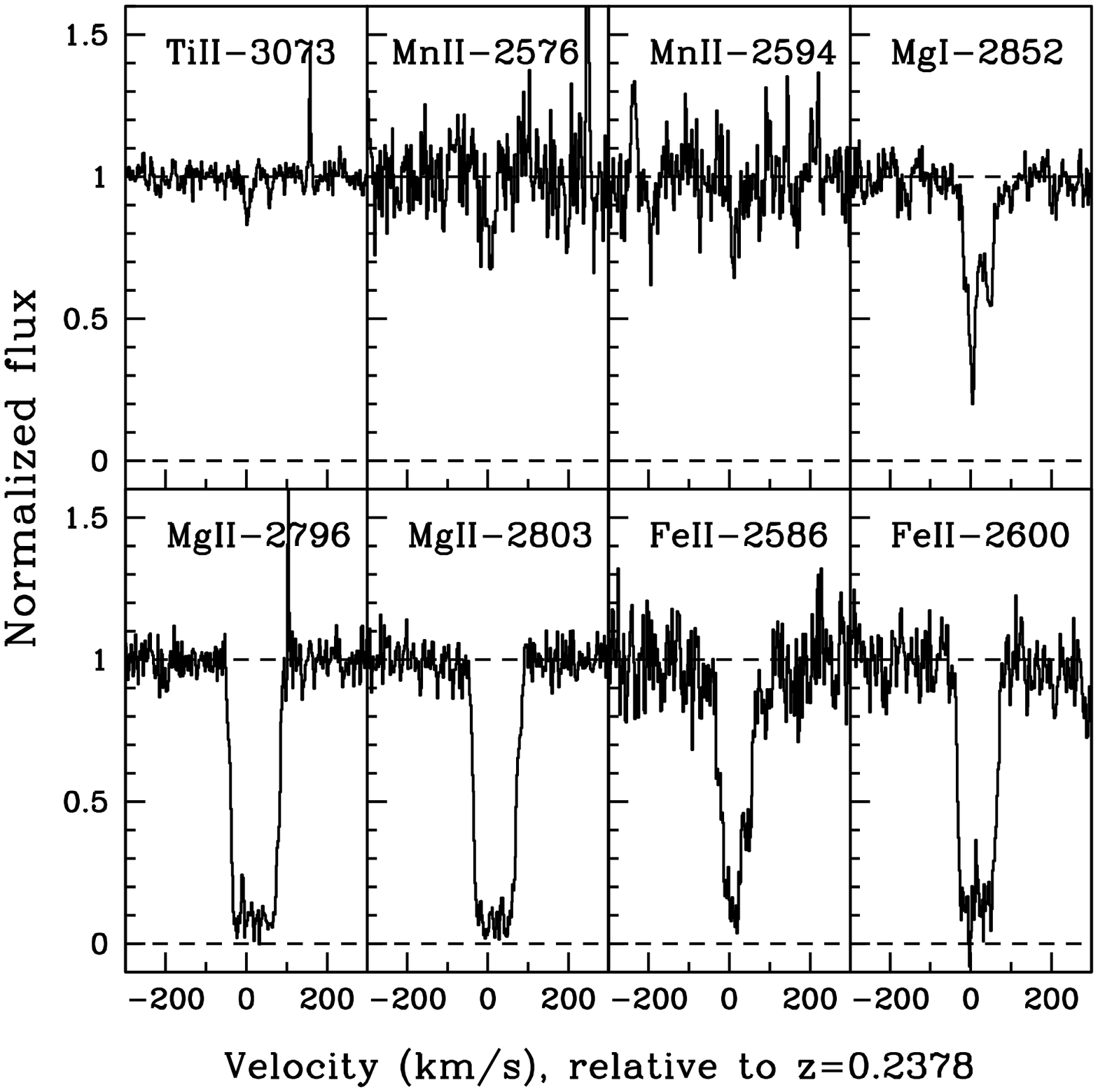,height=3.3in}
\caption{Low-ionization metal absorption profiles from the $z = 0.2378$ DLA 
towards 0952+179, from the VLT-UVES spectrum. The x-axis contains velocity, 
in \kms, relative to the DLA redshift. }
\label{fig:0952}
\end{figure}

\begin{figure}
\centering
\epsfig{file=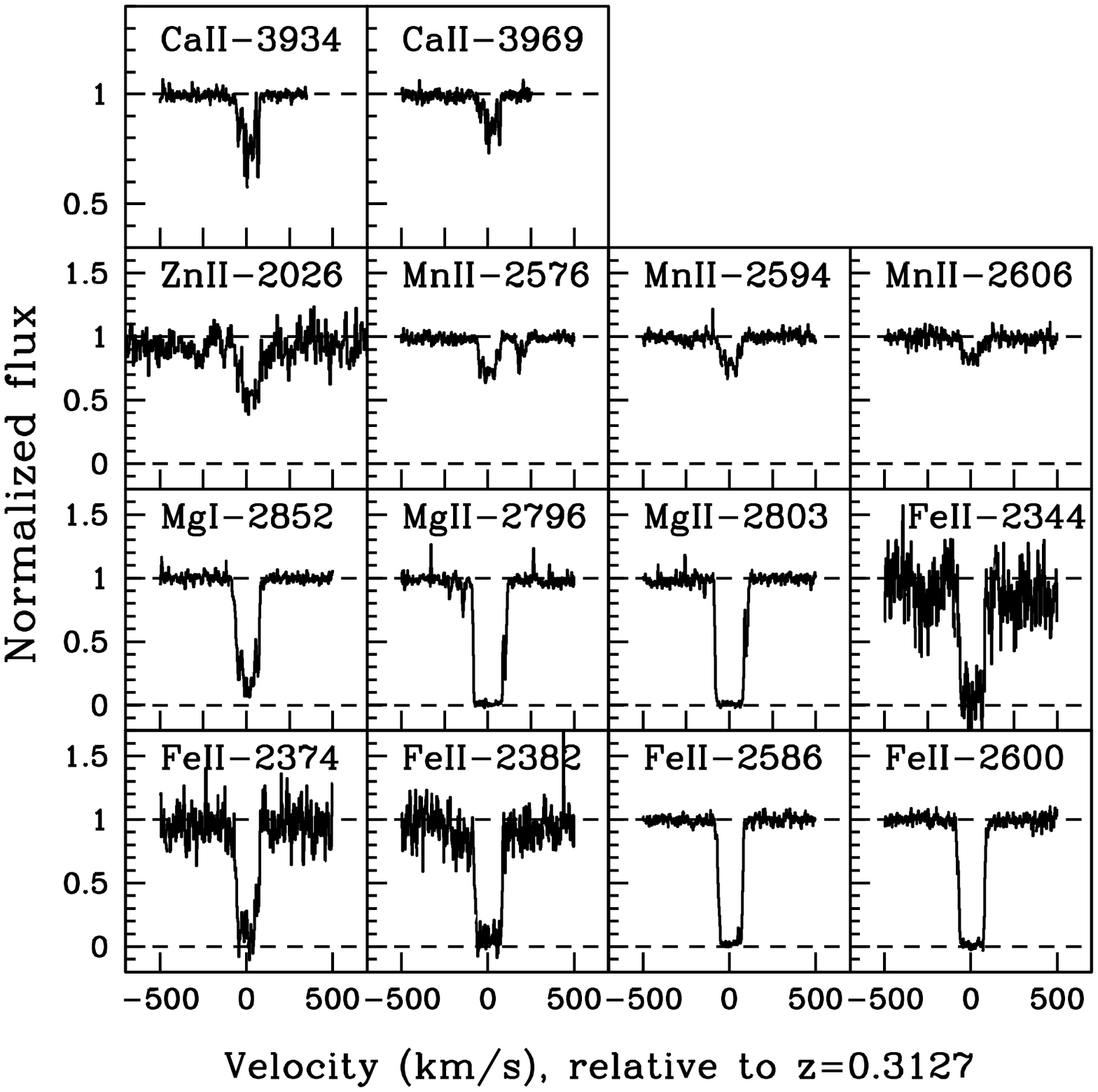,height=3.3in}
\caption{Low-ionization metal absorption profiles from the $z = 0.3127$ DLA 
towards 1127$-$145, from the VLT-UVES and HST-STIS spectra. The x-axis contains velocity, 
in \kms, relative to the DLA redshift.}
\label{fig:1127}
\vskip -0.1in
\end{figure}

\begin{figure}
\centering
\epsfig{file=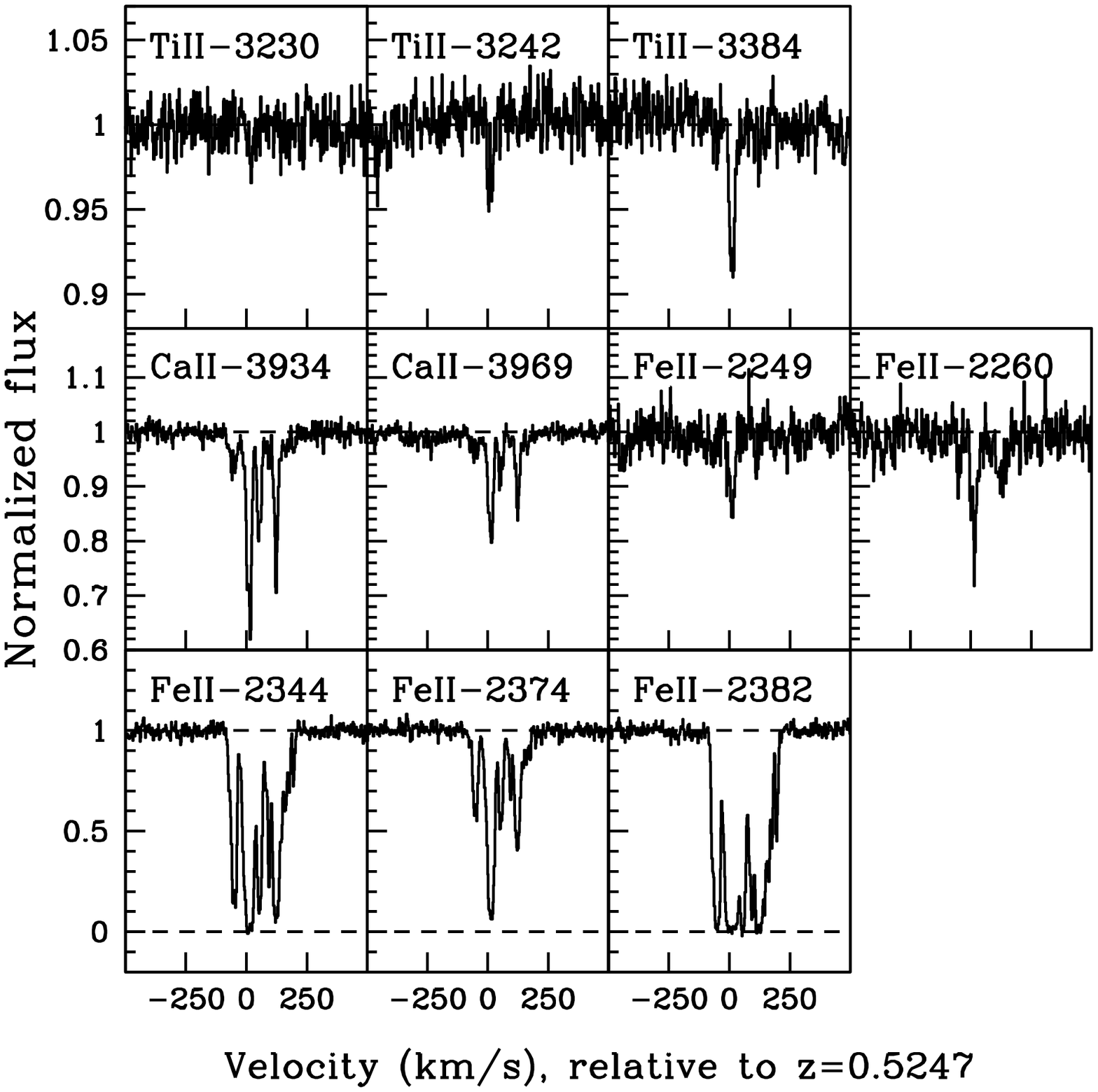,height=3.3in}
\caption{Low-ionization metal absorption profiles from the $z = 0.5247$ DLA 
towards 0827+243, from the VLT-UVES spectrum. The x-axis contains velocity, 
in \kms, relative to the DLA redshift. }
\label{fig:0827}
\vskip -0.1in
\end{figure}

\begin{figure}
\centering
\epsfig{file=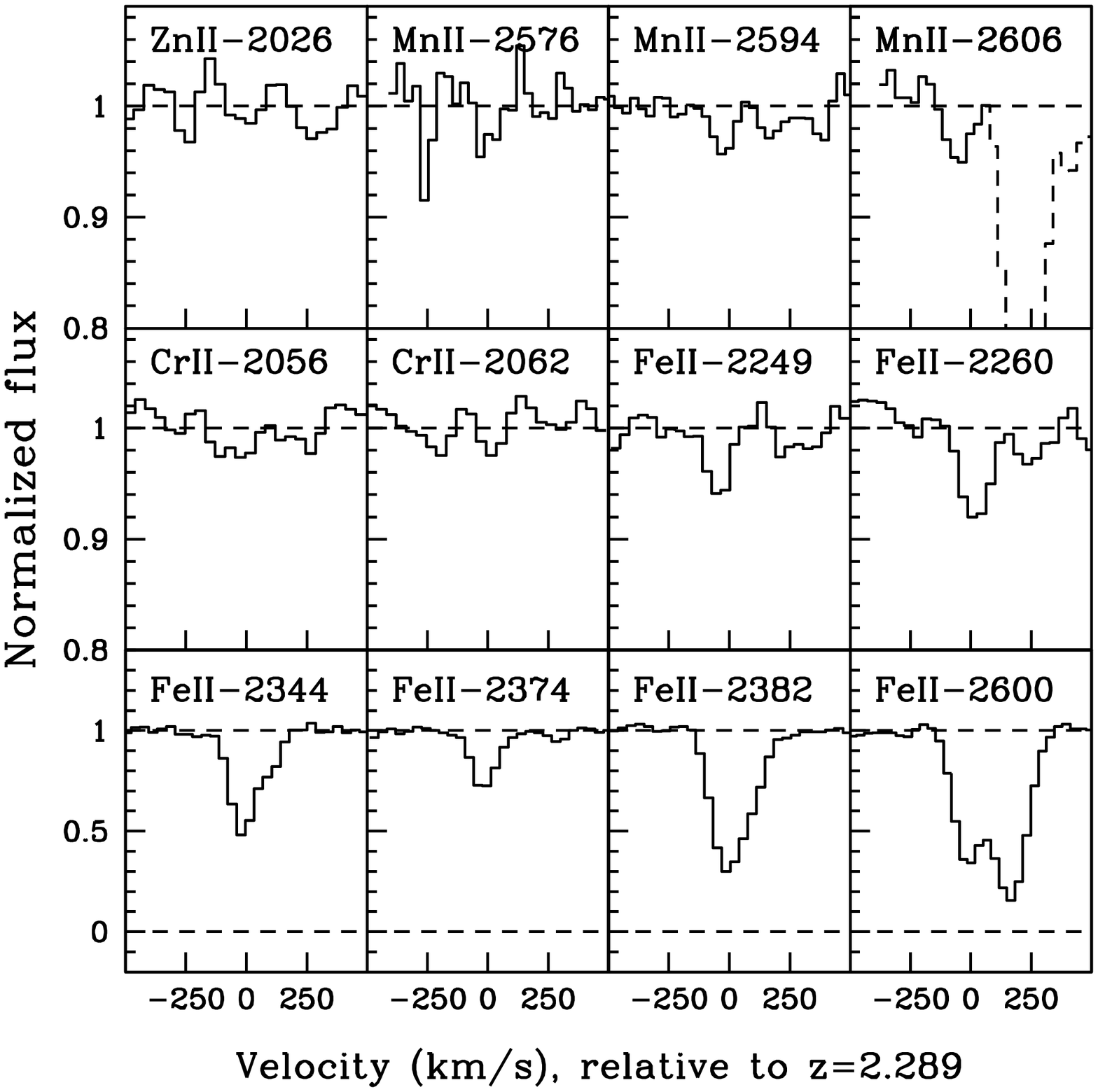,height=3.3in}
\caption{Low-ionization metal absorption profiles from the $z = 2.2890$ DLA 
towards 0311+430, from the GMOS-N spectrum. The x-axis contains velocity, 
in \kms, relative to the DLA redshift. The dashed part of the spectrum indicates
absorption from a different redshift, which is slightly blended with the Mn{\sc ii}$\lambda$2606
line from the $z=2.289$0 DLA. }
\label{fig:0311}
\end{figure}

Table~\ref{table:abundances} lists the column densities derived for various species in the six DLAs, 
along with the \hi\ column density of the absorber. The abundances, relative to the solar abundance, 
on the solar scale of \citet{asplund09}, are summarized for each system in Section~\ref{sec:notes}. 
The metal lines detected in each spectrum (and, in some cases, a few undetected transitions) are 
plotted in Figs.~\ref{fig:0738lowz}~$-$~\ref{fig:0311}.

Finally, we measured the velocity width at 90\% optical depth, $\Delta V_{\rm 90}$,
of the low-ionization metal lines detected in our Keck- HIRES  or VLT-UVES spectra of a 
subset of the absorbers of our sample following the criteria and procedures described in 
\citet{prochaska97}. In the majority of cases, we analyzed an unsaturated, low-ion 
transition observed at high S/N. For $\approx 5$ systems, however, this did not prove 
possible (e.g.\ the DLA at $z=0.2212$ toward 0738+313). In these cases, we measured 
$\Delta V_{\rm 90}$ from a modestly saturated line or a transition from a non-dominant 
ion (e.g. \mgone).  Inspection of the full set of line profiles for 
these systems indicate that the results should be representative of the gas kinematics.
The derived $\Delta V_{\rm 90}$ values are listed in Table~\ref{table:main}, along with 
the transition used for the estimate.

\section{Spin temperatures, covering factors and velocity widths}
\label{sec:tspin}

For a DLA towards a radio-loud background quasar, the \hi\ column density 
$\nhi$, the \hii\ optical depth $\tau_{\rm 21cm}$ and the spin temperature $\ts$ 
are related by the expression \citep{rohlfs06}
\begin{equation}
\label{eqn:tspin0}
{\rm N_{HI}} = C_0 \times \ts \times \int \tau_{\rm 21cm} {\rm dV} \;\;,
\end{equation}
\noindent where $\nhi$ is in \cm, $\ts$ is in K, the integral $\int \tau_{\rm 21cm} {\rm dV}$ 
is over the observed line profile in velocity space, in \kms, and the constant 
$C_0 = 1.823 \times 10^{18}$~\cm~K$^{-1}$~km$^{-1}$~s. Note that the \hii\ optical depth
is given by $\tau_{\rm 21cm} = -\log\left[ 1 - \Delta S/S \right]$, where $\Delta S$ is the line depth 
and $S$ is the flux density of the background quasar.

Equation~\ref{eqn:tspin0} implicitly assumes that the all quantities are measured at the same 
spatial resolution. However, in the case of DLAs, the \hi\ column density is measured from the 
redshifted Lyman-$\alpha$ absorption line towards the optical quasar, i.e. along a narrow pencil 
beam through the absorber. Conversely, the \hii\ optical depth is measured against the 
more-extended radio emission. Ideally, the \hii\ absorption studies should be carried out with 
very long baseline interferometry, so that the radio and optical sightlines have similar transverse 
extents in the absorber. Unfortunately, current VLBI arrays such as the VLBA or the European 
VLBI Network have both very poor low-frequency coverage and low sensitivity. Further, even the 
1000~km baselines of the next-generation Square Kilometre Array would only give a spatial resolution 
of $\approx 1$~kpc at the redshift of a DLA at $z \approx 2$ \citep[e.g.][]{kanekar04}, far worse 
than the ``effective'' resolution of hundreds of AU allowed by optical absorption spectroscopy 
(arising from the size of the background quasar). As a result, \hii\ absorption studies are 
likely to continue to probe conditions on far larger spatial scales than optical absorption studies 
in the foreseeable future. One must hence attempt to correct for resolution effects, as has been done 
in the present and earlier studies \citep[e.g.][]{briggs83,carilli96,kanekar09a,ellison12}.
This is discussed in detail below.
 
There are two effects that must be taken into account in the case of \hii\ absorption studies at 
low angular resolution. First, for extended radio structure in the quasar, the foreground absorber may 
cover only part of the radio emission, but the angular extent of the emission may be smaller than the 
telescope beam and hence unresolved. In such cases, if the (unresolved) total quasar flux density 
measured by the observer is $S$~Jy, but only a fraction $f$ of this emission is covered by the 
foreground absorber, the true \hii\ optical depth is larger than the observed optical depth. The 
\hii\ optical depth corrected for covering factor effects is given by 
$\tau_{\rm 21cm} = -\log \left[1 - \Delta S/(fS) \right]$ and equation~(\ref{eqn:tspin0})
is modified to
\begin{equation}
\label{eqn:tspin1}
{\rm N_{HI}} = 1.823 \times 10^{18} \times \ts \times \int -\log \lsb 1 - \frac{\Delta S}{fS} \rsb {\rm dV} \;\;,
\end{equation}
\noindent where $f$ is referred to as the DLA covering factor. For most DLAs, the \hii\ optical depth 
is low, $\tau_{\rm 21cm} << 1$, and the expression 
reduces to 
\begin{equation}
\label{eqn:tspin}
{\rm N_{HI}} = 1.823 \times 10^{18} \times \lsb \ts/f \rsb \int \frac{\Delta S}{S} {\rm dV} \;\;.
\end{equation}

\noindent The covering factor $f$ is included in the above equation to account for the fact that 
low-frequency radio emission is often very extended, implying that the foreground DLA may not 
cover the entire background radio emission. The angular resolution of redshifted \hii\ absorption 
studies (with single dishes or short-baseline interferometers like the GMRT or WSRT) is typically 
quite poor ($\gtrsim 10''$ for even the GMRT, i.e.  $\gtrsim 80$~kpc for a DLA at $z \sim 3$), 
and the measured \hii\ optical depth is hence a lower limit, because some of the radio emission 
may not be covered by the absorber. \citet{curran05} have emphasized unknown covering factors as 
a critical issue in interpreting the high spin temperatures typically obtained in \hii\ absorption 
studies of high-$z$ DLAs. This is because an apparently-low \hii\ optical depth could arise 
due to an unknown low covering factor, a high spin temperature or an over-estimated 
\hi\ column density. 

The covering factor can be estimated from very long baseline interferometric (VLBI) 
studies close to the redshifted \hii\ line frequency \citep{briggs83,kanekar09a}, to 
measure the compact flux density arising from the radio core at this frequency. 
The ratio of the core flux density to the total source flux density gives the quasar 
core fraction. In the following, we will  assume that this quantity is a reasonable 
estimate of the covering factor. We note that this is not formally a measurement of the 
covering factor for an individual absorber because (1)~it is possible that the absorber 
covers some fraction of the extended emission, in 
addition to the core emission, and (2)~it is also possible that the absorber does not entirely 
cover the radio core. The latter is unlikely if the core size is significantly smaller than 
the size of a typical galaxy at the absorber redshift. Thus, for core sizes~$\lesssim 1$~kpc, the 
ratio of the core flux density to the total source flux density should provide at least a 
lower limit to the covering factor. 

We note, in passing, that the results of the paper are derived under the assumption that 
the DLA covering factor is equal to the quasar core fraction. However, in 
Section~\ref{sec:f-core}, we consider the possibility that the core fraction and covering 
factor are entirely unrelated, and use Monte Carlo simulations to show that the hypothesis 
that they are equal is actually not required for the results described below.

Until quite recently, there were very few estimates of DLA covering factors from low-frequency 
VLBI studies. This is because such studies are technically challenging due to propagation 
effects in the atmosphere, which can result in decorrelation of the signals on the long 
VLBI baselines. However, over the last few years, covering factors have been estimated for 
a large number of high-$z$ DLAs, mostly with the low-frequency receivers of the VLBA 
\citep[e.g.][; this work]{kanekar09a,ellison12,srianand12,kanekar13}.

The second issue that must be taken into account in the case of \hii\ absorption studies 
at low angular resolution is the fact that, for DLAs, the \hi\ column density in 
equation~(\ref{eqn:tspin}) is determined from the Lyman-$\alpha$ profile. Using this to 
derive the spin temperature involves the assumption that the \hi\ columns along the optical 
and radio sightlines are the same. Since the optical quasar is far smaller than the radio 
core (typically, sizes of $100-1000$~AU at optical wavebands and $\gtrsim 10$~pc in the radio), 
the \hi\ column densities can, in principle, be different along the two sightlines. This 
assumption will be examined in detail in Section~\ref{sec:lmc}.

Next, for neutral gas along the line of sight at different temperatures, the spin 
temperature derived from equation~(\ref{eqn:tspin}) is the column-density-weighted 
harmonic mean of the spin temperatures of the different phases, i.e.
\begin{equation}
\label{eqn:harmonic}
\frac{1}{\ts} = \Sigma_i  \frac{n_i}{{\rm T_{s,i}}} \;\;,
\end{equation}
\noindent where $n_i = N_i/\nhi$ is the fraction of the total \hi\ column density in the $i^{\rm th}$
cloud, and ${\rm T_{s,i}}$ is its spin temperature. Thus, a measurement of $\ts = {\rm T}$ along a sightline 
could arise due to the presence of either a single intervening \hi\ ``cloud'' with this spin 
temperature or a combination of clouds with spin temperatures above and below the measured value, 
${\rm T}$. For example, 
a measured spin temperature of $\approx 1000$~K does not rule out the classic ``two-phase'' model 
of the neutral Galactic ISM \citep{field69,mckee77}, as this would arise naturally for sightlines
with two types of \hi\ ``clouds'', at temperatures of $\approx 100$~K and $\approx 8000$~K.

It should be emphasized that the spin temperature along multi-phase sightlines 
is biased towards cold gas. Specifically, a sightline with \hi\ equally divided between cold
and warm phases, at spin temperatures of $\ts = 100$~K and $\ts = 8000$~K, respectively, 
would yield a derived spin temperature of $\approx 180$~K. Conversely, a sightline with 
90\% of the \hi\ at $\ts = 8000$~K, and 10\% at $\ts = 100$~K would yield a derived spin 
temperature of $\approx 900$~K. Further, collisions drive the spin temperature towards the 
gas kinetic temperature \citep[$\approx 40-200$~K;]{wolfire95} in the cold neutral medium (CNM).
However, in the warm neutral medium (WNM), the number density is too low to thermalize the 
transition and the spin temperature is typically significantly lower than the kinetic temperature
$\tk$ \citep[e.g. $\ts \approx 1000-4000$~K for $\tk \approx 5000-8000$~K;][]{liszt01}. The 
result of both these effects is that a high measured spin temperature ($\gtrsim 1000$~K) can 
only be explained by a preponderance of neutral gas in the WNM. The only caveat to this statement 
is if the \hi\ column density measured towards the optical quasar is significantly larger than
that towards the radio core, which, as we will discuss in Sec.~\ref{sec:lmc}, may occur on 
individual sightlines but is very unlikely to arise in a systematic manner.

For DLAs that do not show detectable \hii\ absorption, one can use the upper limit on the 
\hii\ optical depth in equation~(\ref{eqn:tspin}) to obtain a lower limit to the spin temperature.
The derived limit depends on the assumed shape of the velocity profile; further, the 
detection sensitivity worsens with increasing line width ($\propto \sqrt{\Delta V}$, where 
$\Delta V$ is the line FWHM), implying that significant non-thermal broadening makes it 
harder to detect \hii\ absorption. For non-detections, we will in all cases assume a 
gaussian profile of FWHM=$15$~\kms, corresponding to thermally-broadened \hi\ at a kinetic
temperature of 5000~K (i.e. in the WNM range). This is a conservative strategy as it allows 
for significant non-thermal broadening of cold \hi\ along the line of sight. We also note that
individual \hii\ spectral components in DLAs with detected \hii\ absorption typically have 
FWHMs of $5-15$~\kms.

\section{Proximate DLAs and sub-DLAs}
\label{sec:pdlas}

Two sub-classes of damped absorbers, proximate DLAs (``PDLAs'') and sub-DLAs, merit special 
mention \citep[see also][]{ellison12}. PDLAs are defined as absorbers with redshifts within 
$\approx 3000$~\kms\ of the quasar redshift, i.e. arising in gas associated with the quasar 
host galaxy
\citep[e.g.][]{moller98,ellison02,prochaska08b}. For such systems, the population distribution in the hyperfine 
levels, and thus the spin temperature, will be significantly influenced by the quasar radiation 
field at the \hii\ line frequency \citep[e.g.][]{field58,wolfe75}. Since this complicates 
the interpretation of the derived spin temperatures, we will exclude PDLAs from the 
later analysis and discussion.

Sub-DLAs are absorbers with \hi\ column densities below the defining DLA column density of 
$\nhi = 2 \times 10^{20}$~\cm, but which still show damping wings in the Lyman-$\alpha$ 
profile \citep[e.g.][]{peroux03}. Numerical estimates of self-shielding against the UV background 
in DLAs and sub-DLAs suggest that the absorbers are physically distinct: these studies find 
that most of the gas in DLAs is neutral, while that in sub-DLAs is predominantly ionized and at 
high temperatures \citep{viegas95,prochaska99b,wolfe05,milutinovic10}. In keeping with the above, recent Galactic 
\hii\ absorption studies have shown that the CNM fraction increases sharply at 
$\nhi \approx 2 \times 10^{20}$~\cm, with very low CNM fractions below this threshold 
\citep{kanekar11b}. Sub-DLAs are thus likely to show significantly higher spin temperatures 
than DLAs; we will hence exclude them from the full sample and the later analysis.

\section{Notes on individual sources}
\label{sec:notes}

The \hii\ optical depths listed in Table~\ref{table:21cm} and the covering factors of Table~\ref{table:vlba} 
and \citet{kanekar09a} were used to derive spin temperatures for the DLAs and sub-DLAs of our sample whose \hii\ 
data were not affected by RFI. The elemental abundances of Table~\ref{table:abundances} were used to 
complement and, in some cases, to improve the metallicities given in the literature. This section briefly 
summarizes the results for each absorber for which new data have been presented in this paper, again in 
order of increasing right ascension.

\begin{figure}
\centering
\epsfig{file=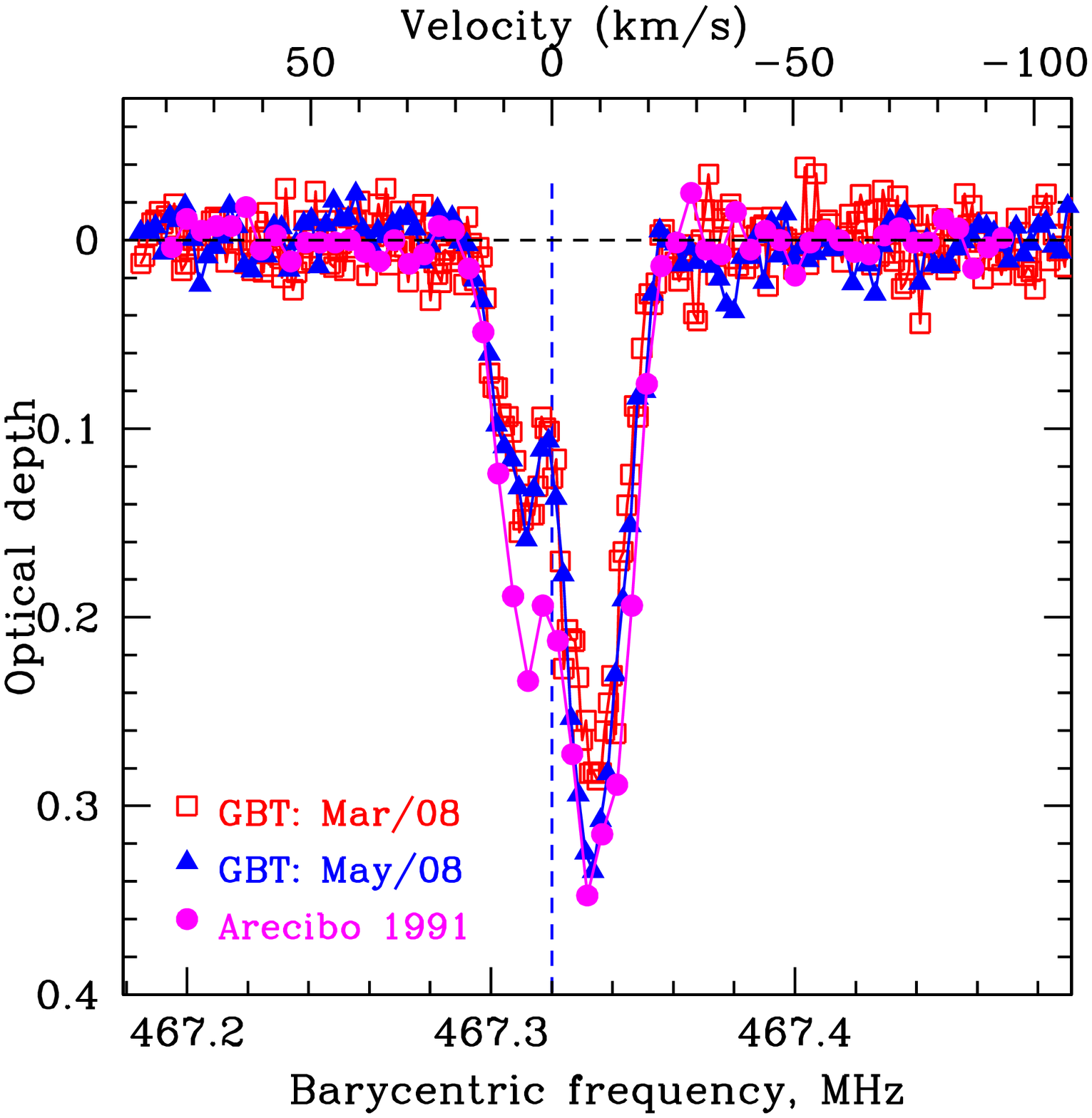,height=3.4truein,width=3.4truein}
\caption{\label{fig:0458} 
Comparison between the \hii\ absorption profiles obtained from the $z = 2.0395$ DLA towards 0458$-$020 
with the Arecibo telescope (in 1991) and the GBT telescope (in March and May 2008).}
\end{figure}

\setcounter{table}{3}
\begin{table*}
\caption{\label{table:main} The main sample of DLAs with \hii\ absorption studies and 
low-frequency VLBI estimates of the covering factor. See Section~\ref{sec:sample}
 for a description of the columns of the table.}
\begin{center}
\begin{tabular}{|c|c|c|c|c|c|c|c|c|c|c|c|c|}
\hline
    &               &               &       &         &     &       &       &     &    &                &      & \\
QSO & $z_{\rm QSO}$ & $z_{\rm abs}$ & log[$\nhi$/\cm] & $f$ & $\ts$ & [Z/H] &  Z & [Z/Fe] & Dust & $\Delta V_{\rm 90}^a$ & Line & Refs.$^b$ \\
    &               &               &                 &     &  (K)    &       &     &        &      & (\kms)               &      & \\
\hline
           &       &        &                  &      &                    &                       & & & & & & \\
0738+313  & 0.635 & 0.0912 & $21.18 \pm 0.06$ & 0.98 & $    775 \pm 100 $ & $  -1.21 \pm 0.16$ &  Fe$^c$ & $ <   0.46        $ &  Fe  &  42 &  Ti{\sc ii}$\lambda$3384 &  2-5   \\
0738+313   & 0.635 & 0.2212 & $20.90 \pm 0.08$ & 0.98 & $    870 \pm 160 $ & $  <  -0.72        $ &  Zn & $ <   0.71~        $ &  Cr  &  60 &  Mg{\sc i}$\lambda$2852  &  2-6 \\
0952+179   & 1.472 & 0.2378 & $21.32 \pm 0.05$ & 0.66 & $   6470 \pm 965 $ & $  -1.89 \pm 0.06 $ &  Fe$^c$ & $ <   1.27        $ &  Fe  &  98 &  Mg{\sc i}$\lambda$2852  &  1,5,7-9  \\
1127$-$145   & 1.184 & 0.3127 & $21.70 \pm 0.08$ & 0.90 & $    820 \pm 145 $ & $  -0.76 \pm 0.10 $ &  Zn & $ <   1.23~      $ &  Fe  & 123 &  Ca{\sc ii}$\lambda$3934 &  1,6,7,9,10  \\
1229$-$021   & 1.043 & 0.3950 & $20.75 \pm 0.07$ & 0.42 & $     95 \pm  15 $ & $  -0.45 \pm 0.15 $ &  Zn & $ >   0.81~      $ &  Fe  & 122 &  Mn{\sc ii}$\lambda$2576 &  9,11-13  \\
0235+164     & 0.940 & 0.5242 & $21.70 \pm 0.09$ & 1.00 & $    210 \pm  45 $ & $  -0.14 \pm 0.17 $ &  XR$^d$ & $  1.71 \pm 0.45 $ &  Fe  &  -  &  -         &  14,15,16 \\
0827+243   & 0.939 & 0.5247 & $20.30 \pm 0.04$ & 0.70 & $    330 \pm  65 $ & $  -0.51 \pm 0.05   $ &  Fe$^c$ & $      -           $ &   -  & 188 &  Fe{\sc ii}$\lambda$2374 &  5,7-9,16 \\
1429+400   & 1.215 & 0.6039 & $21.20 \pm 0.10$ & 0.32 & $     90 \pm  23 $ & $  -0.80 \pm 0.21 $ &  Zn & $  0.70 \pm 0.24 $ &  Fe  &  70 &  Mn{\sc ii}$\lambda$2576 &  17 \\
1122$-$168   & 2.400 & 0.6819 & $20.45 \pm 0.05$ & 0.04 & $ >  100         $ & $  -0.90 \pm 0.29 $ &  Fe$^c$ & $ <  -0.17        $ &  Fe  & 144  & Fe{\sc ii}$\lambda$2586 &  1,18  \\
1331+305   & 0.849 & 0.6922 & $21.25 \pm 0.02$ & 0.90 & $    965 \pm 105 $ & $  -1.35 \pm 0.05 $ &  Zn & $  0.26 \pm 0.06 $ &  Fe  &  26 &  Mn{\sc ii}$\lambda$2576 & 12,19,20   \\
0454+039   & 1.350 & 0.8596 & $20.69 \pm 0.06$ & 0.50 & $ > 1380         $ & $  -0.99 \pm 0.12 $ &  Zn & $ -0.02 \pm 0.12 $ &  Fe  & 100 &  Fe{\sc ii}$\lambda$2260 &  1,12,21,22  \\
2149+212   & 1.538 & 0.9115 & $20.70 \pm 0.10$ & 0.02 & $ >   55         $ & $  <  -0.93        $ &  Zn & $      -           $ &   -  &   - &   -        &  1,23,24  \\
2355$-$106   & 1.639 & 1.1727 & $21.00 \pm 0.10$ & 1.00 & $   2145 \pm 570 $ & $  -0.87 \pm 0.20 $ &  Zn & $  0.50 \pm 0.23 $ &  Fe  & 126 &  Mn{\sc ii}$\lambda$2576 &  17,25,26 \\
1621+074     & 1.648 & 1.3367 & $21.35 \pm 0.10$ & 0.34 & $    460 \pm 105 $ & $  -1.07 \pm 0.14 $ &  Zn & $  0.45 \pm 0.15 $ &  Fe  &  55 &  Fe{\sc ii}$\lambda$2260 &  17,26  \\
2003$-$025   & 1.457 & 1.4106 & $20.54 \pm 0.20$ & 0.81 & $    485 \pm 195 $ & $       -           $ &  -  & $      -           $ &   -  &   - &   -        &  1,24,25 \\
1331+170   & 2.084 & 1.7764 & $21.18 \pm 0.07$ & 0.72 & $    625 \pm 115 $ & $  -1.27 \pm 0.09 $ &  Zn & $  0.74 \pm 0.09 $ &  Fe  &  72 &  Si{\sc ii}$\lambda$1808 &  9,27-30 \\
1157+014   & 2.000 & 1.9436 & $21.80 \pm 0.07$ & 0.63 & $   1015 \pm 255 $ & $  -1.44 \pm 0.10 $ &  Zn & $  0.35 \pm 0.10 $ &  Fe  &  84 &  Ni{\sc ii}$\lambda$1741 &  9,31-34 \\
0458$-$020   & 2.286 & 2.0395 & $21.78 \pm 0.07$ & 1.00 & $    560 \pm  95 $ & $  -1.27 \pm 0.08 $ &  Zn & $  0.45 \pm 0.10 $ &  Fe  &  84 &  Cr{\sc ii}$\lambda$2056 &  27,35-37  \\
2039+187   & 3.056 & 2.1920 & $20.70 \pm 0.09$ & 0.35 & $    160 \pm  35 $ & $       -           $ &  -  & $      -           $ &   -  &   - &   -        &  38,39  \\
1048+347   & 2.520 & 2.2410 & $20.54 \pm 0.06$ & 0.69 & $ > 2155         $ & $       -           $ &  -  & $      -           $ &   -  &   - &   -        &  38,39  \\
0311+430   & 2.870 & 2.2890 & $20.30 \pm 0.11$ & 0.52 & $    72 \pm  18  $ & $  -0.49 \pm 0.13 $ &  Fe$^c$ & $ >   0.33        $ &  Fe  &  -  &  -         &  1,38-40  \\
0432$-$440   & 2.649 & 2.3020 & $20.78 \pm 0.11$ & 0.38 & $ >  555         $ & $  -1.09 \pm 0.13 $ &  Si & $  0.34 \pm 0.15 $ &  Fe  & 125 &  Fe{\sc ii}$\lambda$1608 &  1,41,42  \\
0438$-$436   & 2.863 & 2.3474 & $20.78 \pm 0.11$ & 0.59 & $    900 \pm 250 $ & $  -0.69 \pm 0.12 $ &  Zn & $  0.59 \pm 0.16 $ &  Fe  &  55 &  Fe{\sc ii}$\lambda$2260 &  1,41-43  \\
0229+230   & 3.420 & 2.6200 & $20.30 \pm 0.11$ & 0.30 & $ >  140         $ & $       -           $ &  -  & $      -           $ &   -  &   - &   -        &  38,39  \\
0229+230   & 3.420 & 2.6830 & $20.70 \pm 0.09$ & 0.30 & $ >  270         $ & $       -           $ &  -  & $      -           $ &   -  &   - &   -        &  38,39  \\
1402+044   & 3.215 & 2.7076 & $21.01 \pm 0.23$ & 0.34 & $ >  890         $ & $       -           $ &  -  & $      -           $ &   -  &   - &   -        &  1,44-46  \\
0913+003   & 3.074 & 2.7434 & $20.74 \pm 0.11$ & 0.54 & $ >  925         $ & $  -1.44 \pm 0.16 $ &  Si & $  0.13 \pm 0.19 $ &  Fe  & 105 &  Si{\sc ii}$\lambda$1526 &  1,41,42  \\
2039+187   & 3.056 & 2.7520 & $20.70 \pm 0.09$ & 0.35 & $ >  790         $ & $       -           $ &  -  & $      -           $ &   -  &   - &   -        &  38,39  \\
1354$-$170   & 3.147 & 2.7799 & $20.30 \pm 0.15$ & 0.97 & $ > 1030         $ & $  -1.83 \pm 0.16 $ &  Si & $  0.54 \pm 0.18 $ &  Fe  &  30 &  Fe{\sc ii}$\lambda$1608 &  1,47 \\
0800+618   & 3.033 & 2.9603 & $20.50 \pm 0.02$ & 0.63 & $ >  570         $ & $       -           $ &  -  & $      -           $ &   -  &   - &   -        &  1,48  \\
0537$-$286   & 3.110 & 2.9746 & $20.30 \pm 0.11$ & 0.47 & $ >  520         $ & $  <  -0.40        $ &  Zn & $      -           $ &   -  &  -  &  -         &  9,41,42,49  \\
2342+342   & 3.053 & 2.9084 & $21.10 \pm 0.10$ & 0.71 & $ > 2200         $ & $  -1.06 \pm 0.10 $ &  S  & $  0.47 \pm 0.12 $ &  Fe  & 100 &  Si{\sc ii}$\lambda$1808 &  1,9,47,50  \\
0336$-$017   & 3.197 & 3.0621 & $21.20 \pm 0.10$ & 0.68 & $ > 8890         $ & $  -1.36 \pm 0.10 $ &  S  & $  0.39 \pm 0.11 $ &  Fe  & 108 &  Si{\sc ii}$\lambda$1250 &  1,9,30,51  \\
0335$-$122   & 3.442 & 3.1799 & $20.78 \pm 0.11$ & 0.62 & $ > 1850         $ & $  -2.53 \pm 0.16 $ &  Si & $  0.07 \pm 0.19 $ &  Fe  &  39 &  Al{\sc ii}$\lambda$1670 &  9,41,42,49  \\
0201+113   & 3.639 & 3.3869 & $21.25 \pm 0.07$ & 0.76 & $   1050 \pm 175 $ & $  -1.19 \pm 0.17 $ &  S  & $  0.16 \pm 0.22 $ &  Fe  &  64 &  Ni{\sc ii}$\lambda$1317 &  9,52-55  \\
1239+376   & 3.819 & 3.4110 & $20.40 \pm 0.09$ & 0.54 & $ > 2330         $ & $       -           $ &  -  & $      -           $ &   -  &   - &   -        &  38,39  \\
1418$-$064   & 3.689 & 3.4482 & $20.40 \pm 0.10$ & 0.69 & $ >  930         $ & $  -1.45 \pm 0.14 $ &  Si & $  0.25 \pm 0.17 $ &  Fe  &  41 &  Fe{\sc ii}$\lambda$1608 &  1,9,41,42 \\
    &               &               &       &      &       &       &     &        &    &                &      & \\
\hline
\end{tabular}
\end{center}
\tablenotetext{a}{Most of the $\Delta V_{\rm 90}$ measurements were carried out by us; a few values 
are taken from \citet{prochaska99} and \citet{ledoux06}.}
\tablenotetext{b}{References for $\nhi$, metallicity, covering factor and spin temperature estimates:
(1)~This work, (2)~\citet{rao98}, (3)~\citet{chengalur99}, (4)~\citet{lane00}, (5)~\citet{kulkarni05}, 
(6)~\citet{lane98}, (7)~\citet{rao00}, (8)~\citet{kanekar01a}, (9)~\citet{kanekar09a}, 
(10)~\citet{chengalur00}, (11)~\citet{brown79}, (12)~\citet{boisse98}, (13)~\citet{lanzetta92},
(14)~\citet{roberts76}, (15)~\citet{junkkarinen04}, (16)~\citet{chen05}, (17)~\citet{ellison12},
(18)~\citet{delavarga00}, (19)~\citet{wolfe08b}, (20)~\citet{brown73}, (21)~\citet{pettini00}, 
(22)~\citet{steidel95}, (23)~\citet{nestor08}, (24)~\citet{rao06}, (25)~\citet{kanekar09b}, 
(26)~\citet{gupta09}, (27)~\citet{wolfe79}, (28)~\citet{carswell11}, (29)~\citet{prochaska99}, 
(30)~\citet{prochaska01b}, (31)~\citet{wolfe81}, (32)~\citet{ledoux03}, (33)~\citet{dessauges07}, 
(34)~\citet{dessauges06}, (35)~\citet{wolfe85}, (36)~\citet{briggs89}, (37)~\citet{moller04},
(38)~\citet{ellison08}, (39)~\citet{kanekar13}, (40)~\citet{york07}, (41)~\citet{ellison01}, 
(42)~\citet{akerman05}, (43)~\citet{kanekar06}, (44)~\citet{prochaska05}, (45)~\citet{noterdaeme09}, 
(46)~\citet{curran10}, (47)~\citet{prochaska03b}, (48)~\citet{jorgenson06}, (49)~\citet{kanekar03},
(50)~\citet{prochaska07}, (51)~\citet{ledoux06}, (52)~\citet{white93}, (53)~\citet{kanekar07}, 
(54)~\citet{srianand12}, (55)~\citet{ellison01b}, (56)~\citet{moller93}, 
(57)~\citet{lu96}, (58)~\citet{srianand10}, (59)~\citet{storrie00}, (60)~\citet{lopez03}, 
(61)~\citet{ledoux02}, (62)~\citet{bergeron86}. (63)~\citet{roy13b}.}
\tablenotetext{c}{In five DLAs, at $z=0.0912$ towards 0738+313, $z=0.2378$ towards 0952+179, $z = 0.5247$ 
towards 0827+243, $z=0.6819$
towards 1122$-$168 and $z=2.2890$ towards 0311+430, the metallicity was estimated using [Z/H]~$=$~[Fe/H]~$+ 0.4$, 
following \citet{prochaska03a}. However, in four of the five cases (excluding the DLA towards 0827+243), 
limits are also available on [Zn/H] or [Si/H]; these limits were used along with the [Fe/H] measurements to 
derive the dust depletion [Z/Fe].} 
\tablenotetext{d}{In the case of the $z = 0.5242$ DLA towards 0235+164, the metallicity is from an 
X-ray absorption study \citep{junkkarinen04}.}
\end{table*}

\setcounter{table}{4}
\begin{table*}
\caption{\label{table:sub} DLAs or sub-DLAs with \hii\ absorption studies, but not included in the ``main'' sample}
\begin{center}
\begin{tabular}{|c|c|c|c|c|c|c|c|c|c|c|c|c|}
\hline
    &               &               &       &      &       &       &     &        &    &                &      & \\
QSO & $z_{\rm QSO}$ & $z_{\rm abs}$ & log[$\nhi$/\cm] & $f$ & $\ts$ & [Z/H] &  Z  & [Z/Fe] & Dust & $\Delta V_{\rm 90}$ & Line & Refs.$^b$ \\
    &               &               &       &      &  (K)     &       &     &        &    & (\kms)               &      & \\
\hline
\multicolumn{13}{|c|}{\bf PDLAs, with the absorber within 3000~\kms\ of the background quasar} \\
\hline
    &               &               &       &      &       &       &     &        &    &                &      & \\
0105$-$008   & 1.374 & 1.3708 & $21.70 \pm 0.15$ & 0.32 & $    305 \pm  45 $ & $  -1.40 \pm 0.16 $ &  Zn & $  0.16 \pm 0.16 $ &  Fe  &  33 &  Fe{\sc ii}$\lambda$2249 &  9,17,26  \\
0824+355   & 2.249 & 2.2433 & $20.30     $ & 0.20 & $ >  320         $ & $       -           $ &  -  & $      -           $ &   -  &   - &   -        &  1,48  \\
0405$-$331   & 2.570 & 2.5693 & $20.60 \pm 0.11$ & 0.44 & $ > 1220         $ & $  -1.37 \pm 0.16 $ &  Si & $  0.35 \pm 0.19 $ &  Fe  & 261 &  Si{\sc ii}$\lambda$1526 &  1,41,42  \\
1013+615   & 2.805 & 2.7681 & $20.60 \pm 0.15$ & 0.81$^\star$ & $ >  945 $ & $       -           $ &  -  & $      -           $ &   -  &   - &   -        &  1,44,54  \\
0528$-$250   & 2.813 & 2.8110 & $21.35 \pm 0.07$ & 0.94 & $ > 2103         $ & $  -0.89 \pm 0.10 $ &  Zn & $  0.45 \pm 0.13 $ &  Fe  & 304 &  S{\sc ii}$\lambda$1253  &  1,54,56,57   \\
1354$-$107   & 3.006 & 2.9660 & $20.78 \pm 0.11$ &  -   & $ >  615         $ & $  -1.28 \pm 0.16 $ &  Si & $  0.24 \pm 0.19 $ &  Fe  &  59 &  Si{\sc ii}$\lambda$1808 &  41,42,49 \\
J1337+3152 & 3.174 & 3.1745 & $21.36 \pm 0.10$ & 1.00$^\star$ & $ 600 \pm 297 $ & $  -1.73 \pm 0.28 $ &  Zn & $  0.17 \pm 0.29 $ &  Fe  &   - &   -        &  58 \\
    &               &               &       &      &       &       &     &        &    &                &      & \\
\hline
\multicolumn{13}{|c|}{\bf Sub-DLAs, with $\nhi < 2 \times 10^{20}$~\cm} \\
\hline
2128$-$123   & 0.501 & 0.4297 & $19.37 \pm 0.08$ &  -   & $ >  980         $ & $       -           $ &  -  & $      -           $ &   -  &   - &   -        &  49,61  \\
0215+015     & 1.715 & 1.3439 & $19.89 \pm 0.09$ &  -   & $ > 1020         $ & $  -0.66 \pm 0.22 $ &  Fe & $      -           $ &  -   &  -  &  -         &  49,62  \\
0237$-$233   & 2.223 & 1.6724 & $19.78 \pm 0.07$ & 0.90 & $    390 \pm 125 $ & $  -0.57 \pm 0.12~ $ &  Zn & $  0.09 \pm 0.12~ $ &  Fe  &  43 &  Fe{\sc ii}$\lambda$2374 &  17,25 \\
1402+044     & 3.215 & 2.4850 & $20.20 \pm 0.20$ & 0.35 & $ >  380         $ & $       -           $ &  -  & $      -           $ &   -  &   - &   -        &  1,44$-$46 \\
J1406+3433   & 2.566 & 2.4989 & $20.20 \pm 0.20$ & 0.76 & $ >  210 $ & $       -           $ &  -  & $      -           $ &   -  &   - &   -        &  44,45,54  \\
J0733+2721   & 2.938 & 2.7263 & $20.25 \pm 0.20$ &  -   & $ >  690         $ & $       -           $ &  -  & $      -           $ &   -  &   - &   -        &  45,54  \\
\hline
\multicolumn{13}{|c|}{\bf DLAs with covering factor estimates at frequencies significantly higher ($> 1.5$~times) the redshifted \hii\ line frequency}  \\
\hline
    &               &               &       &      &       &       &     &        &    &                &      & \\
0620+389   & 3.469 & 2.0310 & $20.30 \pm 0.11$ & 0.79 & $ > 400  $ & $       -           $ &  -  & $      -           $ &   -  &   - &   -        &  38,39  \\
J0852+2431 & 3.617 & 2.7902 & $20.70 \pm 0.20$ & 0.49 & $ >  420 $ & $       -           $ &  -  & $      -           $ &   -  &   - &   -        &  45,54  \\
J0816+4823 & 3.573 & 3.4358 & $20.80 \pm 0.20$ & 0.60 & $ >  145 $ & $       -           $ &  -  & $      -           $ &   -  &   - &   -        &  45,54    \\
    &               &               &       &      &       &       &     &        &    &                &      & \\
\hline
\multicolumn{13}{|c|}{\bf Absorbers without covering factor estimates} \\
\hline
    &               &               &       &      &       &       &     &        &    &                &      & \\
J0011+1446   & 4.967 & 3.4523 & $21.65 \pm 0.15$ &  -   & $ >  3040$        &    -        &  - & $      -           $ &  -   &  -  &  -         &  63  \\
0347$-$211   & 2.944 & 1.9470 & $20.30 \pm 0.11$ &  -   & $ >  380         $ & $  <  -0.55       $ &  Zn & $      -           $ &  -   &  -  &  -         &  1,41,42  \\
1614+051   & 3.215 & 2.5200 & $20.40         $ &  -   & $ >  450         $ & $       -           $ &  -  & $      -           $ &  -   &  -  &  -         &  1,59  \\
J0407$-$4410 & 3.020 & 2.5950 & $21.05 \pm 0.10$ &  -   & $ >  380         $ & $  -1.00 \pm 0.10 $ &  Zn & $  0.35 \pm 0.10 $ &  Fe  &  79 &  Si{\sc ii}$\lambda$1808 & 51,54,60   \\
J0407$-$4410 & 3.020 & 2.6214 & $20.47 \pm 0.10$ &  -   & $ >   80         $ & $  -1.99 \pm 0.12 $ &  Si & $  0.33 \pm 0.12 $ &  Fe  &   - &   -        &  54,60  \\
J0801+4725 & 3.276 & 3.2235 & $20.70 \pm 0.15$ &  -   & $ > 2820 $ & $       -           $ &  -  & $      -           $ &   -  &   - &   -        &  45,54,63  \\
J1435+5435 & 3.811 & 3.3032 & $20.30 \pm 0.20$ &  -   & $ >  420         $ & $       -           $ &  -  & $      -           $ &   -  &   - &   -        &  45,54  \\
           &       &        &                  &      &                    &                       & & & & & & \\
\hline
\end{tabular}
\end{center}
\begin{flushleft}
References: See Table~\ref{table:main}.\\
$^\star$~These PDLAs also have covering factor estimates at frequencies $> 1.5$~times higher than the redshifted \hii\ line frequency.
\end{flushleft}
\end{table*}

\begin{enumerate}
\item{0201+113, $z =   3.3869$: The $z = 3.3869$ DLA towards 0201+113 was detected by \citet{white93} and 
its \hi\ column density was measured to be $\nhi = (1.8 \pm 0.3) \times 10^{21}$~\cm\ by \citet{ellison01b}. After 
inconclusive results from a number of \hii\ absorption studies with different radio telescopes \citep{debruyn96,briggs97,kanekar97}, 
\hii\ absorption was finally detected from this absorber with the GMRT \citep{kanekar07}. The DLA covering factor 
is $f = 0.76$, from a 327~MHz VLBA imaging study \citep{kanekar09a}, yielding a spin temperature 
of $\ts = (1050 \pm 175) \times (f/0.76) $~K. } 
\item{0311+430, $z = 2.2890$: The $z = 2.2890$ DLA towards 0311+430 was found by \citet{ellison08} and has 
an \hi\ column density of $(2.0 \pm 0.5) \times 10^{20}$~\cm\ and a spin temperature of 
$\ts = (72 \pm 18) \times (f/0.52)$~K \citep{york07,kanekar13}. The Si{\sc ii}$\lambda$1808 line, detected in the 
original low-resolution GMOS spectrum, yields [Si/H]~$> -0.56$, assuming the line to be unsaturated \citep{york07,ellison08}.
We do not detect the Zn{\sc ii}$\lambda$2026 line in our new GMOS spectrum (see Table~\ref{table:abundances}), 
implying $\log[N_{\rm ZnII}/{\rm cm}^{-2}] < 12.5$ and [Zn/H]~$< -0.43$. The metallicity of the $z = 2.2890$ DLA thus lies in 
the range $-0.56 < {\rm [Z/H]} < -0.43$, from the Zn{\sc ii} upper limit and the Si{\sc ii} lower limit. 
However, we clearly detect the Fe{\sc ii}$\lambda$2249, $\lambda$2260 and $\lambda$2374 transitions in the new GMOS 
spectrum, with $\log[N_{\rm FeII}/{\rm cm}^{-2}] = 14.91 \pm 0.05$, $14.80 \pm 0.10$ and $> 14.32$, respectively. 
Note that the Fe{\sc ii}$\lambda$2374 line yielded a lower column density than the other two lines, probably due 
to saturation effects. We have hence listed this as a lower limit, and used the former two lines to obtain 
$\log[N_{\rm FeII}/{\rm cm}^{-2}] = 14.86 \pm 0.06$,
and [Fe/H]~$= -0.89 \pm 0.13$. This yields  a metallicity of [Z/H]=[Fe/H]+0.4=$-0.49 \pm 0.13$ for the $z = 2.2890$ DLA.}
\item{0336$-$017, $z = 3.0621$: The $z = 3.0621$ DLA towards 0336$-$017 was found by \citet{lu93}, with 
an \hi\ column density of $\nhi = (1.5 \pm 0.3) \times 10^{21}$~\cm\ \citep{prochaska01b}. The DLA covering factor 
is $f = 0.68$ \citep{kanekar09a}; our upper limit on the \hii\ optical depth then yields the lower limit $\ts > 8890 
\times (f/0.68)$~K on the DLA spin temperature, assuming an FWHM of 15~\kms\ for the \hii\ line. We note that this 
$\ts$ limit is extremely high, at the upper end of the range of WNM kinetic temperatures. The low-ionization metal 
lines of this DLA have a relatively large velocity spread, $\Delta V_{\rm 90} \approx 108$~\kms (see Table~\ref{table:main}). However, 
even assuming that the \hii\ line has a comparable velocity FWHM (perhaps due to non-thermal broadening) would only 
lower the spin temperature limit to $\ts \gtrsim 8890 \times \sqrt(15/108)$, i.e. to $\ts \gtrsim 3315$~K. The \hi\ 
content in this DLA thus appears to be dominated by warm gas, unless there is significant small-scale structure along the 
sightline and the \hi\ column along the radio sightline is much lower than that towards the optical quasar. }
\item{0347$-$211, $z = 1.9470$: The $z = 1.9470$ DLA towards 0347$-$211 was detected by \citet{ellison01} in the
CORALS survey, and has an \hi\ column density of $\nhi = (2.0 \pm 0.5) \times 10^{20}$~\cm. No low-frequency VLBI 
images of the background quasar are currently available; we hence do not have a reliable estimate of its covering factor. 
Our upper limit on the \hii\ optical depth yields $\ts > (380 \times f)$~K.}
\item{0405$-$331, $z = 2.5693$: The PDLA towards 0405$-$331 has an \hi\ column density of $\nhi = (4 \pm 1) 
\times 10^{20}$~\cm\ \citep{ellison01} and a covering factor of $f = 0.44$ \citep{kanekar09a}, yielding $\ts > 1220 
\times (f/0.44)$~K.}
\item{0432$-$440, $z = 2.3020$: The $z = 2.3020$ DLA towards 0432$-$440 has an \hi\ column density of $\nhi 
= (6.0 \pm 1.5) \times 10^{20}$~\cm\ \citep{ellison01}. The VLBA 327~MHz image of this DLA shows a weak extension, but
we were unable to obtain a stable 2-component fit (probably due to the weakness of the extension); we hence used a 
single-Gaussian model here, obtaining $f = 0.38$. Combining this with our upper limit on the \hii\ optical depth yields 
$\ts > 555 \times (f/0.38)$~K.}
\item{0438$-$436, $z = 2.3474$: This DLA has an \hi\ column density of $\nhi = (6.0 \pm 1.5) \times 10^{20}$~\cm\
\citep{ellison01}, and a covering factor of $f = 0.59$ \citep{kanekar09a}. The GBT detection of \hii\ absorption then
yields $\ts = (900 \pm 250) \times (f/0.59)$~K \citep{kanekar06}.}
\item{0454+039, $z =   0.8597$: The $z = 0.8597$ DLA towards 0454+039 has an \hi\ column density of $\nhi = (5.0 \pm 0.7) 
\times 10^{20}$~\cm\ (\citealp{boisse98}; see also \citealt{steidel95}) and a covering factor $f = 0.53$ \citep{kanekar09a}. 
Our GBT upper limit to the \hii\ optical depth then yields $\ts >  990 \times (f/0.53)$~K.}
\item{0458$-$020, $z = 2.0395$: There are multiple estimates of the \hi\ column density of this DLA, 
$\nhi = (4.5-6) \times 10^{21}$~\cm\ \citep{wolfe93,pettini94,moller04,heinmuller06}, from different telescopes 
and spectrographs, all consistent with each other within the measurement errors. We will use the value 
$\nhi = (6 \pm 1) \times 10^{21}$~\cm\ \citep{moller04}, as this was obtained from a high-sensitivity VLT-FORS1 
spectrum (the other results use either echelle spectroscopy or 4m-class telescopes). \\
\hii\ absorption in the $z = 2.0395$ DLA was originally detected by \citet{wolfe85}, with two distinct absorption 
components \citep[see also][]{briggs89}. We observed the \hii\ line with the GBT on two separate occasions, 
in March and May 2008; these spectra are shown in Fig.~\ref{fig:0458}, overlaid on an old \hii\ spectrum obtained 
with the Arecibo telescope (kindly provided by Art Wolfe). The GBT and Arecibo spectra are clearly different, 
with the secondary \hii\ component significantly weaker in the GBT spectra. We note that part of this 
discrepancy might arise due to flux scale errors (and perhaps bandpass calibration issues) in the Arecibo spectrum, 
given that the source is at the edge of the Arecibo declination range. However, while the secondary component has 
the same depth in the two GBT runs, a statistically-significant difference is seen in the primary component.
Indeed, the integrated \hii\ optical depths obtained at the GBT are $\int \tau_{\rm 21cm} {\rm dV} = 
(5.534 \pm 0.080)$~\kms\ in March~2008 and $\int \tau_{\rm 21cm} {\rm dV} = (6.198 \pm 0.088)$~\kms\ in May~2008; 
the difference has $\approx 5\sigma$ significance, indicating that the \hii\ absorption has changed on a 
timescale of $\approx 2$~months. No significant narrow-band interference was observed on either observing run and 
the doppler shift between the runs is $\approx 20$~\kms, comparable to the FWHMs of the components. We conclude 
that the spectra show evidence for variability in the \hii\ profile, on timescales of months, and possibly of years. 
Such line variations have been seen earlier, on timescales as short as a few days, in two low-$z$ DLAs, at $z \sim 0.5242$ 
towards AO0235+164 \citep{wolfe82} and $z \sim 0.3127$ towards 1127$-$145 \citep{kanekar01c}. They can be explained
by a number of scenarios, including refractive interstellar scintillation, gravitational microlensing, transverse 
motion of compact components in the quasar radio jet, etc \citep[e.g.][]{briggs83b,kanekar01c,gwinn01,macquart05}, 
all typically 
requiring small-scale structure in the quasar radio emission, the ISM of the absorber and/or the ISM of the Galaxy. \\
The DLA covering factor has been estimated to be $\approx 1$ from a VLBI study in the \hii\ line \citep{briggs89}. 
The average integrated \hii\ optical depth from the two GBT runs is $\int \tau_{\rm 21cm} {\rm dV} = (5.866 \pm 0.059)$~\kms;
we then obtain $\ts = (560 \pm 95)$~K, assuming $f = 1$. Note that the \hii\ optical depth is fairly high, $\tau \approx 0.3$, 
implying that the low optical depth approximation of equation~(\ref{eqn:tspin}) should not be used.}
\item{0528$-$250, $z = 2.8112$: The $z = 2.8112$ PDLA towards 0528$-$250 has an \hi\ column density 
of $(2.24 \pm 0.05) \times 10^{21}$~\cm\ \citep{moller93} and a covering factor of unity \citep{kanekar09a}. 
Our WSRT non-detection of \hii\ absorption then yields $\ts > 925 \times (f/1.0)$~K. We note that \citet{srianand12}
present a GBT spectrum of this absorber, obtaining an upper limit of $\int \tau_{\rm 21cm} {\rm d}V < 0.58$~\kms\ and 
$\ts > 2103$~K, using the b-parameter of the detected H$_2$ absorption for the line width. We have quoted this 
result in Table~\ref{table:sub}, due to its higher sensitivity.}
\item{0738+313, $z =   0.0912$: This is the lowest redshift DLA of the sample, with 
$\nhi = (1.5 \pm 0.21) \times 10^{21}$~\cm\ \citep{rao98} and $\ts = (775 \pm 100) \times (f/0.98)$~K \citep{chengalur99,lane00}.
Note that this is one of the few DLAs whose spin temperature has been directly measured from a VLBA \hii\ 
absorption study \citep{lane00}. \citet{kulkarni05} detected Fe{\sc ii} and Cr{\sc ii} absorption from this system 
in an HST-STIS spectrum, yielding [Fe/H]~$=-1.61 \pm 0.16$ and [Cr/H]~$= -1.55 \pm 0.23$; their non-detection of 
Zn{\sc ii}$\lambda$2026 absorption gives [Zn/H]~$< -1.15$. Our Keck-HIRES spectrum yields 
$\log[N_{\rm TiII}/{\rm cm}^{-2}] = 12.51 \pm 0.04$ and thus a titanium abundance [Ti/H]~$= -1.58 \pm 0.08$. 
This is consistent with the original Ti{\sc ii} detection of \citet{khare04}, from a medium-resolution 
spectrum ($\log[N_{\rm TiII}/{\rm cm}^{-2}] = 12.53^{+0.14}_{-0.06}$).}
\item{0738+313, $z =   0.2212$: The second DLA towards 0738+313 was also detected by \citet{rao98}, 
with $\nhi = (7.9 \pm 1.5) \times 10^{20}$~\cm; this has $\ts = (870 \pm 160) \times (f/0.98)$~K 
\citep{lane98,chengalur99,kanekar01b}. Zn{\sc ii} absorption was not conclusively detected in the 
HST-STIS spectrum of \citet{kulkarni05}, giving [Zn/H]~$< -0.72$. However, these authors did detect 
Cr{\sc ii} absorption from the DLA, with [Cr/H]~$= -1.43 \pm 0.22$. We obtain 
$\log[N_{\rm FeII}/{\rm cm}^{-2}] > 14.30$ and $\log[N_{\rm TiII}/{\rm cm}^{-2}] < 11.90$ from our Keck-HIRES 
spectrum, yielding [Fe/H]~$> -2.05$ and [Ti/H]~$< -1.91$.}
\item{0800+618, $z =   2.9603$: This absorber was detected in the UCSD survey of \citet{jorgenson06}, with 
an \hi\ column density of $\nhi = (3.16 \pm 0.15) \times 10^{20}$~\cm. Using this with our covering factor estimate 
($f = 0.63$) and our limit on the \hii\ optical depth yields $\ts > 570 \times (f/0.63)$~K.}
\item{0824+355, $z =   2.2433$: The PDLA towards 0824+355 has $\nhi = 2 \times 10^{20}$~\cm\ \citep{jorgenson06}.
The VLBA 327~MHz image shows three clear components, apparently a core with an extended jet, and possibly a fourth 
weak component. We fit a 3-component model to the image, obtaining a core covering factor of $f = 0.20$. Our non-detection of 
\hii\ absorption then yields $\ts > 320 \times (f/0.20)$~K. }
\item{0827+243, $z =   0.5247$: This is another DLA from the HST sample of \citet{rao00}, with $\nhi = (2.00 \pm 0.18) 
\times 10^{20}$~\cm. \hii\ absorption was detected in this absorber by \citet{kanekar01a} and its covering factor 
measured by \citet{kanekar09a}; the DLA spin temperature is $\ts = (330 \pm 65) \times (f/0.7)$~K. 
\citet{kulkarni05} give weak upper limits on the Zn and Cr abundances from their 
HST-STIS non-detections of absorption, with [Zn/H]~$< -0.04$ and [Cr/H]~$< 0.43$. Conversely, detections of 
Fe{\sc ii} and Mn{\sc ii} absorption in a medium-resolution spectrum \citet{khare04} yielded 
[Fe/H]~$=-1.01 \pm 0.11$ and [Mn/H]~$=-0.86 \pm 0.35$ \citep{chen05}. Our high-resolution
VLT-UVES spectrum gives a slightly higher Fe{\sc ii} column density (albeit consistent within the errors), 
with [Fe/H]=$-0.91 \pm 0.05$, and also yields [Ti/H]~$=-1.39 \pm 0.06$. The DLA metallicity is then 
[Z/H]~$=$~[Fe/H]~$+ 0.4$~$= -0.51 \pm 0.05$.}
\item{0913+003, $z =   2.7434$: The DLA towards 0913+003 has $\nhi = (5.5 \pm 1.4) \times 10^{20}$~\cm\
\citep{ellison01} and $f = 0.54$. Our GBT non-detection of \hii\ absorption then yields $\ts > 925 \times (f/0.54)$~K.}
\item{0952+179, $z =   0.2378$: The $z = 0.2378$ DLA towards 0952+179 was detected by \citet{rao00}, 
with $\nhi = (2.09 \pm 0.24) \times 10^{21}$~\cm. \citet{kanekar01a} detected \hii\ absorption 
in this absorber, and its covering factor was measured to be $f=0.66$ by \citet{kanekar09a}, giving 
$\ts = (6470 \pm 965) \times (f/0.66)$~K. \citet{kulkarni05} report an upper limit on the Zn{\sc ii} 
column density (with [Zn/H]~$< -1.02$) and a detection of Cr{\sc ii} absorption (with [Cr/H]~$=-1.64 \pm 0.16$.
Our VLT-UVES spectrum yields $\log[N_{\rm FeII}/{\rm cm}^{-2}] = 14.52 \pm 0.03$, and [Fe/H]~$=2.29 \pm 0.06$.
This gives a metallicity of [Z/H]~$ =$~[Fe/H]$+0.4$~$=-1.89 \pm 0.06$. We also obtain [Ti/H]~$=-1.61 \pm 0.07$ 
and [Mn/H]~$=-2.22 \pm 0.07$.}
\item{1013+615, $z =   2.7670$: \citet{prochaska05} obtain $\nhi = (4.0 \pm 0.9) \times 10^{20}$~\cm\ for this PDLA 
from an SDSS spectrum [see also \citet{jorgenson06} and \citet{srianand12}]. No VLBI studies of this quasar have been 
carried out at frequencies similar to that of the redshifted \hii\ line. While \citet{srianand12} estimate $f = 0.86$, 
this is from a 1.4~GHz VLBA image. Our GBT non-detection of \hii\ absorption yields $\ts > 1100 \times f$~K,
about a factor of 1.5 more sensitive than the non-detection of \citet{srianand12}. }
\item{1122$-$168, $z = 0.6819$: The $z = 0.6819$ DLA towards 1122$-$168 has $\nhi = (2.82 \pm 0.99) 
\times 10^{20}$~\cm\ \citep{delavarga00}. The VLBA image of the background quasar shows at least two components,
with a total flux density of $\approx 50$~mJy. We identify the more compact component with the quasar core; 
this has a total 1.4~GHz flux density of $12$~mJy, yielding a covering factor of $f \approx 0.04$. If both 
components are covered by the foreground DLA, the covering factor would be $f \approx 0.17$. Our GBT upper limit 
to the \hii\ optical depth yields $\ts > 100 \times (f/0.04)$~K; the weak limit is due to the extremely 
low covering factor.}
\item{1127$-$145, $z =   0.3127$: With $\nhi = (5.01 \pm 0.92) \times 10^{21}$~\cm\ \citep{rao00}, this 
is one of the highest column density DLAs at low redshifts, $z \lesssim 0.5$. \hii\ absorption from this DLA 
was originally detected by \citet{lane98} \citep[see also][]{kanekar01c} and the covering factor measured 
by \citet{kanekar09a}; the DLA spin temperature is $(820 \pm 145) \times (f/0.9)$~K. Our detection of 
Zn{\sc ii} absorption in the HST-STIS spectrum yields $\log[N_{\rm ZnII}/{\rm cm}^{-2}] = 13.53 \pm 0.13$, and 
[Zn/H]~$=-0.80 \pm 0.16$.
We also obtain $\log[N_{\rm MnII}/{\rm cm}^{-2}] = 13.26 \pm 0.03$ and $\log[N_{\rm FeII}/{\rm cm}^{-2}] > 15.16$, 
implying [Mn/H]~$= -1.92 \pm 0.09$ and [Fe/H]~$> - 2.0$, from the VLT-UVES spectrum.}
\item{1157+014, $z =   1.9436$: \hii\ absorption from this DLA was detected by \citet{wolfe81}, using the 
Arecibo telescope.  The DLA has an \hi\ column density of $\nhi = (6.3 \pm 1.5) \times 10^{21}$~\cm\ 
\citep{wolfe81,ledoux03} and a covering factor of $f = 0.63$ \citep{kanekar09a}. The new GBT \hii\ spectrum, 
shown in Fig.~\ref{fig:det2}[A], confirms the detection of \citet{wolfe81}, although the absorption is 
significantly weaker in our spectrum. We note, however, that strong out-of-band RFI has been found at the 
GBT around the redshifted \hii\ line frequency ($\approx 482.5$~MHz); this could imply significant 
uncertainty in our flux density calibration. We will hence use the original integrated \hii\ optical 
depth of \citet{wolfe81} to estimate the spin temperature; this yields $\ts = (1015 \pm 255) \times (f/0.63)$~K.}
\item{1354$-$170, $z = 2.7799$: This DLA has an \hi\ column density of $\nhi = (2.0 \pm 0.7) \times 10^{20}$~\cm\ 
\citep{prochaska03b} and a covering factor of $\approx 0.97$ (this work). Our GBT non-detection of \hi\ absorption then 
yields $\ts > 1030 \times (f/0.97)$~K.}
\item{1402+044, $z =   2.7076$: The \hi\ column density of this DLA is $\nhi = (1.0 \pm 0.5) \times 10^{21}$~\cm\
(\citealp{noterdaeme09}; see also \citealt{prochaska05}) and its covering factor is $f = 0.34$ (this work). 
Combining these with our upper limit on the \hii\ optical depth yields $\ts > 890 \times (f/0.34)$~K.}
\item{1418$-$064, $z = 3.4483$: This CORALS DLA has an \hi\ column density of $\nhi = (2.5 \pm 0.6) \times 
10^{20}$~\cm\ \citep{ellison01} and a covering factor of $f = 0.69$ \citep{kanekar09a}. Our GMRT non-detection 
of \hii\ absorption then yields $\ts > 930 \times (f/0.69)$~K.}
\item{1614+051, $z =   2.5200$: This DLA has $\nhi = 2.5 \times 10^{20}$~\cm\ \citep{storrie00}, but does not 
have an estimate of the covering factor; we obtain $\ts > 450 \times f$~K.}
\item{2003$-$025, $z = 1.4106$: This DLA was found by \citet{rao06} in their HST survey of strong \mgtwo\ 
absorbers, with $\nhi = (3.5 \pm 1.6) \times 10^{20}$~\cm. \hii\ absorption was later detected in this absorber
by \citet{kanekar09b}. The VLBA image shows a central core with weak extended emission, and a core covering factor
of $f = 0.53$; this yields a spin temperature of $\ts = (485 \pm 195) \times (f/0.53)$~K.}
\item{2149+212, $z =   0.9115$: The $z = 0.9115$ DLA towards 2149+212 was also found by \citet{rao06}.
Unfortunately, its very low covering factor ($f = 0.03$) implies that our GBT upper limit on the \hii\ optical 
depth only yields a weak limit on the spin temperature, $\ts > 55 \times (f/0.03)$~K.}
\item{2342+342, $z =   2.9082$: This DLA was detected by \citet{white93}, and has $\nhi = (1.3 \pm 0.3) 
\times 10^{21}$~\cm\ \citep{prochaska03b} and $f = 0.71$ \citep{kanekar09a}. Our GMRT non-detection of 
\hii\ absorption then yields $\ts > 2200 \times (f/0.71)$~K.}
\end{enumerate}

\begin{figure}
\centering
\epsfig{file=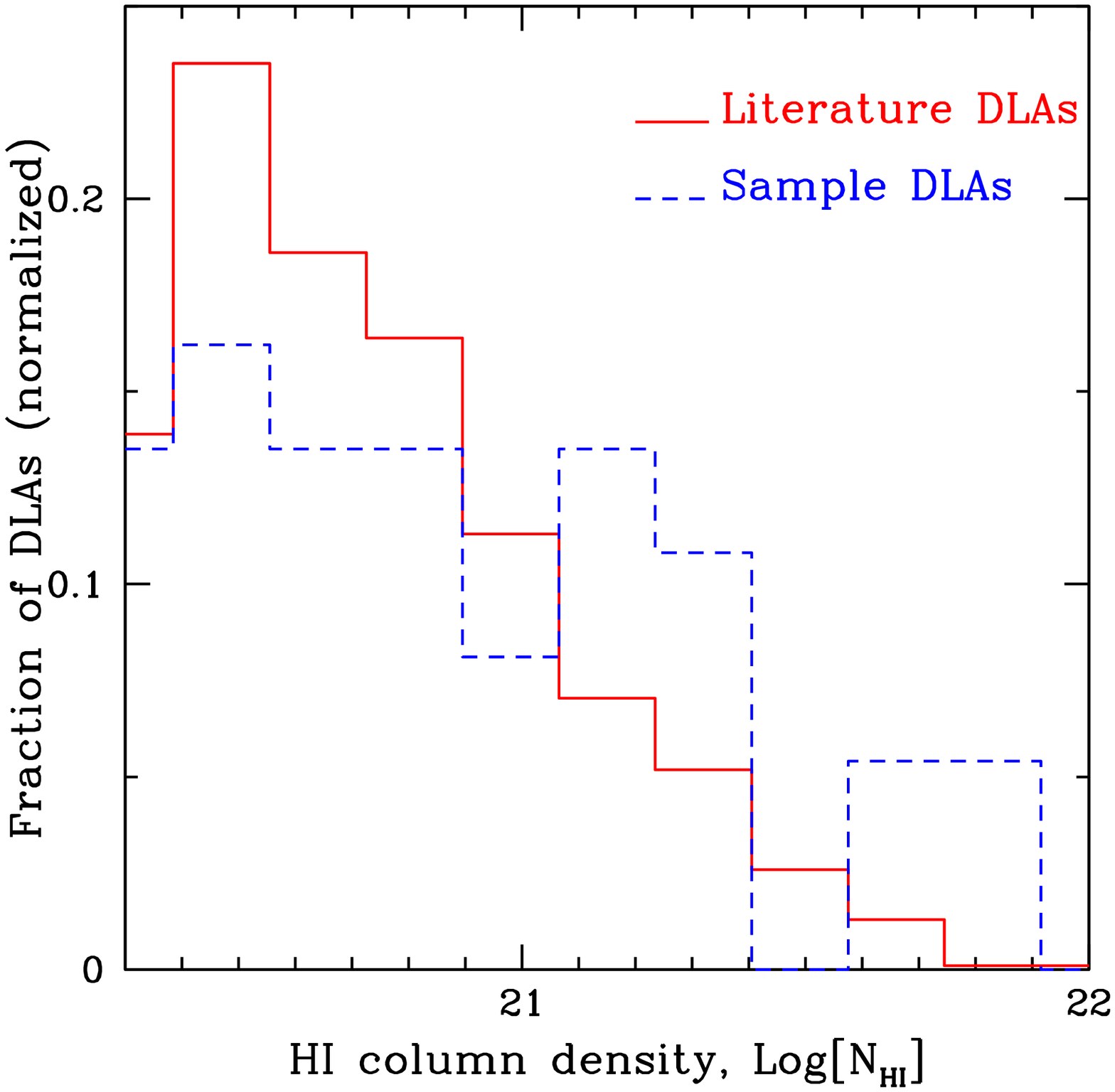,height=3.4truein,width=3.4truein}
\epsfig{file=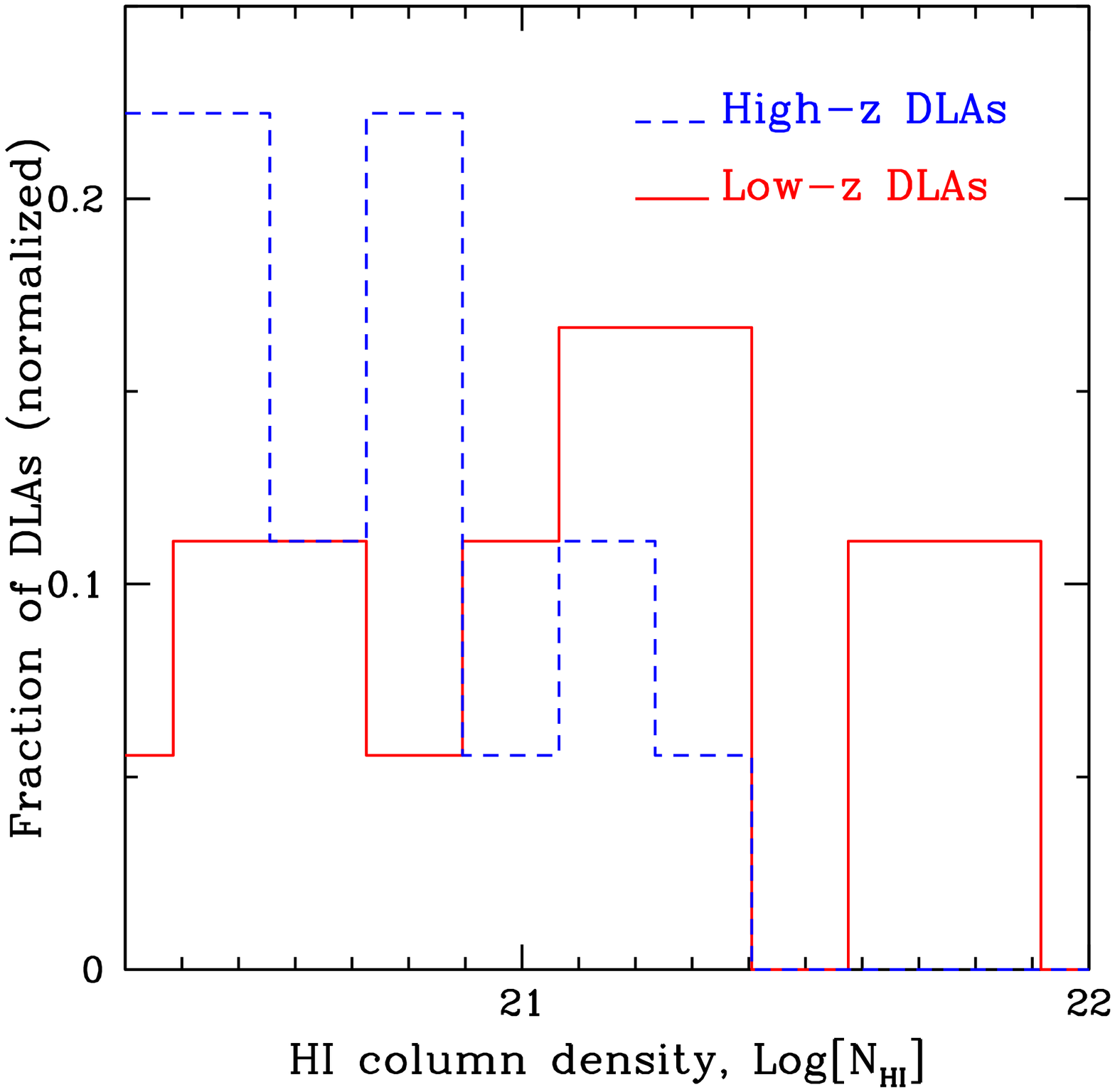,height=3.4truein,width=3.4truein}
\caption{[A]~Top panel: Histograms of the distribution of the \nhi\ values of the 37 DLAs of our sample 
and of a sample of 1080 DLAs from a number of surveys in the literature. A Wilcoxon two-sample test 
finds no statistically-significant evidence against the 
hypothesis that the sample DLAs are drawn from the general population; see main text for discussion.
[B]~Bottom panel: Histograms of the distributions of the \nhi\ values of the low-$z$ (solid) and high-$z$ 
(dashed) DLA sub-samples, separated at the median redshift $z = 2.192$. A Wilcoxon two-sample test
finds that the two \nhi\ distributions are different at $\approx 3\sigma$ significance.}
\label{fig:nhi_hist}
\end{figure}

\section{The full sample}
\label{sec:sample}

Including systems from the literature, the full sample of DLAs and sub-DLAs towards ``compact'' 
radio quasars with estimates of the spin temperature via \hii\ absorption spectroscopy now contains 
60 systems (e.g. \citealp{kanekar03,srianand12,ellison12,kanekar13}; this work).
This includes six sub-DLAs, with $10^{19}$~\cm~$< \nhi < 2 \times 10^{20}$~\cm, and 
seven PDLAs, for which the absorber redshift lies within $3000$~\kms\ of the quasar redshift. 
Note that we have excluded DLAs towards extended radio sources (e.g. at $z = 0.4369$ towards 
3C196 \citep{brown83,briggs01}, $z = 0.5318$ towards 1629+12 \citep{rao00,kanekar03} and 
$z = 0.6561$ towards 3C336 \citep{rao00,curran07c} from the sample, as the optical and 
radio sightlines are very different for these sources.


Seven absorbers of the full sample are PDLAs. As noted earlier, spin temperatures in PDLAs are likely 
to be affected by the proximity of a bright radio source; we will hence exclude these systems from 
our analysis. The six sub-DLAs are also expected to have systematically higher spin temperatures
than DLAs, based on observations in the Galaxy \citep{kanekar11b}; these too will be excluded 
from the analysis. For seven DLAs, the lack of covering factor estimates implies that we only have estimates 
of $\ts/f$ from the \hii\ absorption spectroscopy; we will exclude these systems from the following 
discussion. 37 of the remaining 40 DLAs have covering factor estimates at low frequency 
($\leq 1.4$~GHz) VLBI studies, at frequencies within a factor of $\approx 1.5$ of the redshifted \hii\ 
line frequency \citep[e.g.][; this work]{kanekar09a,ellison12,kanekar13}. The covering factor estimates 
are likely to be reliable in these cases.  However, for three absorbers at $z > 2$, the VLBI images are 
at 1.4~GHz, more than a factor of 3 larger than the redshifted \hii\ line frequency 
\citep{srianand12,kanekar13}. Compact radio cores are likely to have an inverted spectrum 
(due to synchrotron self-absorption), while extended radio structure is expected to have a steep 
spectrum. As a result, measurements of the covering factor at a significantly higher frequency than 
the line frequency could over-estimate the covering factor, and hence, under-estimate the spin temperature.
We will hence also exclude these three systems from our analysis, ending with 37 {\it bona fide} DLAs 
towards compact background quasars with covering factor measurements close to the redshifted 
\hii\ line frequency.

Our final sample of 37 absorbers contains 19 detections of \hii\ absorption, nine at $z < 1$, five 
at $1 < z < 2$, four at $2 < z < 3$ and one at $z > 3$.  Of the 18 remaining systems with 
non-detections of \hii\ absorption, fourteen have strong lower limits on the spin temperature, 
$\ts > 500$~K, while four have weak lower limits, $\ts > 55-270$~K. We emphasize that all systems 
of the sample have estimates of the DLA covering factor from low-frequency VLBI studies. 
These covering factor estimates have been used to derive the DLA spin temperatures.

Detailed information on various properties of the absorbers of the full sample is provided 
in Tables~\ref{table:main} and \ref{table:sub}. The 37 absorbers constituting our final sample are 
listed in Table~\ref{table:main}, while Table~\ref{table:sub} contains the 23 excluded systems: 
(1)~the seven PDLAs, (2)~the six sub-DLAs, (3)~the three DLAs with estimates of $f$ at frequencies 
$> 1.5$ times the redshifted \hii\ line frequency, and (4)~the seven DLAs with no covering factor 
estimates. The columns of both tables contain: (1)~the quasar name, (2)~the quasar 
emission redshift, (3)~the DLA absorption redshift, (4)~the \hi\ column density and error, 
(5)~the covering factor (i.e. the quasar core fraction at low frequencies), (6)~the spin temperature 
or lower limits to $\ts$, after correcting for the absorber covering factor, (7)~the metallicity
[Z/H], (8)~Z, the element used for the metallicity estimate, (9)~the dust depletion [Z/Fe], (10)~``Dust'', 
the element used for the dust abundance, (11)~the velocity spread of the low-ionization metal lines, between 
90\% optical depth points, $\Delta V_{\rm 90}$, in \kms\ \citep{prochaska97}, (12)~the metal-line transition 
on which the $\Delta V_{\rm 90}$ estimate is based, and (13)~references for the \hi\ column density, spin temperature, 
abundances and $\Delta V_{\rm 90}$ values. The metallicities are relative to the solar abundance, on the scale of 
\citet{asplund09}. Following \citet{prochaska03a}, we have used, in order of preference, [Z/H]~$\equiv $~[Zn/H], 
[S/H],~[Si/H] and [Fe/H]+0.4, except for the $z=0.524$ DLA towards AO0235+164, where the metallicity is from 
an X-ray measurement \citep{junkkarinen04}. Most of the $\Delta V_{\rm 90}$ estimates are from our own Keck- HIRES  
or VLT-UVES high-resolution echelle spectra.

\begin{figure*}
\centering
\epsfig{file=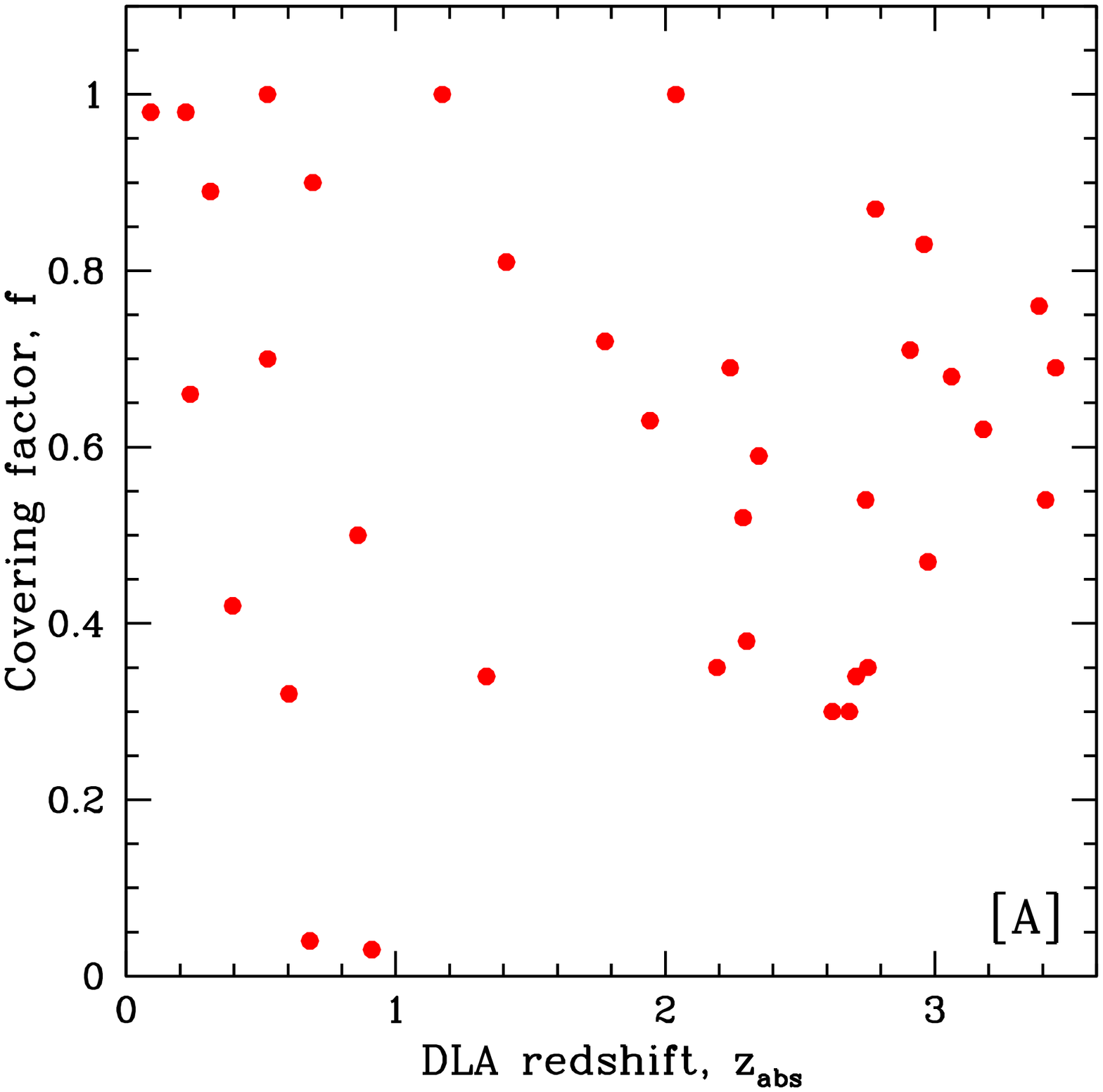,height=3.4truein,width=3.4truein}
\epsfig{file=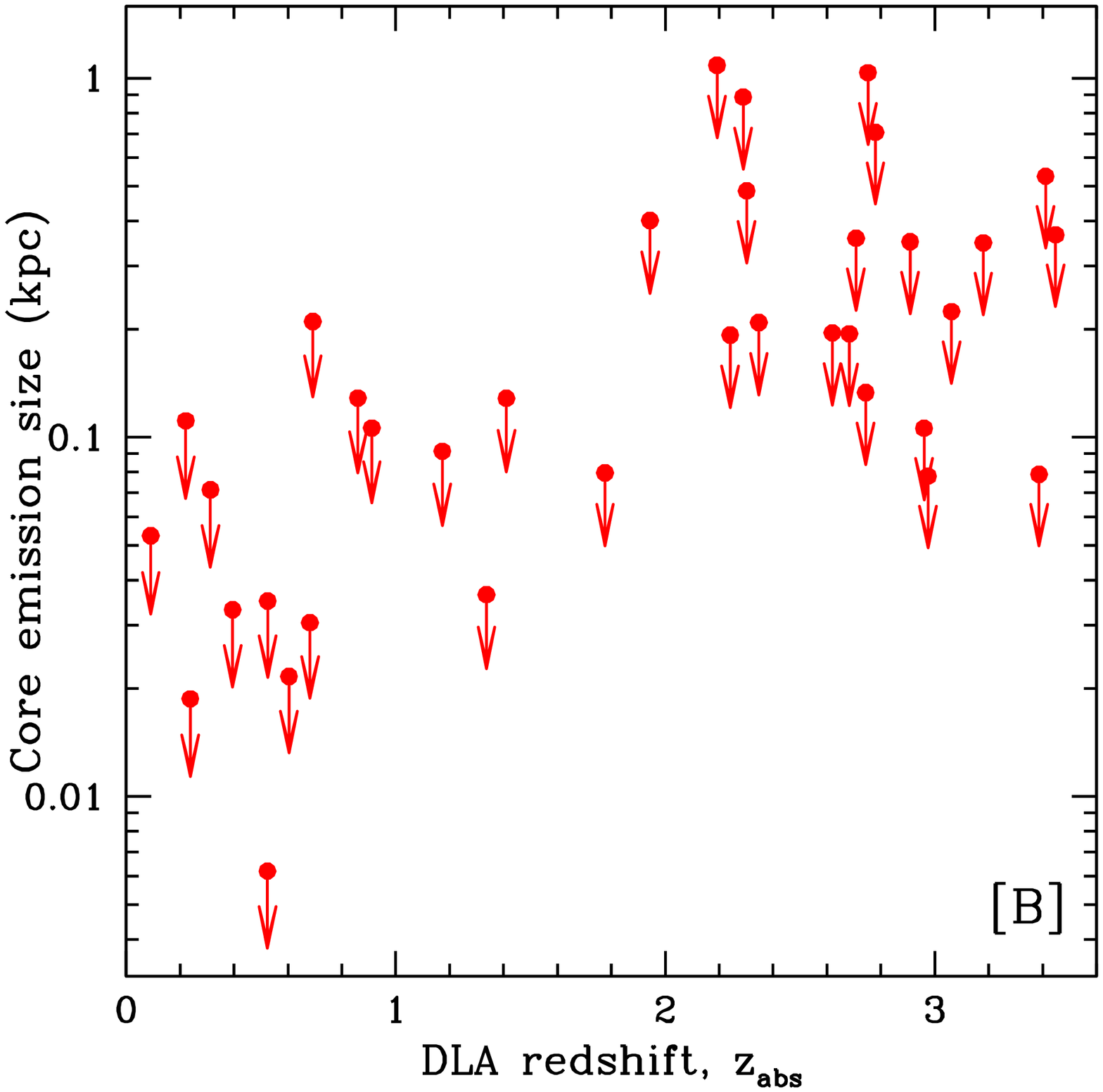,height=3.4truein,width=3.4truein}
\caption{[A]~Left panel: The covering factors of the 37 DLAs of the full sample, plotted 
against absorber redshift, $z_{abs}$. No difference is apparent between the distributions of 
covering factors above and below the median redshift, $z_{med} = 2.192$. [B]~Right panel: The 
spatial extent of the compact radio emission of the background quasars (at the DLA redshift), 
plotted against $z_{abs}$. Note that each point is an upper limit to the size of the radio 
emission, due to the possibility of residual phase errors in the VLBA data. The points have hence 
been shown with downward-pointing arrows. As a result, the limits are much poorer for the DLAs at 
$z \gtrsim 2$, as all these systems were observed with the VLBA at 327~MHz, where the ionospheric 
activity is significantly worse than at higher frequencies. Despite this, the upper limit to the 
core size is $\lesssim 1$~kpc in 36 out of the 37 DLAs of the sample.}
\label{fig:fvsz}
\end{figure*}

\section{Results and Discussion}
\label{sec:ts}

\subsection{The present $\ts$ sample and the general DLA population}

Before proceeding with a discussion of the various properties of our final sample 
of DLAs, it is relevant to test whether the sample is representative of the known 
DLA population, so that inferences drawn from the former can be applied to the 
latter. Since the defining property of a damped Lyman-$\alpha$ system is its 
\hi\ column density, we compared the distribution of \nhi\ values in our sample with 
those of the general DLA population. The comparison sample was made up of 1080 DLAs 
drawn from a variety of surveys
\citep{storrie00,ellison01,jorgenson06,rao06,prochaska05,noterdaeme09}.
The distributions of \nhi\ values in the two samples are shown in Fig.~\ref{fig:nhi_hist}[A],
where the fraction of DLAs plotted on the y-axis has been normalized to unity in 
the bin with the largest fraction to allow for a direct comparison between the two 
distributions. A similar plot in Fig.~\ref{fig:nhi_hist}[B] compares the \nhi\ 
distributions in the high-$z$ and low-$z$ DLA samples.

While the fraction of high-$\nhi$ DLAs appears larger in the \hii\ sample, the number of 
absorbers in this sample is quite small (37 systems), implying large fluctuations from Poisson 
statistics. The literature sample is dominated by the SDSS DLAs, at $z \ge 2.2$; conversely, 
the median absorber redshift in our sample is $z_{med} = 2.192$, so our sample contains 
a significantly higher fraction of low-$z$ DLAs than the literature sample (e.g. 
$\approx 40$\% at $z < 1.5$ in our sample, against $\approx 3$\% for the literature sample). 
While the samples have different redshift distributions, a number of studies have 
found no evidence for redshift evolution in the \hi\ column density distribution function
\citep[e.g.][]{prochaska05,zwaan05}. We find that the \nhi\ distributions of our sample and 
the literature DLAs agree within $1.8 \sigma$ significance in a Wilcoxon two-sample test\footnote{The 
statistical tests described here were mostly carried out using the 
Astronomical Survival analysis {\sc asurv} package \citep{isobe86}.}. We hence conclude 
that there is no statistically significant evidence against the null hypothesis that 
the DLAs of our sample are representative of the general DLA population.

\subsection{The covering factor}
\label{sec:f}

\begin{figure}
\centering
\epsfig{file=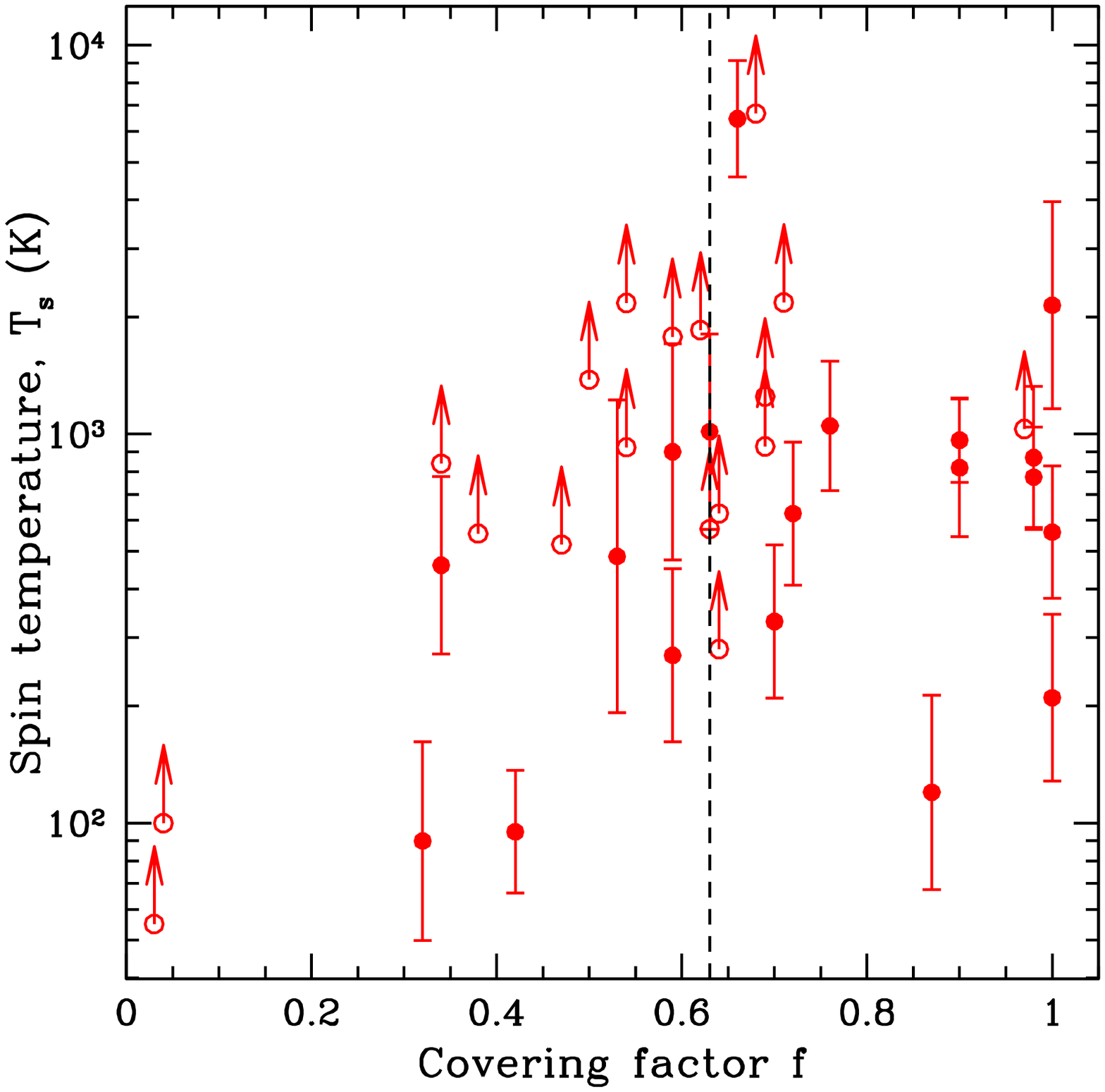,height=3.4truein,width=3.4truein}
\caption{The spin temperature $\ts$ plotted against the covering factor for the 37 DLAs 
of the sample; non-detections of \hii\ absorption are indicated by lower limits 
to the spin temperature. The median covering factor $f_{med} = 0.63$ is indicated by the 
vertical dashed line. No evidence was found that either the detection probability or the spin 
temperature depends on the absorber covering factor.}
\label{fig:tsvsf}
\end{figure}

It has been suggested by \citet{curran06} \citep[see also][]{curran12} that 
unknown covering factors are the main cause of the high spin temperature 
estimates in high-$z$ DLAs. These authors argue that this is due to a geometric 
effect, owing to the fact that the angular diameter distances of absorbers 
at $z \gtrsim 1$ are comparable to those of the background quasars (at $z > 1$), 
implying that higher-redshift absorbers are ``less efficient'' at obscuring the 
background radio emission and hence have lower covering factors. While this may or may 
not be applicable for systems which do not have estimates of the low-frequency 
covering factor $f$, all 37 DLAs of our sample have estimates of the covering factor at 
frequencies within a factor of $1.5$ of the redshifted \hii\ line frequency. The estimated 
covering factors are plotted versus redshift in Fig.~\ref{fig:fvsz}[A]: no evidence was 
found for a difference between the distributions of covering factors for the sub-samples 
of absorbers with $z < z_{med}$ and $z > z_{med}$ (note that $z_{med}=2.192$); the distributions 
agree within $0.7\sigma$ significance in a Wilcoxon two-sample test. 

Fig.~\ref{fig:fvsz}[B] plots the upper limit to the spatial extent of the core radio 
emission at the DLA redshift versus the DLA redshift. While the spatial extent does 
appear to be larger at high redshifts, it should be emphasized that the core spatial 
extents plotted in Fig.~\ref{fig:fvsz}[B] are {\it upper limits}, due to the possibility 
of residual 
phase errors in the VLBA data. This issue is especially important at low frequencies 
(327~MHz and 606~MHz), due to ionospheric effects. As a result, the size estimates of 
Fig.~\ref{fig:fvsz}[B] are systematically larger for the DLAs at $z \gtrsim 2$, as 
these were all observed with the 327~MHz VLBA receivers. 

Fig.~\ref{fig:fvsz}[B] shows that the background radio cores have a transverse size of 
$\lesssim 1$~kpc at the absorber redshift in 36 of the 37 absorbers, significantly smaller 
than the size of even dwarf galaxies. For DLAs at $z > 2$, the median upper limit (due to 
the possibility of residual phase errors) to the transverse size of the core emission is $\approx 350$~pc.
This indicates that the radio cores are likely to be entirely covered by all the foreground 
absorbers. The sole exception is 0458$-$020, where the agreement between VLBI and single-dish 
\hii\ absorption profiles led \citet{briggs89} to argue that the $z = 2.0395$ DLA covers both 
the radio core and the extended lobes, implying an absorber size $\gtrsim 16$~kpc; note that 
this system lies off the scale in Fig.~\ref{fig:fvsz}[B]. 

We also examined whether the probability of detecting \hii\ absorption or 
the estimated spin temperatures depend on the absorber covering factors. Fig.~\ref{fig:tsvsf} 
plots the spin temperature versus covering factor for the 37 DLAs of the sample. 
The median covering factor of the absorbers in the sample is $f_{med} = 0.64$. For sightlines 
with $f > f_{med}$, there are 12 detections and 5 non-detections of \hii\ absorption, while 
sightlines with $f < f_{med}$ have 7 detections and 11 non-detections. The detection
probabilities are $71^{+27}_{-20}$\% for the high-$f$ sample and $37^{+20}_{-13}$\%
for the low-$f$ sample.\footnote{The quoted errors are $1\sigma$ Gaussian confidence intervals, 
based on small-number Poisson statistics \citep{gehrels86}. Note that the error does 
not scale linearly with confidence interval, e.g. the $2\sigma$ lower confidence interval is 
not twice the $1\sigma$ lower confidence interval, and care is hence needed while comparing
detection rates for different sub-samples. In all cases, we have used two approaches to test 
whether the detection rates for different sub-samples are in agreement: (1)~naively
assuming that the errors scale linearly with confidence interval, and (2)~checking whether
the $2\sigma$ confidence intervals of the detection rates of the sub-samples overlap with 
each other. The quoted differences in the detection rates are from the first approach.}
Thus, while the detection rate appears higher for the 
high-$f$ sample (as would be expected on physical grounds), the difference is not 
statistically significant (agreeing within $\approx 1.2\sigma$ significance) in the current 
sample. Similarly, a Peto-Prentice generalized Wilcoxon two-sample test (for censored data) 
finds no evidence for a difference between the distributions of $\ts$ values for the 
high-$f$ and low-$f$ samples; the two $\ts$ distributions agree within $0.1\sigma$ 
significance. We thus find no statistically significant evidence that the probability 
of detecting \hii\ absorption or the spin temperature estimates depend on the 
absorber covering factor.

The fact that we do not detect a significant dependence of the detection rate of \hii\ 
absorption on covering factor may appear surprising. This is likely to be at least partly 
due to the fact that the sample is still relatively small, with only $\approx 18$ 
systems apiece in the high-$f$ and low-$f$ sub-samples. Further, the sensitivity to \hii\ 
absorption is not uniform across the sample: the optical depth sensitivity is typically 
better for brighter background sources.

We will revisit the arguments of \citet{curran06} regarding angular diameter distances 
in Section~\ref{sec:ts_or_f}.

\subsection{Differing \hi\ columns along optical and radio sightlines}
\label{sec:lmc}

\begin{figure}
\centering
\epsfig{file=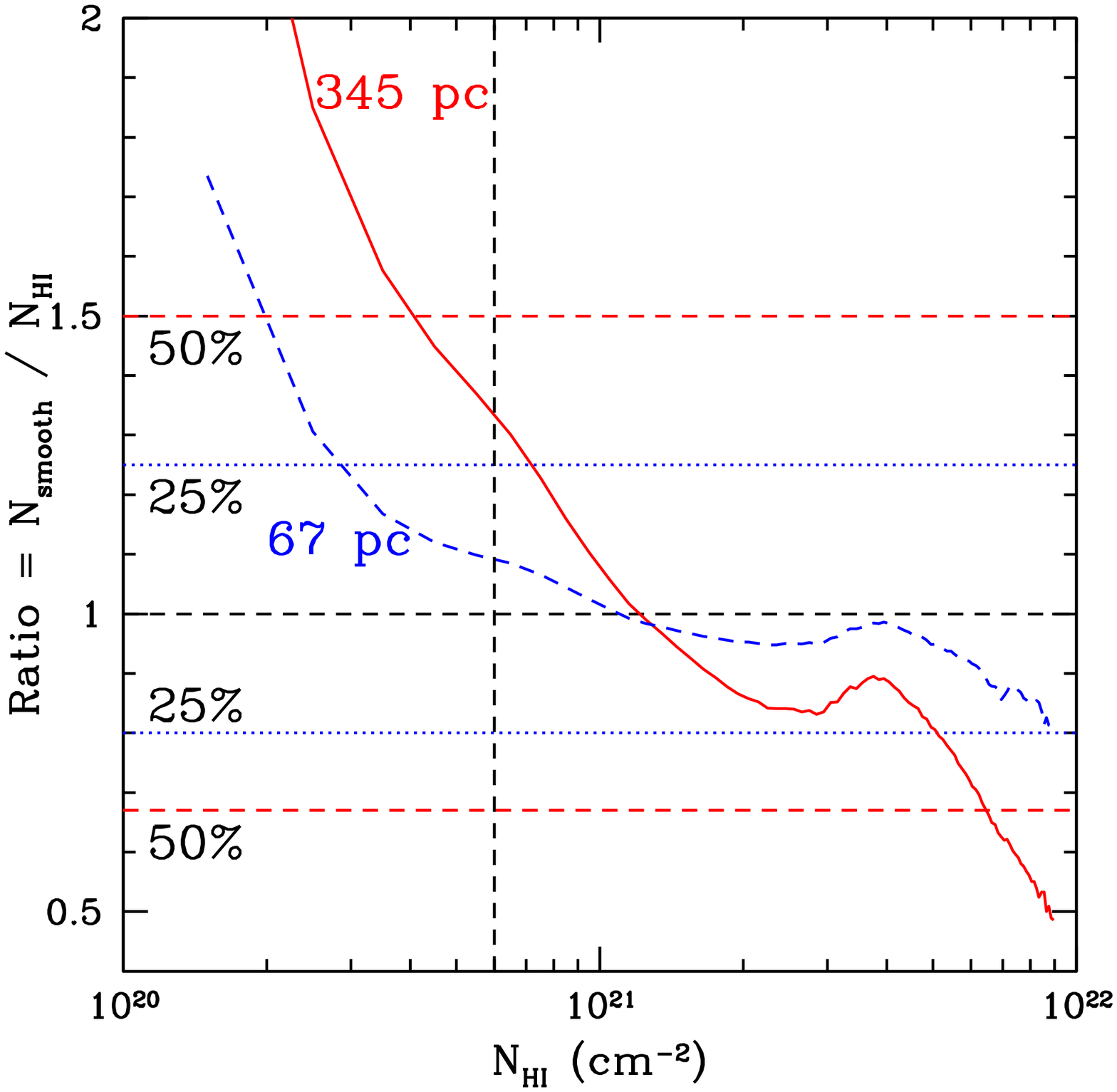,height=3.4truein,width=3.4truein}
\caption{\hi\ in the LMC: the ratio of the smoothed \hi\ column density N$_{\rm smooth}$ 
at spatial resolutions of 67~pc (dashed curve) and 345~pc (solid curve) to the \hi\ 
column density at the original (15~pc) resolution N$_{\rm HI}$ plotted versus N$_{\rm HI}$. 
The horizontal dotted and dashed lines respectively demarcate regions where the ratio 
is within 25\% and 50\% of unity. The dashed vertical line indicates $\nhi = 
6 \times 10^{20}$~\cm, the median \hi\ column density of the DLA sample.}
\label{fig:lmc}
\end{figure}

As noted in Section~\ref{sec:tspin}, a critical assumption in estimating spin temperatures 
in DLAs is that the \hi\ column densities measured from the damped Lyman-$\alpha$ profile 
can be used in equation~(\ref{eqn:tspin}), i.e. that the $\nhi$ values along the optical 
and radio sightlines are the same. If the \hi\ columns are systematically different 
along the radio and the optical sightlines (e.g. if there is significant spatial structure in 
the \hi\ distribution), it would lead to systematic errors in the derived spin temperatures. 
For example, \citet{wolfe03b} suggest that the average \hi\ column towards the radio core 
of 0201+113 may be significantly lower than that towards the optical quasar, thus resulting 
in an incorrectly high estimate of the spin temperature. Note that this effect is not the 
same as the issue of the geometric covering factor discussed in the preceding section, although 
the two are often confused in the literature.

Absorption spectroscopy of lensed quasars provides a interesting tool to measure \hi\ column 
density along multiple sightlines through a DLA, and to thus test for structure in the 
\hi\ distribution \citep[e.g.][]{smette95,zuo97,lopez05,monier09}. For example, 
\citet{smette95} measured \hi\ column densities of 
${\rm log} [\nhi/cm^{-2}] = 20.6$ and ${\rm log} [\nhi/cm^{-2}] = 17.6$ at $z = 1.6616$ 
towards images A and B of quasar HE~1104$-$1805, respectively, at a separation of 
$\approx 3''$ ($\approx 26$~kpc at the redshift). Conversely, \citet{lopez05} 
found the \hi\ column densities in a $z = 0.9313$ galaxy towards images A and B of the 
quasar HE~0552$-$3315 to be in excellent agreement, at a separation of $0.644''$
($\approx 5.2$~kpc at the absorber redshift). Unfortunately, there are as yet only a 
handful of DLAs known towards lensed quasars.

The most direct way of testing whether the \hi\ columns are different along the radio and 
optical sightlines is to compare \hi\ column densities 
measured from pencil-beam (i.e. ultraviolet or optical) studies and studies at coarser 
angular resolution. This has been done in the Galaxy \citep{dickey90,wakker11}, by 
comparing $\nhi$ values obtained at the same or neighbouring locations from Lyman-$\alpha$ 
absorption studies of background quasars and \hii\ emission studies. \citet{wakker11} 
carried out such a comparison along 59 sightlines, using HST-STIS Lyman-$\alpha$ 
absorption spectra and \hii\ emission spectra at angular resolutions of $9' - 36'$ 
from the GBT and the 140-foot telescope. For the former, the $\nhi$ values were derived 
from the usual Voigt profile fits to the damping wings of the Lyman-$\alpha$ line, as 
is done for high-$z$ DLAs. For the latter, the $\nhi$ values were inferred under the 
assumption that the \hii\ emission is optically-thin, in which case the observed
\hii\ emission brightness temperature is directly proportional to the \hi\ column density 
\citep{rohlfs06}. \citet{wakker11} found the \hi\ column densities obtained from the two 
methods to be in excellent agreement, with the ratio having a mean value of unity and a 
dispersion of 10\%. Note that one does not know the distance to the absorbing clouds, and 
hence cannot estimate the spatial resolution of the \hii\ emission spectra. If the clouds are at an 
average distance of $\approx 1$~kpc, it would imply that the \hi\ column density averaged 
over regions of size $\approx 10$~pc is similar to that on scales of $100-1000$~AU.

Similarly, \citet{welty12} compared \hi\ column densities in the Large Magellanic 
Cloud (LMC) measured from the Lyman-$\alpha$ absorption profile towards stars in the LMC 
and \hii\ emission spectroscopy at a spatial resolution of 15~pc. They found good 
agreement between the \nhi\ values by excluding \hii\ emission from gas lying behind 
the stars (based on the emission velocity). This too indicates that the \hi\ column 
density on scales of $\approx 100-1000$~AU is similar to that on scales of $\approx 15$~pc.

However, Fig.~\ref{fig:fvsz}[B] shows that the spatial extent subtended by the radio core
at the DLA redshift may be significantly larger than 10~pc, with a median value of $\approx 
350$~pc for DLAs at $z > 2$. Although, as noted in the preceding section, these transverse
sizes are upper limits, most of the background quasars are likely to have radio cores 
of spatial extent greater than a few tens of parsecs at the redshift of the foreground 
DLA \citep[see][for the $z = 3.387$ DLA towards 0201+113]{wolfe03b}. This is because 
the angular size of the radio core must be sufficiently large at low frequencies to have 
a brightness temperature lower than either the equipartition limit or the inverse 
Compton limit \citep[$\approx 10^{11} - 10^{12}$~K;][]{kellermann69,readhead94,singal09},
unless the radio emission is relativistically beamed towards us. Relativistic beaming is 
unlikely to be important for most of the background quasars of our sample, as few of them 
have highly variable flux densities. Thus, for most of the absorbers of the sample, the radio core 
emission is likely to subtend a transverse size of tens of pc, and perhaps as large as a 
few hundred pc, at the DLA redshift. 

One thus has to hence examine whether the \hi\ column density averaged over spatial scales of 
a few hundred pc is systematically different from that measured along a pencil beam. 
A direct comparison between the \hi\ column densities measured from the Lyman-$\alpha$ 
line and from \hii\ emission has so far been carried out in a single DLA, at $z = 0.009$ towards 
SBS~1549+593 \citep{bowen01a}. For this system, \citet{chengalur02} measured $\nhi \approx (4.9 \pm 0.6) 
\times 10^{20}$~\cm\ from a GMRT \hii\ emission study with a spatial resolution of $\approx 5.3$~kpc 
at the DLA redshift, while \citet{bowen01a} obtained a slightly lower value, 
$\nhi \sim (2.2 \pm 0.5) \times 10^{20}$~\cm, from the damped Lyman-$\alpha$ profile (albeit 
with unknown systematic errors due to the extended wings of the STIS G140L grating line spread 
function and blending with the Lyman-$\alpha$ line of the Galaxy; \citealt{bowen01a}). Note that 
the spatial resolution of the \hii\ study here is far larger than the typical transverse scale 
subtended by the radio cores at the redshifts of the DLAs of our sample.

An alternative approach is to compare the \hi\ column density measured at high spatial resolution 
along sightlines in an \hii\ emission cube of a nearby galaxy with the $\nhi$ values obtained along 
the same sightlines on smoothing the cube to lower spatial resolutions. Such a comparison was carried 
out by \citet{ryanweber05}, between $\nhi$ values measured at 15~pc and 3.6~kpc resolution.
These authors found that the fine structure --- both low and high column density --- is washed out 
on smoothing to 3.6~kpc resolution, resulting in sightlines with original $\nhi$ values in 
the sub-DLA regime returning higher values and gas with $\nhi >10^{21}$~\cm\ yielding values with 
$\nhi < 10^{21}$~\cm. 

We follow the same approach as \citet{ryanweber05}, analysing a high-spatial-resolution \hii\ emission 
image of the LMC, obtained by combining data from the Australia Telescope Compact Array and the 
Parkes multi-beam receiver \citep{kim03b}. The spectral cube has a spatial resolution of 
$\approx 15$~pc at the distance of the LMC. We have smoothed the cube to 
different spatial resolutions, up to a maximum of 1~kpc, and measured the \hi\ column density 
at the same position at each resolution, assuming the \hii\ emission to be optically thin in
all cases. Finally, we averaged the results from different spatial locations with the same 
value of the original, unsmoothed \hi\ column density, using uniform $\nhi$ bins.

The results are shown in Fig.~\ref{fig:lmc}, which plots the ratio of the \hi\ column density measured 
at two different spatial resolutions (67~pc and 345~pc) to that at the original resolution (15~pc) 
versus the unsmoothed $\nhi$. These spatial resolutions were chosen because the median transverse sizes 
of quasar radio cores at the absorber redshift are $\approx 70$~pc and $\approx 350$~pc for DLAs 
at $z < 2$ and $z > 2$, respectively, allowing a direct comparison. 

It is clear that the behaviour depends on the absolute value of the \hi\ column density: at both 
smoothed resolutions, the ratio is larger than unity (i.e the smoothed \hi\ column density is 
larger than the original) for $\nhi \lesssim 10^{21}$~\cm, and lower than unity (i.e. the 
smoothed $\nhi$ is lower than the original) for $\nhi \gtrsim 10^{21}$~\cm. However, smoothing to 
a resolution of $\approx 67$~pc does not significantly alter the \hi\ column density over most 
of the range: the mean value of the ratio is within $\approx 10$\% of unity for $5.0 \times 
10^{20}$~\cm~$ < \nhi < 6.3 \times 10^{21}$~\cm. Even for $2 \times 10^{20}$~\cm~$< \nhi 
< 5 \times 10^{20}$~\cm, the ratio is $< 1.5$.

The situation is similar in the case of smoothing to a spatial resolution of $345$~pc, 
although with somewhat larger deviations from unity. Here, the ratio is within 25\% of unity for 
$\nhi = (0.7 - 5.1) \times 10^{21}$~\cm, within 50\% of unity over the ranges 
$\nhi = (5.1 - 6.5) \times 10^{21}$~\cm\ and $\nhi = (4 - 7) \times 10^{20}$~\cm, and lower than 
2 for $\nhi = (2 - 4) \times 10^{20}$~\cm.

The above results can be used to draw inferences about the derived spin temperatures in DLAs, 
{\it assuming that the spatial distribution of \hi\ in DLAs is similar to that in the LMC}.
From the above results, using the \hi\ column density measured from the damped Lyman-$\alpha$ 
profile in the equation for the \hii\ optical depth to derive the spin temperature would not 
result in a {\it systematically higher or lower} $\ts$ estimate in all DLAs. The spin temperature 
would be slightly under-estimated for DLAs with $\nhi < 1 \times 10^{21}$~\cm, and over-estimated for 
DLAs with $\nhi > 1 \times 10^{21}$~\cm. 

All DLAs of the sample have \hi\ column densities in the range $2 \times 10^{20} - 6.3 \times 10^{21}$~\cm.
For absorbers at $z < 2$ (where the median transverse size of the radio core is $\approx 69$~pc), 
the derived spin temperature would be within 10\% of the ``true'' value for all $\nhi$ values 
except the range $2 \times 10^{20}$~\cm~$< \nhi < 5 \times 10^{20}$~\cm, where the derived 
$\ts$ could be lower than the ``true'' value by up to a factor of 1.5. Conversely, for 
DLAs at $z > 2$ (where the median transverse size of the radio core is $\approx 350$~pc),
the derived $\ts$ would be within 50\% of the correct value for all $\nhi$ values except for 
the range $\nhi = (2 -4) \times 10^{20}$~\cm. For the latter range, 
the $\ts$ estimate could be lower than the ``true'' value by up to a factor of 2.

The median \hi\ column density of the present DLA sample is $\nhi \approx 6 \times 10^{20}$~\cm, 
with 23 absorbers having $\nhi \leq 1 \times 10^{21}$~\cm. Of the remaining 14~systems, twelve
have $\nhi < 5.1 \times 10^{21}$~\cm\ and should thus, on the average, have systematic errors lower 
than 25\% on the spin temperature. There are thus only two absorbers, at $z = 1.9436$ towards 
1157+014 and $z = 2.0395$ towards 0458$-$020, whose \hi\ column densities are in the regime where 
the spin temperature might be over-estimated by $\approx 50$\%. Conversely, there are 10 DLAs in 
the sample with $\nhi < 4 \times 10^{20}$~\cm, seven of which are at $z > 2$, for which the spin 
temperature could be under-estimated, on the average, by using the optical $\nhi$ in 
equation~(\ref{eqn:tspin}). 

For clarity, we emphasize that the above statements are only valid in a statistical sense; 
it is certainly possible that the \hi\ column densities along the radio and optical sightlines
are very different in individual absorbers. However, on the average, the use of the \hi\ column 
density derived from the Lyman-$\alpha$ profile is likely to result in {\it under-estimating} 
the spin temperature for most DLAs of the present sample, by up to a factor of two at the lowest 
\hi\ column densities. The spin temperature is likely to be over-estimated for only a few DLAs 
with the highest $\nhi$ values, $\nhi > 5 \times 10^{21}$~\cm.

\subsection{\hii\ optical depths, DLA spin temperatures and \hi\ column densities}
\label{sec:ts2}

\begin{figure*}
\centering
\epsfig{file=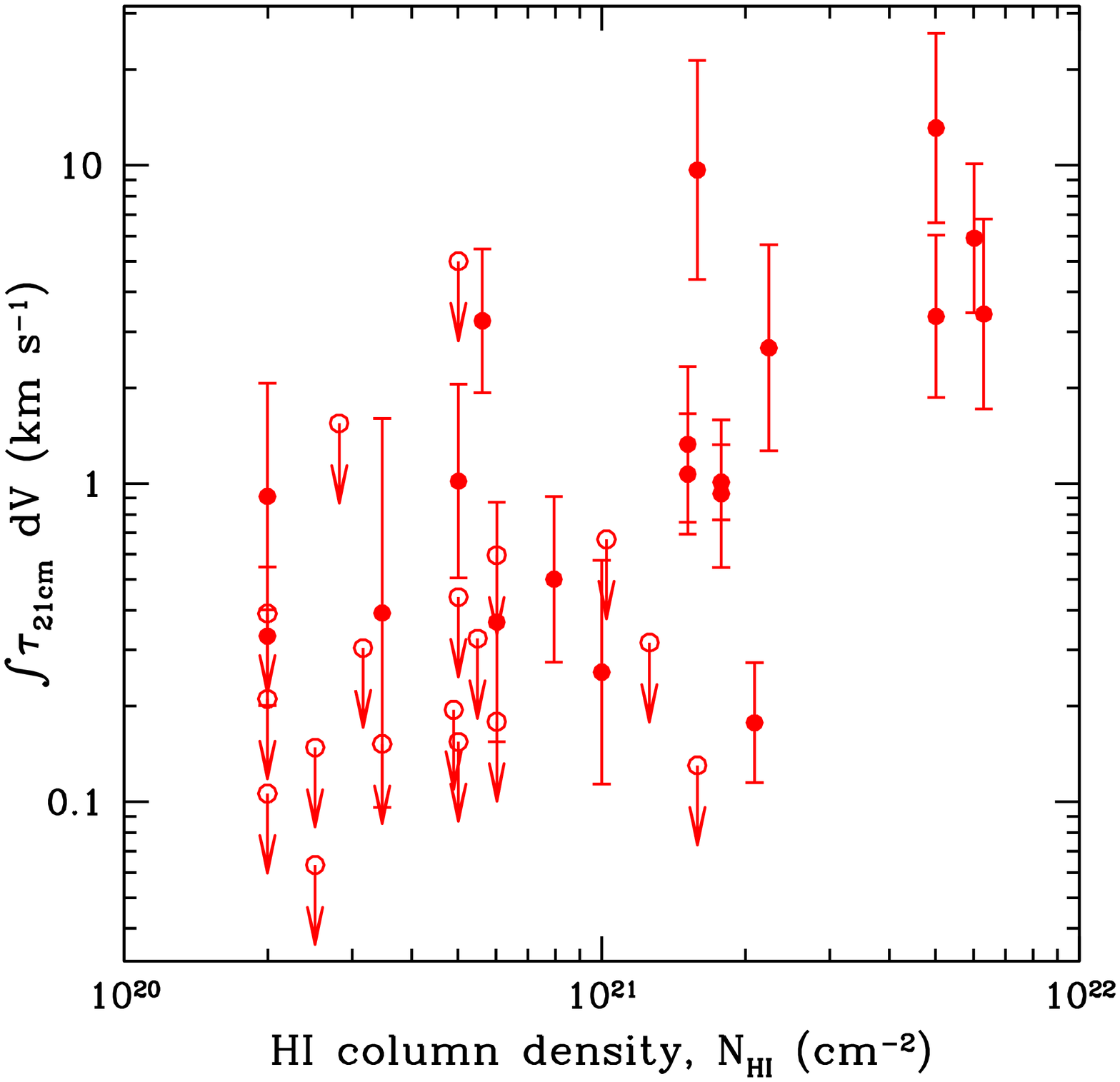,height=3.4truein,width=3.4truein}
\epsfig{file=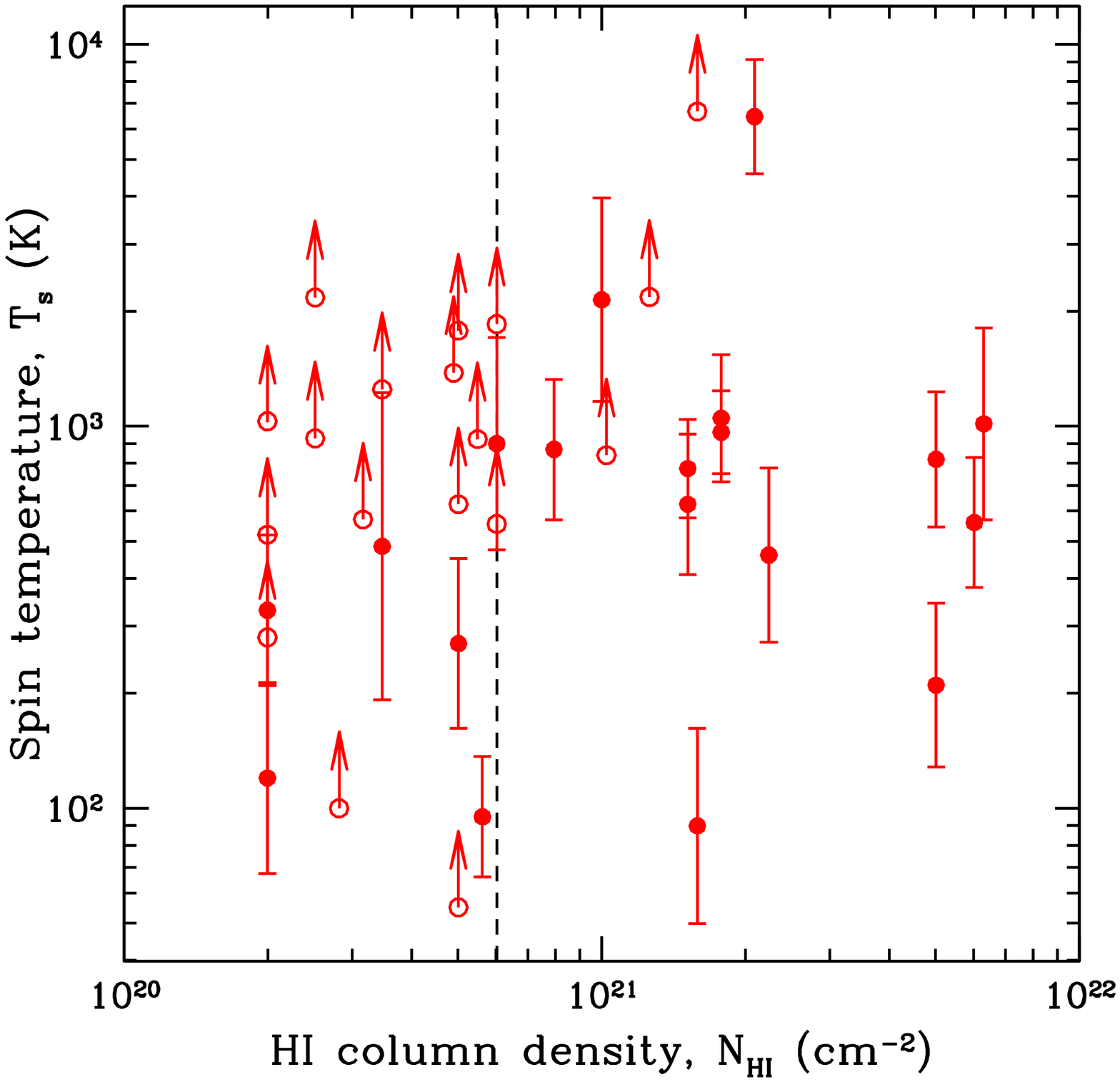,height=3.4truein,width=3.4truein}
\caption{[A]~(left panel) The integrated \hii\ optical depth $\int \tau_{\rm 21cm} {\rm d}V$, 
corrected for DLA covering factor, and [B]~(right panel) the spin temperature $\ts$, 
plotted against the \hi\ column density $\nhi$ for the 37 DLAs of the sample. Non-detections 
of \hii\ absorption are plotted as open circles, with arrows indicating upper limits to the 
\hii\ optical depth and lower limits to the spin temperature. The dashed vertical line
in the right panel indicates the median value of the \hi\ column density, $\nhi \approx 6 \times 10^{20}$~\cm.
See text for discussion.}
\label{fig:tdvnhi}
\end{figure*}

It is well known that the strength of the \hii\ absorption in the Galaxy and M31 shows a 
rough power-law dependence on the \hi\ column density \citep[e.g.][]{braun92,carilli96,kanekar11b}.
A similar trend has been seen in earlier studies of DLAs, using smaller samples and without 
covering factor measurements \citep{carilli96,kanekar01a}. For the current sample of 37 DLAs,
Fig.~\ref{fig:tdvnhi}[A] shows the integrated \hii\ optical depth ($\int \tau_{\rm 21cm} {\rm d}V$), 
corrected for absorber covering factor, plotted against the \hi\ column density $\nhi$. It is clear 
that the strength of the \hii\ absorption does correlate with $\nhi$; the correlation is detected at 
$\approx 3.8\sigma$ significance in a non-parametric generalized Kendall-tau test \citep{brown74,isobe86}.

Fig.~\ref{fig:tdvnhi}[B] shows the spin temperature  plotted versus \hi\ column density 
for the 37 DLAs of the sample; no evidence is seen for a relation between the two quantities. 
We further tested whether the DLA spin temperatures are systematically different in absorbers
with low and high \hi\ column densities, below and above the median $\nhi \approx 6 \times 10^{20}$~\cm.
A Peto-Prentice test finds that the spin temperature distributions of the sub-samples 
with $\nhi$ values above and below the median are in agreement within $\approx 0.8\sigma$ significance. 

We also considered the possibility that the $\nhi$ values along the radio and optical 
sightlines might be somewhat different at low and high \hi\ column densities, as inferred from the 
results in Section~\ref{sec:lmc} for the LMC. This was done by using the ratio of the \hi\ 
column densities measured along the high-resolution and the smoothed sightlines in the LMC 
(from Fig.~\ref{fig:lmc}) to convert the measured $\nhi$ values towards the optical QSO to 
that towards the radio core. For DLAs at $z < 2$ and $z > 2$, we used the ratios measured at 
smoothed resolutions of $67$~pc and 345~pc, respectively, in each case equal to the median size
of the core at the DLA redshift. We also corrected the spin temperature for the new \nhi\ value.
Again, no significant difference was obtained in the spin temperature distributions for the 
sub-samples with low and high \hi\ column densities; the distributions agree at $1.4\sigma$
significance in a Peto-Prentice test.

We thus find no evidence that DLA spin temperatures have a systematic dependence on the \hi\ 
column density (see also \citealt{srianand12}). Note that the spin temperatures of the Galactic 
sample also do not show a statistically significant dependence on the \hi\ column density: a 
similar Peto-Prentice test finds that the Galactic spin temperature distributions above and below 
the median \hi\ column density ($1.48 \times 10^{21}$~\cm) are in agreement within 
$\approx 1.3 \sigma$ significance.

In order to consider whether the detection rate of \hii\ absorption depends on \hi\ column
density, we restricted the sample to the 34 DLAs with either detections of \hii\ absorption 
or strong lower limits on the spin temperature ($\ts > 500$~K). This was done to avoid the 
possibility that our conclusions might be biased by DLAs with weak limits on the \hii\ 
optical depth. The median \hi\ column density of this sub-sample is $\nhi = 6 \times 10^{20}$~\cm.
The detection rates of \hii\ absorption are $\approx 81^{+19}_{-22}$\% and 
$\approx 33^{+23}_{-14}$\% for absorbers with \hi\ column densities higher and lower than 
$6 \times 10^{20}$~\cm, respectively. While the detection rate appears higher in the high-$\nhi$ 
sample, the difference has only $1.3\sigma$ significance. At present, we find no statistically 
significant evidence that the detection rate of \hii\ absorption in DLAs depends on the \hi\ column density
(although we note that this too might arise due to the relatively small size of the two sub-samples).

\subsection{Redshift evolution of DLA spin temperatures}
\label{sec:ts3}

\begin{figure}
\centering
\epsfig{file=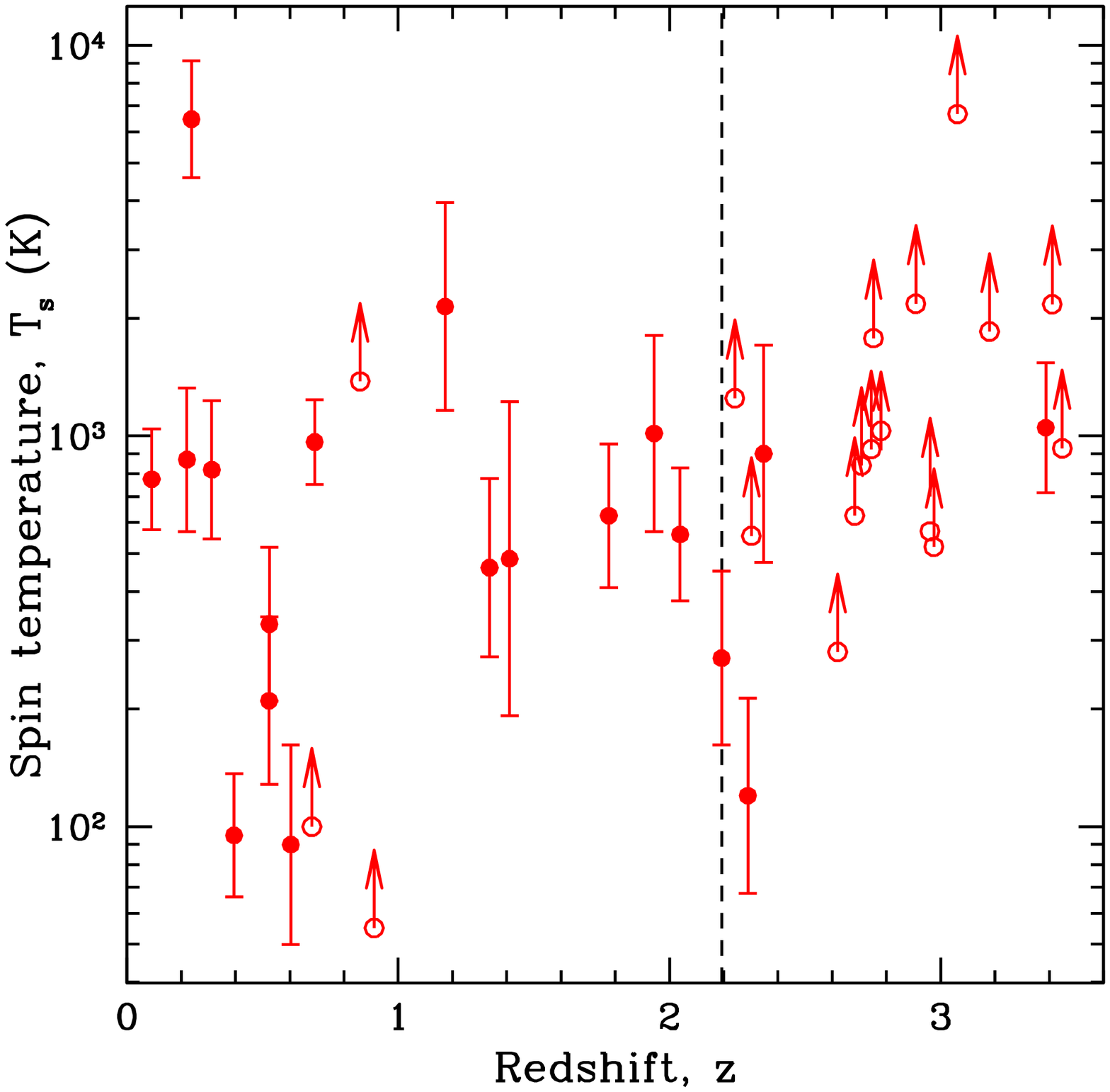,height=3.4truein,width=3.4truein}
\caption{The spin temperature $\ts$ plotted against redshift for the 39 DLAs (filled circles) 
and sub-DLAs (open stars) of the sample;
non-detections of \hii\ absorption are indicated by lower limits to the spin temperature. 
The median absorption redshift, $z_{med} = 2.192$, is indicated by the vertical dashed 
line. A generalized Wilcoxon two-sample test finds that the distributions of spin temperatures 
are different in the high-$z$ and low-$z$ sample; see text for discussion.}
\label{fig:tsvsz}
\end{figure}

The high spin temperatures obtained in DLAs have been an issue of interest ever since 
the earliest studies of \hii\ absorption in high-$z$ DLAs \citep{wolfe79,wolfe81}. There are
two separate questions here: (1)~whether DLA spin temperatures are systematically different 
from those measured in the Milky Way and local spiral galaxies, and (2)~whether DLA spin 
temperatures evolve with redshift (i.e. whether low-$z$ and high-$z$ DLAs have different
distributions of spin temperatures). While a number of studies have indicated that high-$z$ DLAs 
tend to have higher spin temperatures than those seen in both low-$z$ DLAs and the Galaxy 
\citep[e.g.][]{carilli96,chengalur00,kanekar01a,kanekar03,srianand12}, the results have been 
uncertain due to a combination of small sample size and lack of covering factor measurements.

Fig.~\ref{fig:tsvsz} shows the spin temperature plotted versus redshift for the 37 DLAs of our 
sample; the median absorber redshift is $z_{med} = 2.192$. We emphasize that 
all systems have estimates of the absorber covering factor and that this is by far the largest 
sample that has been used for such studies. There are 3 detections and 15 non-detections of 
\hii\ absorption in DLAs at $z > 2.192$, and 15 detections and 3 non-detections of \hii\ 
absorption in systems at $z< 2.192$. The detection fractions are $(17^{+16}_{-9})$\% 
($z > 2.192$) and $(83^{+17}_{-21})$\% ($z < 2.192$). The detection fraction thus appears higher 
in the low-$z$ sample, albeit only at $\approx 2.5\sigma$ significance \citep[see also][]{kanekar09b}. 
However, a comparison between the two spin temperature distributions via a Peto-Prentice test
finds that the $\ts$ distributions are different at $\approx 3.5\sigma$ significance. Note there 
is nothing special about the median redshift, $z = 2.192$, chosen to demarcate the high-$z$ and 
low-$z$ DLA sub-samples. For example, a comparison between the $\ts$ distributions of absorber 
at redshifts above and below $z = 2.4$ (23 systems in the low-$z$ sample and 14 in the high-$z$ 
sample), yields a difference with $4.0\sigma$ statistical significance. We conclude that there 
is statistically-significant evidence for evolution in the spin temperatures of DLAs from high 
redshifts to low redshifts, with a larger fraction of low-$\ts$ DLAs in the low redshift sample as 
well as a higher detection fraction at low redshifts.

It should be mentioned that DLAs in the low-$z$ and high-$z$ samples were not gathered via the 
same selection criteria. Most of the DLAs in the low-$z$ sample were initially targetted due to 
the presence of strong \mgtwo\ absorption in the quasar spectrum, with follow-up Lyman-$\alpha$
spectroscopy resulting in the detection of the DLA \citep[e.g.][]{rao98,rao00,rao06,ellison12}. 
Conversely, DLAs in the high-$z$ sample were obtained from direct absorption surveys in the 
Lyman-$\alpha$ absorption line. While the \mgtwo\ rest equivalent width is correlated with the 
absorber metallicity \citep{murphy07b}, there is no evidence that the \mgtwo\ selection criterion 
\citep[at an \mgtwo\ rest equivalent width threshold of $0.5$\AA;][]{rao06} preferentially yields 
DLAs with higher metallicity. It thus appears unlikely that the different selection methods could 
give rise to differing spin temperature distributions in the low-$z$ and high-$z$ samples.

The distributions of \hi\ column densities of the low-$z$ and high-$z$ DLA samples (again 
separated at the median redshift, $z=2.192$) do show a weak difference in a Wilcoxon two-sample test
(at $\approx 2.7\sigma$ significance). Histograms of the \nhi\ distributions of the low-$z$ and
high-$z$ sub-samples are plotted in Fig.~\ref{fig:nhi_hist}[B]. While it should be noted that the
sizes of the two sub-samples are small (with only 18 systems in each sub-sample), there appear to 
be more high-$\nhi$ DLAs in the low-$z$ sub-sample, and more low-$\nhi$ DLAs in the high-$z$ sub-sample.
Since the strength of the \hii\ absorption correlates with the \hi\ column density, this could 
contribute to the larger fraction of detections of \hii\ absorption in the low-$z$ sample. However, 
this would be the case only if the \hii\ studies were carried out at uniform optical depth 
sensitivity. In reality, the \hii\ absorption studies have been targetted at detecting gas at 
high spin temperatures, by aiming for higher optical depth sensitivity in DLAs with lower \hi\ 
column densities. It thus appears unlikely that the difference in \hi\ column density distributions 
in the low-$z$ and high-$z$ DLA samples significantly influences the detection rates of \hii\ 
absorption. Of course, it was shown in Section~\ref{sec:ts2} that the spin temperature does not 
correlate with the \hi\ column density, either in DLAs or in the Galaxy. The difference in \hi\ 
column density distributions should hence also not yield a difference in the spin temperature 
distributions of the high-$z$ and low-$z$ samples.

\begin{figure}
\centering
\epsfig{file=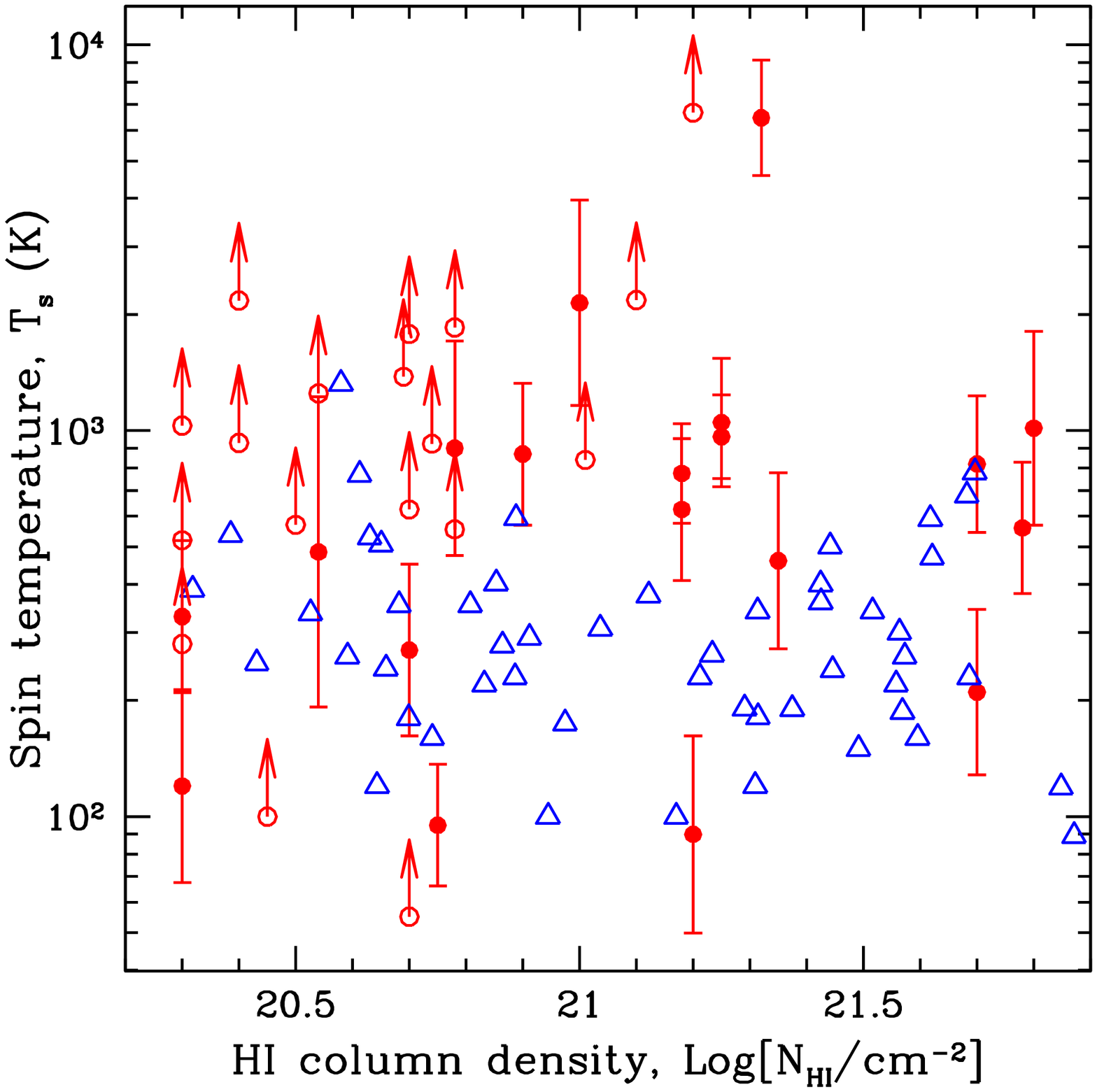,height=3.4truein,width=3.4truein}
\caption{A comparison between the spin temperature distributions in the 37 DLAs of the full 
sample (filled circles) and sightlines in the Milky Way (open triangles), with both 
quantities plotted against \hi\ column density. Note that the Milky Way spin temperatures were 
estimated by comparing \hii\ emission and \hii\ absorption studies on neighbouring sightlines, 
while the DLA spin temperatures were estimated by comparing the \hii\ optical depth with
the \hi\ column density measured from the Lyman-$\alpha$ absorption profile. The figure shows 
that the majority of DLA spin temperature lie at or above the upper edge of the envelope of 
Galactic $\ts$ values \citep[see also;][]{carilli96}. Two-sample tests yield strong evidence that the 
spin temperature distributions in DLAs and in the Galaxy are very different.}
\label{fig:tsvsnhi}
\end{figure}

Fig.~\ref{fig:tsvsnhi} shows the spin temperatures of the 37 DLAs of the sample (filled circles) 
and the $\ts$ values measured along 51 Galactic sightlines towards compact radio 
sources with $\nhi \ge 2 \times 10^{20}$~\cm\ (open triangles) plotted versus \hi\ column density; 
the Galactic data are from \citet{colgan88} and \citet{kanekar11b}. As was originally pointed out by 
\citet{carilli96} (albeit for integrated \hii\ optical depths), the majority of the DLAs lie 
at or above the upper edge of the distribution of the Galactic $\ts$ values. While the \hi\ 
column densities of the two samples are consistent with being drawn from the same distribution,
a Peto-Prentice test finds that the distributions of spin temperatures in DLAs 
and the Galaxy are different at $\approx 6.0\sigma$ significance. We conclude that there is clear 
evidence that conditions in the neutral ISM in DLAs are very distinct from those measured along 
sightlines through the Galaxy, with significantly higher spin temperatures in most DLAs than are 
observed in the Milky Way.

\subsection{Spin temperature or covering factor as the cause for $\ts$ evolution ?}
\label{sec:ts_or_f}

\citet{curran06} have argued that the apparent difference in spin temperature distributions 
between low-$z$ and high-$z$ DLA samples arises mainly due to covering factor effects 
\citep[see also][]{curran12}. These authors argue that high-$z$ DLAs are inefficient at 
covering the quasar radio emission because the angular diameter distances of the absorber 
(DA$_{\rm DLA}$) and the quasar (DA$_{\rm QSO}$) are essentially the same for DLAs at 
$z \gtrsim 1$. Conversely, low-$z$ ($z < 1$) DLAs are typically more efficient at covering 
the radio emission because many low-$z$ DLAs would have DA$_{\rm DLA}$~$<$~DA$_{\rm QSO}$
\citep{curran12}.

As discussed in Section~\ref{sec:f}, our VLBA estimates of the covering factor have shown 
that the median upper limit to the transverse size of the radio core emission in DLAs at 
$z > 2$ is $350$~pc. This is significantly lower than the size of even a dwarf galaxy, making
it likely that all the core radio emission is covered in all cases: covering factor 
estimates should thus not be a serious issue for DLAs of the sample, especially since we have
shown in Section~\ref{sec:f} that the covering factor does not correlate with redshift. However, the 
increased sample size allows a simple test of whether the argument of \cite{curran12}, based on the 
similar angular diameter distances of foreground DLAs and background quasars, is indeed
tenable. For this purpose, we restrict the base sample to the 25 DLAs at $z \geq 1$,
the redshift above which the angular diameter distances to both the foreground DLAs and the 
background quasars are roughly equal. We divide these 25 absorbers into two sub-samples with 
redshifts above and below the median redshift of DLAs with $z \geq 1$ ($z_{\rm med} = 2.683$), 
and compare the spin temperature distributions of the two sub-samples.  Since all 25 DLAs have 
DA$_{\rm DLA} \approx $~DA$_{\rm QSO}$, the issue of angular diameter distances cannot affect 
the results. 

There are 12 DLAs each in the low-$z$ ($1.0 < z < 2.683$) and high-$z$ ($z > 2.683$) sub-samples 
(one absorber is at $z=2.683$). The low-$z$ sample has 9 detections and 3 non-detections 
of absorption, while the high-$z$ sample has 11 non-detections and a single detection.
We used a Peto-Prentice generalized Wilcoxon test to compare the spin temperature distributions
(again appropriately taking into account the limits on $\ts$) and find that the two $\ts$ 
distributions differ at $\approx 3.5\sigma$ significance. There are both far more non-detections 
and higher spin temperatures in the high-$z$ DLA sample.

Thus the spin temperature distributions are clearly different even within the sub-sample 
of absorbers at $z \geq 1$, i.e. with DA$_{\rm DLA} \approx $~DA$_{\rm QSO}$. We conclude that 
angular diameter distances (and hence, DLA covering factors) do not play a significant role
in the low \hii\ optical depths measured in high-$z$ DLAs, and that it is indeed the spin 
temperature that shows redshift evolution.

\subsection{The CNM fraction in DLAs}
\label{sec:cnmfrac}

As noted in Section~\ref{sec:tspin}, for sightlines containing a mixture of neutral gas in 
different temperature phases, the spin temperature derived from equation~(\ref{eqn:tspin}) 
is the column-density-weighted harmonic mean of the spin temperatures of different phases. 
One can hence use the measured spin temperatures in DLAs and assumptions about the spin 
temperatures in different gas phases to estimate the fraction of cold ($\leq 200$~K) 
gas along each sightline.

For the standard two-phase medium models of the ISM in the Milky Way, the kinetic temperatures
of the CNM and WNM are $\approx 40-200$~K and $\approx 5000-8000$~K, respectively 
\citep{field69,wolfire95}. While the actual temperature range depends on local conditions 
(e.g. the metallicity, pressure, the constituents of the ISM, etc), detailed studies have shown 
that typical CNM kinetic temperatures are $\approx 100$~K in both the Milky Way and external 
galaxies like M31 \citep[e.g.][]{braun92,heiles03a}. Further, the high densities in the CNM imply that the 
\hii\ hyperfine transition is thermalized by collisions and the spin temperature in the CNM is 
hence approximately equal to the kinetic temperature \citep[e.g.][]{field59,liszt01}. 

On the other hand, there are few reliable estimates of the gas kinetic temperature in 
the WNM, even in the Milky Way, and there is even evidence that significant amounts of 
gas are in the unstable phase \citep[e.g.][]{heiles03a,kanekar03b}. In DLAs, there have so far 
been two estimates of the WNM kinetic temperature, in the $z = 0.0912$ and $z = 0.2212$ DLAs 
towards 0738+313: \citet{lane00} obtained $\tk \approx 5500$~K in the former system, while 
\citet{kanekar01b} measured $\tk \approx 7600$~K in the latter, both in agreement with the 
temperature ranges predicted by the two-phase models. Note that the spin temperature is 
expected to be lower than the kinetic temperature in the WNM, as the low gas density in WNM
implies that collisions are insufficient to thermalize the \hii\ line here \citep[e.g.][]{liszt01}.

\begin{figure}
\centering
\epsfig{file=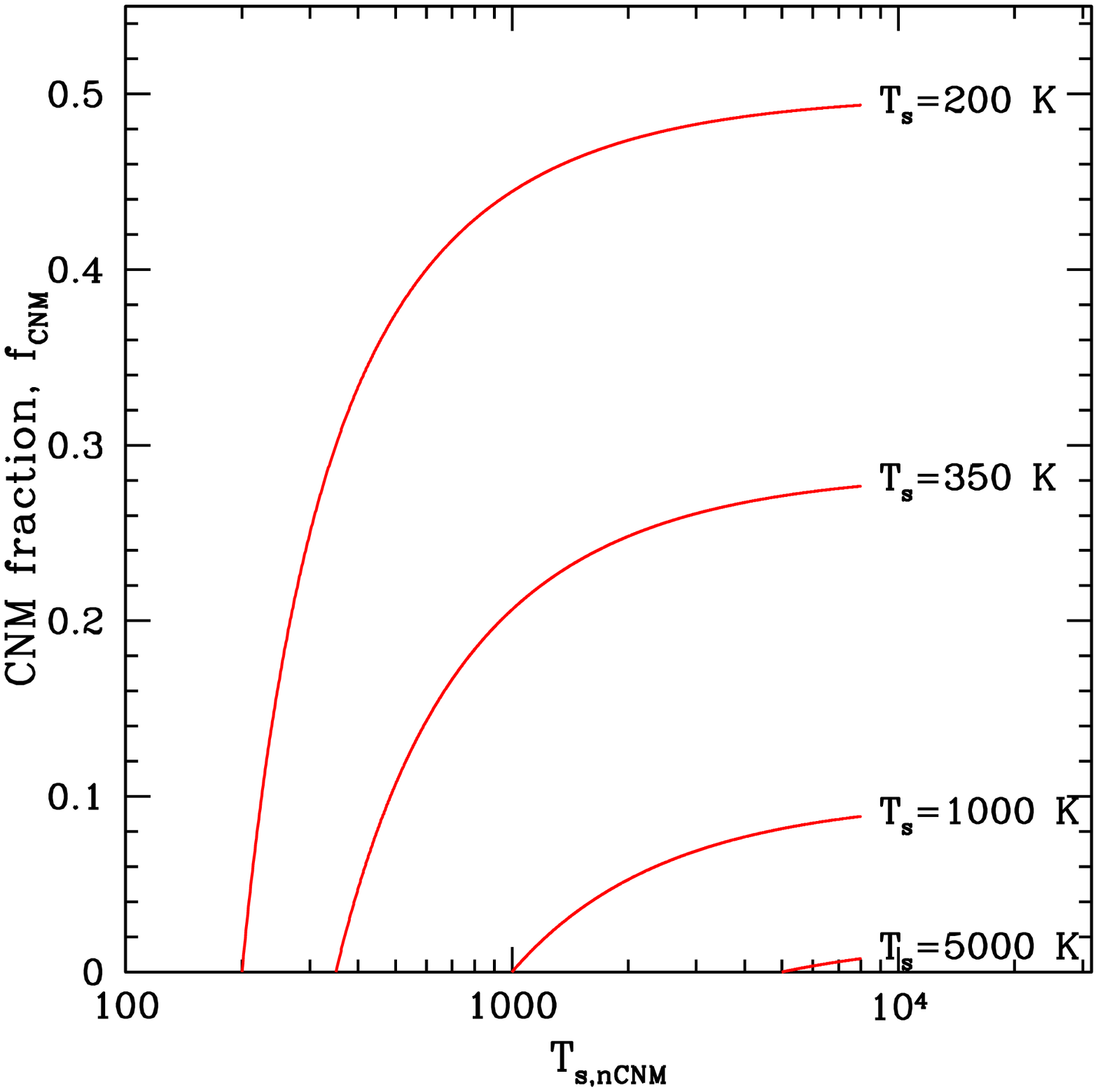,height=3.4truein,width=3.4truein}
\caption{The CNM fraction plotted against the column-density weighted harmonic mean 
spin temperature of gas that is {\it not} in the CNM phase $\tr$, for different DLA 
spin temperatures $\ts$ and an assumed CNM spin temperature of 100~K. Note that the DLA 
spin temperature $\ts$ is the value inferred from equation~\ref{eqn:tspin}, when the possible
multiphase nature of the gas is not taken into account. The {\it highest} CNM fraction 
is obtained when $\tr = 8000$~K (not shown in the figure).}
\label{fig:cnmfrac}
\end{figure}

\citet{kanekar03} and \citet{srianand12} assumed specific values for the CNM and WNM spin 
temperatures in order to infer the CNM fraction in their absorber samples. However, while it 
is reasonable to assume that the CNM spin temperature is equal to its 
kinetic temperature, this assumption is unlikely to be valid for the WNM. Given this, 
we will use a conservative approach to estimating the fraction of cold gas in the DLAs of 
our sample. We assume that each DLA contains some fraction of neutral gas ($f_{\rm CNM}$) in 
the CNM phase and that the remaining gas is not in the CNM phase but at higher kinetic temperatures, 
with a fraction $f_{\rm nCNM} = 1-f_{\rm CNM}$. Note that both fractions include gas at different 
kinetic (and spin) temperatures, in the range $\ts \approx \tk = 40-200$~K for the CNM and at 
higher temperatures for the remaining gas in the non-CNM phase. We also assume that the 
column-density weighted harmonic mean spin temperature in the CNM is ${\rm T_{s, CNM}} 
\approx 100$~K (the results do not change significantly if one assumes a higher CNM spin 
temperature, $\approx 200$~K). This allows us to rewrite equation~(\ref{eqn:harmonic}) for 
the measured DLA spin temperature as 

\begin{equation}
\label{eqn:cnm1}
\frac{1}{\ts} = \frac{f_{\rm CNM}}{100} + \frac{1 - f_{\rm CNM}}{\tr} \:\:,
\end{equation}
where $\ts$ is the DLA spin temperature and $\tr$ is the column-density weighted 
harmonic mean spin temperature for the gas that is {\it not} CNM. One then obtains
\begin{equation}
\label{eqn:cnm2}
f_{\rm CNM} = \lsb \frac{1}{(\tr/100)  - 1}\rsb \lsb { \frac{\tr}{\ts} - 1} \rsb \:\:.
\end{equation}

Fig.~\ref{fig:cnmfrac} shows the CNM fraction $f_{\rm CNM}$ plotted against $\tr$ for 
different values of the DLA spin temperature, $\ts$. It is clear that, for a given DLA spin 
temperature, the {\it highest} value of $f_{\rm CNM}$ will be obtained when $\tr$ is 
large. We can thus obtain an upper limit on the CNM fraction for each DLA by assuming 
$\tr=8000$~K in equation~(\ref{eqn:cnm2}), along with the DLA spin temperature. 
Note that this is equivalent to assuming that all the remaining (non-CNM) gas is in the WNM,
with $\ts = 8000$~K. This is a stringent assumption for two reasons: (1)~8000~K is the 
upper limit of the range of stable WNM kinetic temperatures \citep{wolfire95}, and 
(2)~the spin temperature is expected to be lower than the kinetic temperature in the WNM 
\citep{liszt01}. 


\begin{figure}
\centering
\epsfig{file=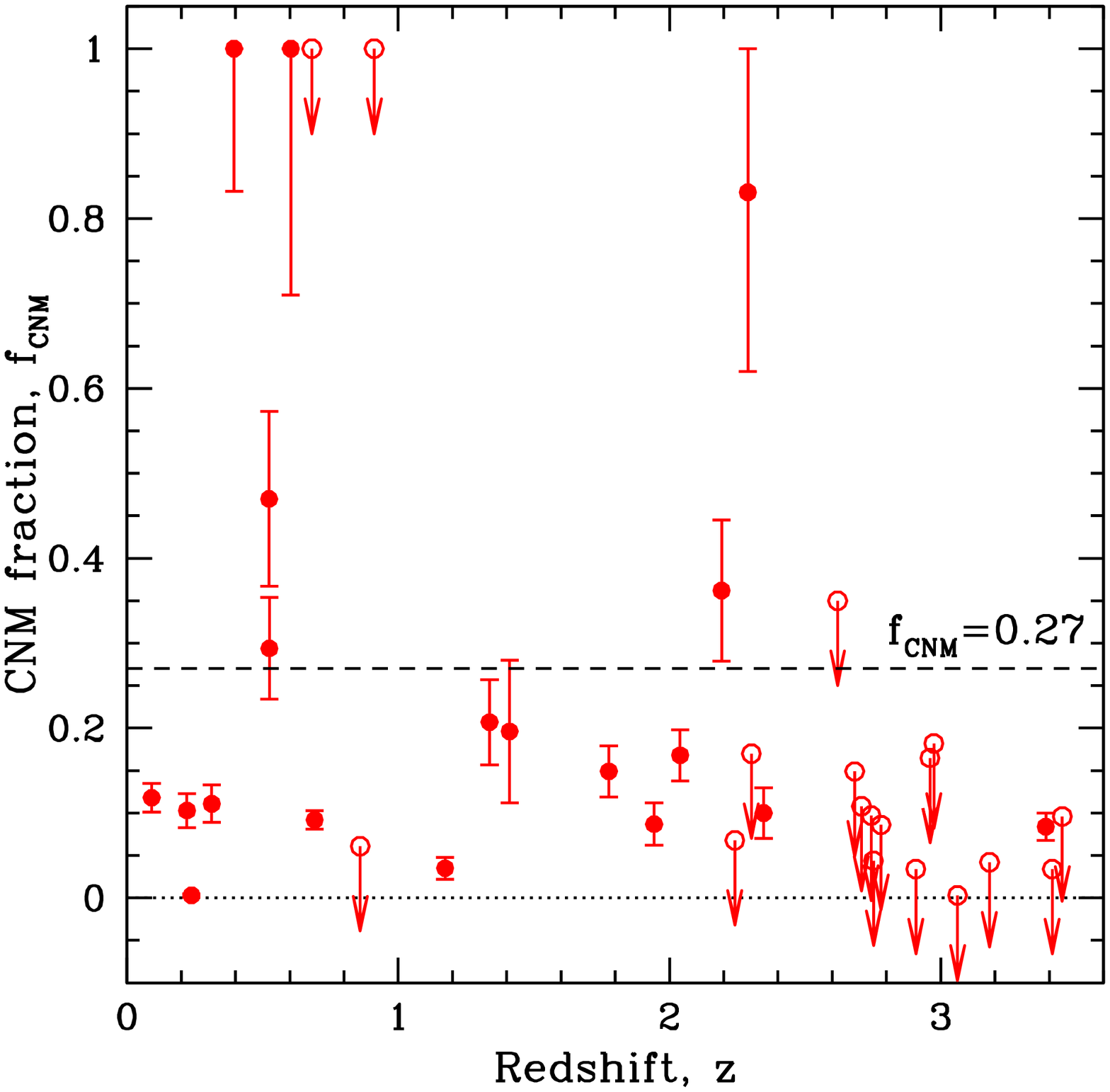,height=3.4truein,width=3.4truein}
\caption{The CNM fraction plotted against redshift for the 37 DLAs of the sample 
(i.e. excluding the two sub-DLAs); the dashed horizontal line shows the median CNM 
fraction in the Galaxy ($f_{\rm CNM} = 0.27$). The column-density weighted harmonic mean 
spin temperatures are assumed to be 100~K in the CNM and 8000~K in the non-CNM phases; this 
ensures that the derived CNM fraction is an upper limit to the ``true'' CNM fraction. 
The majority of DLAs, especially at high redshifts, have low CNM fractions, $\le 20$\%.}
\label{fig:cnm}
\end{figure}

Fig.~\ref{fig:cnm} shows the CNM fraction estimated from the above approach plotted 
versus DLA redshift for the 37 DLAs of the full sample. The downward arrows in the figure 
indicate non-detections of \hii\ absorption, and thus lower limits on the DLA spin 
temperature and upper limits on the CNM fraction. The figure shows that the 
CNM fractions in DLAs are typically quite low, $\lesssim 20$\%, at all redshifts, with 
only a few absorbers having comparable CNM and WNM fractions. At high redshifts, $z > 1.7$, 
only two of the 23 absorbers have CNM fractions greater than 27\%, with one of the upper
limits also in this range. This is significantly lower than the CNM fraction along Galactic 
sightlines, where the median CNM fraction is $\approx 27$\% \citep{heiles03b}. We emphasize further 
that every point in Fig.~\ref{fig:cnm} is an {\it upper} limit to the CNM fraction in a DLA, 
due to the assumption that gas in the non-CNM phase has $\tr = 8000$~K. We conclude that the 
CNM fractions in high-$z$ DLAs are significantly lower than those typical of the Galaxy. 
Similar conclusions have been drawn based on smaller \hii\ absorption samples by \citet{kanekar03},
\citet{srianand12} and \citet{ellison12}, and from H$_2$ absorption studies by \citet{petitjean00}.

We note that three of the four low-$z$ absorbers with $f_{\rm CNM} \gtrsim 0.27 $ 
have been identified with spiral galaxies; these are the DLAs at $z = 0.3950$ towards 
1229$-$021 ($f=1$), $z = 0.5242$ towards 0235+164 ($f = 0.47$), and $z = 0.5247$ towards 
0827+243 \citep[$f=0.294$;][]{lebrun97,burbidge96,chen05}. We suggest that the fourth system, at 
$z = 0.6019$ towards 1429+400 \citep[$f=1$;][]{ellison12}, as well as the two high-CNM 
absorbers at $z \sim 2.192$ towards 2039+187 \citep[$f=0.36$;][]{kanekar13} and $z \sim 2.2890$ 
towards 0311+430 \citep[$f=0.83$;][]{york07} are also likely to be bright disk galaxies. Imaging 
studies of these DLAs would be of much interest.

\subsection{DLA spin temperatures, metallicities and dust depletions}
\label{sec:tszn}

\begin{figure}
\centering
\epsfig{file=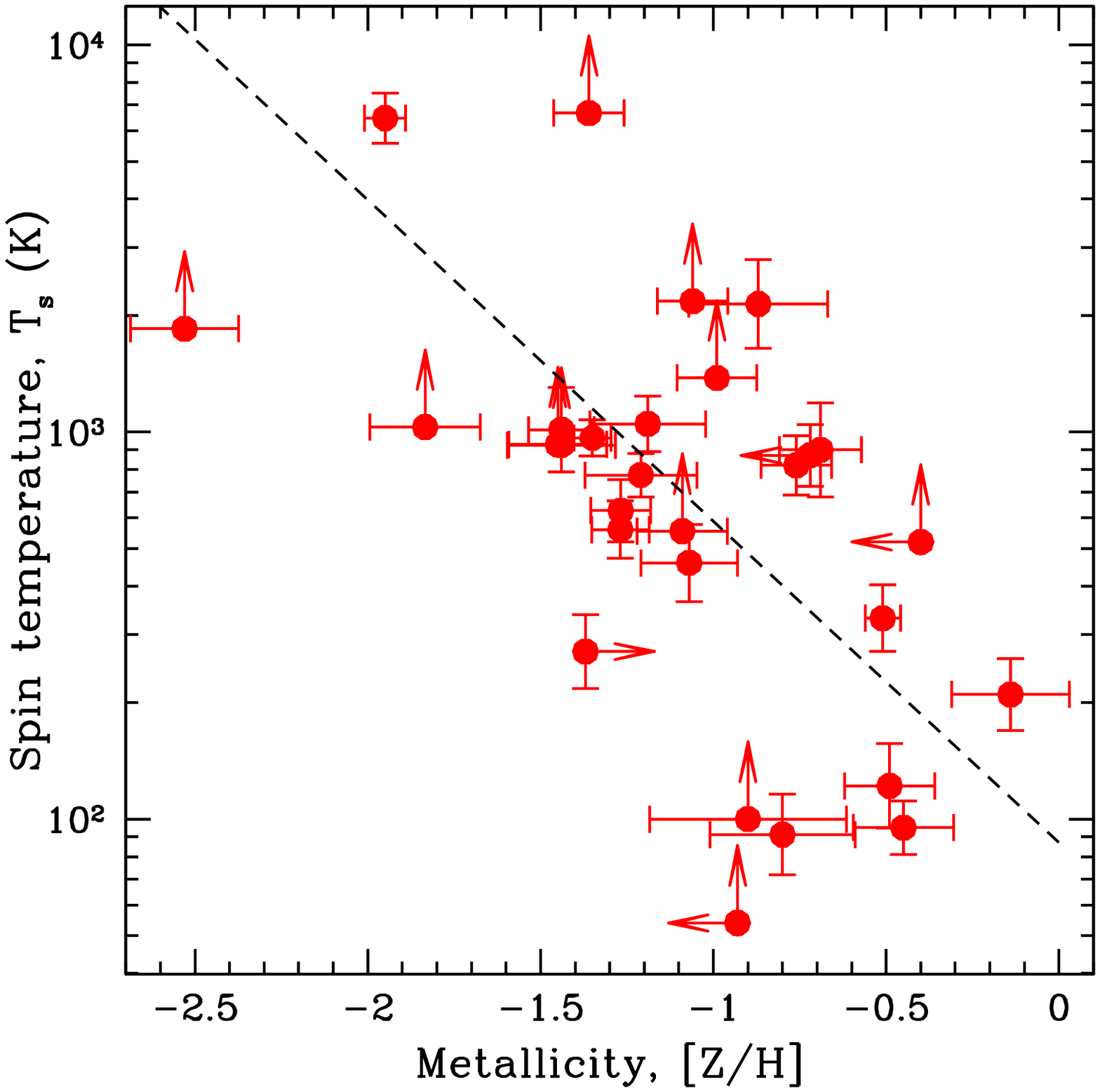,height=3.4truein,width=3.4truein}
\caption{The spin temperature $\ts$ plotted against metallicity [Z/H] for the 29 DLAs of the 
full sample that have measurements of both quantities. The dashed line shows the linear regression 
fit, Log[\ts]~$= (-0.83 \pm 0.16) \times {\rm [Z/H]} + (1.94 \pm 0.20)$, to the 16 measurements 
of both $\ts$ and [Z/H].}
\label{fig:tszn}
\end{figure}

The results of the preceding sections have shown that the majority of DLAs at all redshifts 
have higher spin temperatures and lower CNM fractions than typically seen in the Milky Way 
and M31 \citep[e.g.][]{colgan88,braun92,kanekar11b}, due to lower fractions of the cold phase of 
neutral hydrogen. In the ISM, collisional 
excitation of the fine-structure lines of metals like C{\sc ii} and O{\sc i} dominates the 
cooling below temperatures of $\approx 8000$~K \citep{launay77,wolfire95}. Galaxies with
low metallicities would then be expected to have fewer cooling routes than high-metallicity 
systems like the Milky Way, and hence larger amounts of warm gas. For example, Fig.~6 of 
\citet{wolfire95} shows that, at lower metallicities, a higher central pressure is needed 
to produce the cold phase of \hi. \citet{kanekar01a} used this to argue that the high spin 
temperatures in most high-$z$ DLAs could be naturally explained if the absorbers are typically 
low-metallicity systems with low CNM fractions due to a lack of cooling routes. Conversely, 
DLAs with high metallicities would be expected to have higher CNM fractions and low spin 
temperatures. \citet{kanekar01a} hence predicted an anti-correlation between spin temperature 
and metallicity, if metallicity is indeed the dominant cause in determining the cold gas 
content of a galaxy.

A test of this hypothesis was first carried out by \citet{kanekar09c} \citet[see also][]{kanekar05b}, 
who used $\ts$ and [Z/H] data on 26 DLAs (including 
limits) to detect the predicted anti-correlation at $3.6\sigma$ significance, via 
a non-parametric generalized Kendall test \citep{brown74,isobe86}. The 
anti-correlation was detected at lower significance ($\approx 3\sigma$) in a 
sub-sample of 20~systems with estimates of the covering factor. \citet{ellison12} 
extended this study to 26 DLAs and sub-DLAs with covering factor estimates, using some of the new 
data presented in this paper; this improved the detection significance of the 
anti-correlation to $\approx 3.4\sigma$. Recently, \citet{srianand12} examined the 
possibility of redshift evolution in the relation between $\ts$ and [Z/H], but 
found no evidence in support of this, perhaps due to the small size of their sample.

\begin{figure}
\centering
\epsfig{file=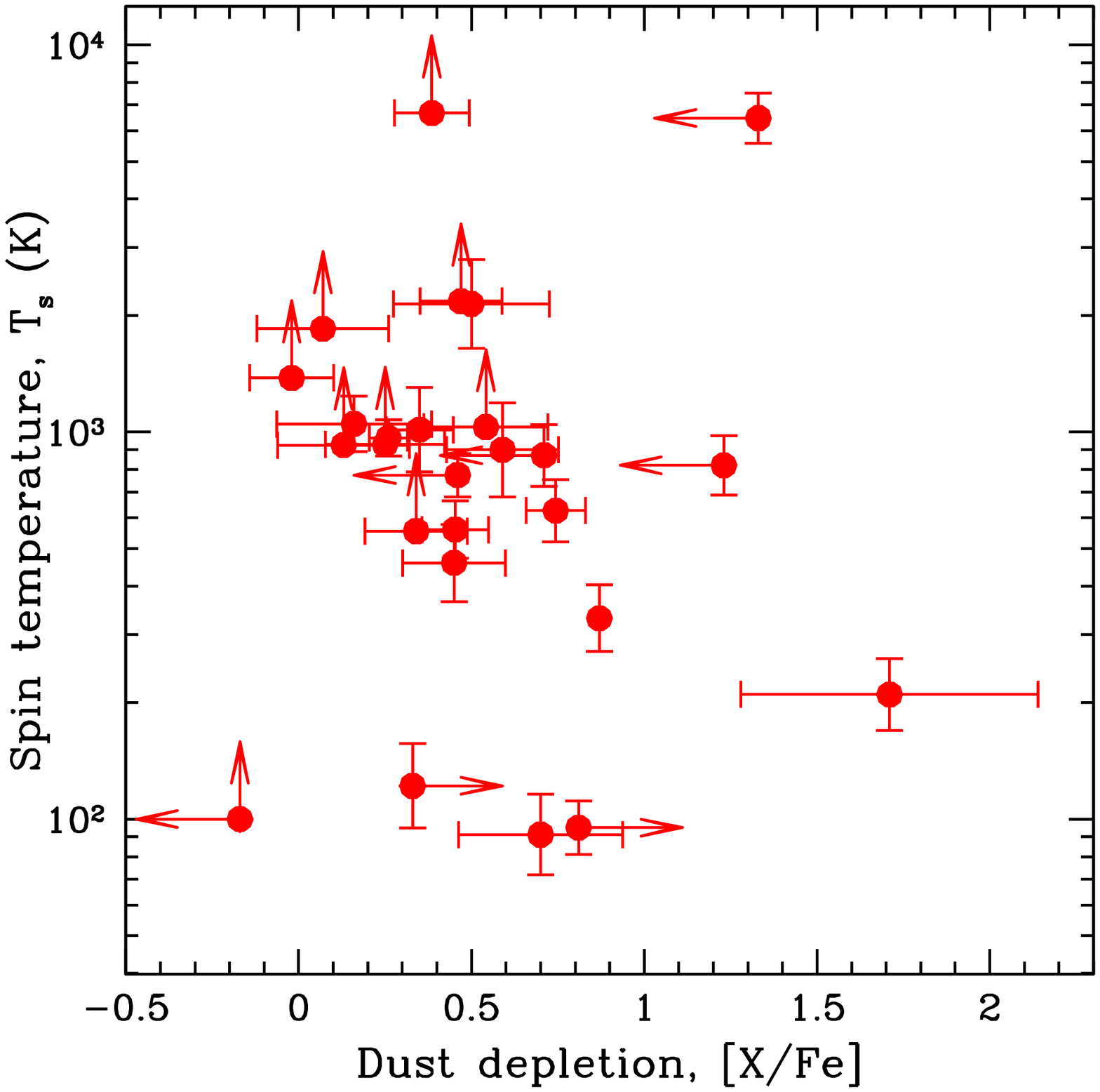,height=3.4truein,width=3.4truein}
\caption{The spin temperature $\ts$ plotted against dust depletion [Z/Fe] for the 26 DLAs 
with low-frequency covering factor estimates that have measurements of both quantities. 
Only weak evidence is found for an anti-correlation between $\ts$ and [Z/Fe].}
\label{fig:tszfe}
\end{figure}

Three of the DLAs in this paper have new VLBA estimates of the covering factor, while 
the detection of \hii\ absorption at $z = 2.192$ towards 2039+187 has recently been 
presented by \citet{kanekar13}. These systems were not included in the analysis of 
\citet{ellison12}. We have also in this paper obtained more accurate estimates of the metallicities
of the $z = 0.3127$ and $z = 0.5247$ DLAs towards 1127$-$145 and 0827+243, respectively.
The sample of DLAs with estimates (including limits) of 
$\ts$ and [Z/H], as well as measurements of the low-frequency covering factor now contains 29
systems. Fig.~\ref{fig:tszn} shows the spin temperature $\ts$ plotted against metallicity [Z/H]
for this sub-sample. We have used the generalized Kendall-tau rank correlation test (the BHK 
statistic; \citealt{brown74}) to detect the anti-correlation between $\ts$ and [Z/H] at 
$\approx 3.5\sigma$ significance, somewhat higher than the result of \citet{ellison12}.
We emphasize that this is a non-parametric test that also takes into account the presence 
of limits in the sample. Further, unlike \citet{ellison12}, we have excluded sub-DLAs from 
the present analysis.

We also tested whether either the high-$z$ or the low-$z$ DLA sub-samples dominate the 
detected anti-correlation between metallicity and spin temperature. The median redshift 
of the 29 DLAs with estimates of both metallicity and spin temperature is $z = 1.7763$,
with 14 DLAs each in the low-$z$ and high-$z$ sub-samples. The anti-correlation between 
$\ts$ and [Z/H] is detected at $2.3\sigma$ and $2.2\sigma$ significance in the low-$z$
and high-$z$ sub-samples, respectively. A similar statistical significance ($2.2\sigma$) 
is obtained for the anti-correlation even on restricting the sample to the 12 DLAs at 
$z < 1$. This indicates that the anti-correlation is not dominated by either the low-$z$ 
or the high-$z$ DLAs. Specifically, the high-$z$ DLAs, where the covering factor is
more uncertain, do not dominate the anti-correlation.

The dashed line in Fig.~\ref{fig:tszn} shows an update of the linear fit of \citet{ellison12} 
to the $\ts$-[Z/H] relation. This uses a regression analysis based on the BCES(Y/X) estimator 
\citep{akritas96}, with [Z/H] as the independent variable, X, applied to the 16 DLAs with 
measurements of {\it both} $\ts$ and [Z/H] \citep{kanekar09c,ellison12}. We obtain Log[\ts]~$= 
(-0.83 \pm 0.16) \times {\rm [Z/H]} + (1.94 \pm 0.20)$, essentially the same as the fit of 
\citet{ellison12}. Note that the sample used for the latter regression analysis included the 
$z = 1.6724$ sub-DLA towards 0237$-$233.

We have also examined the sample for a relation between $\ts$ and dust depletion [Z/Fe]. 26
DLAs with covering factor measurements also have estimates of both quantities (cf. 23 DLAs 
and sub-DLAs in \citealt{ellison12}). Fig.~\ref{fig:tszfe} plots the spin temperature against 
dust depletion for this sub-sample. We find only weak evidence for an anti-correlation between $\ts$ 
and [Z/Fe]; this is detected at $\approx 2.4\sigma$ significance. As argued by \citet{kanekar09c} 
and \citet{ellison12}, it does not appear that the anti-correlation between $\ts$ and 
metallicity is due to an underlying relation between $\ts$ and dust depletion.

We find no evidence that the probability of detecting \hii\ absorption depends 
on the absorber metallicity. For the 25 DLAs with measurements of metallicity (i.e. 
excluding absorbers with only limits on the metallicity), the median metallicity is 
[Z/H]$_{med} = -1.09$. The \hii\ detection rates are $75^{+25}_{-25}$\% and 
$54^{+29}_{-20}$\% for systems above and below the median metallicity, consistent within 
$1\sigma$ confidence intervals.

In passing, we note that the above analysis has used the X-ray metallicity estimate
for the $z = 0.524$ DLA towards QSO~0235+164. As discussed in \citet{ellison12},
excluding this DLA from the sample does not significantly affect the significance
of the anti-correlation between metallicity and spin temperature.

Note that the DLA spin temperatures have been estimated under the assumptions that (1)~the 
\hi\ column density measured along the optical sightline is the same as that towards the 
radio core, and (2)~the quasar core fraction can be interpreted as the DLA covering factor.
Conversely, the DLA metallicities have been estimated from the optical spectra alone, without
any such assumptions. Any breakdown in the assumptions (e.g. due to differences in the \hi\ column densities 
along the optical and radio sightlines) should weaken any underlying relation between $\ts$ and 
[Z/H]. The fact that the predicted anti-correlation between $\ts$ and [Z/H] is yet detected 
in a non-parametric test indicates that such differences between the optical and radio sightlines 
are very unlikely to give rise to the high spin temperatures seen in the majority of high-$z$ DLAs.
We conclude that the assumptions regarding the \hi\ column density and the DLA covering factor 
are likely to be reliable, at least in a statistical sense (unless the observed anti-correlation 
is mere coincidence).

\subsection{On the assumption that the covering factor is equal to the core fraction}
\label{sec:f-core}

The results in Sections~\ref{sec:ts3}--\ref{sec:tszn} are based on the 
assumption that the DLA covering factor is equal to the radio core fraction. While the measured 
radio core sizes are quite small, $< 1$~kpc, far smaller than the size of a typical absorbing galaxy,
it is possible that this assumption is violated for at least some DLAs of the sample. Further,
it is also possible that any detected \hii\ absorption might arise against extended radio structure
and not against the radio core; in such a situation, the covering factor would be uncorrelated
with the core fraction. Unfortunately, it is very difficult to directly measure the covering 
factor of individual DLAs. In this section, we hence use Monte Carlo simulations to test whether our
results critically depend on the assumption that the covering factor is equal to the core fraction.

We first considered the possibility that the covering factor is indeed statistically equal to the 
core fraction, but that there is a random error associated with each covering factor estimate.
We assume that this error is normally distributed with a standard deviation equal to 10\%, 20\%,
50\% and 100\% of the measured core fraction. For each assumed standard deviation, we randomly 
select values of $f$ (constrained to be in the range $0 < f \le 1$), re-compute the spin temperature,
and use the Kendall-tau test to estimate the significance of the anti-correlation between metallicity
and spin temperature, and the Peto-Prentice test to determine the significance of the redshift evolution 
of the spin temperature, as done in Sections~\ref{sec:tszn} and \ref{sec:ts3}, respectively. The process 
was repeated over 10,000 Monte Carlo runs for each assumed standard deviation, and we then computed the 
mean statistical significance of the anti-correlation between $\ts$ and [Z/H] and of the redshift evolution in $\ts$ 
over each set of 10,000 runs. We found that the significance of both results remains high, even if 
the error on the covering factor has a 50\% standard deviation. Specifically, for a standard deviation 
of 50\%, the anti-correlation between $\ts$ and [Z/H] is detected at $\approx 3.2\sigma$ significance, 
while the difference in the $\ts$ distributions in high-$z$ and low-$z$ samples (separated at the median 
redshift, $z_{med} = 2.192$) has $\approx 3.3\sigma$ significance. Indeed, even allowing for a 100\% error 
in the covering factor only reduces the significance of the anti-correlation between $\ts$ and [Z/H] 
to $2.7\sigma$, not a very strong effect, while the difference in $\ts$ distributions has $\approx 3.2\sigma$ 
significance. We conclude that even a large random error in the inferred covering factor does not 
significantly affect our results.

We also considered the worst-case scenario in which the covering factor is entirely uncorrelated 
with the measured radio core fraction. Since the covering factor must take values between 0 and 1, 
we randomly selected values for $f$ in this range (assuming a uniform distribution between 0 and 1), 
independently for each DLA of the sample, and then re-computed the spin temperature for each DLA. 
We then repeated the Monte Carlo simulations for the anti-correlation between spin temperature and 
metallicity and the redshift evolution of the spin temperature (following the approach described in the previous paragraph), 
with the mean statistical significance of the two relations again computed from 10,000 Monte Carlo runs. The 
anti-correlation between $\ts$ and [Z/H] is then detected at $\approx 2.9\sigma$ significance, 
while the difference between the spin temperature distributions in the high-$z$ and low-$z$ DLA samples 
(again separated at the median redshift, $z_{med} = 2.192$) has $\approx 3.1\sigma$ significance,
averaging over 10,000 Monte Carlo runs. Again, while the significances of both results are somewhat lower than
the values obtained on using the assumption that the covering factor is equal to the core fraction 
($\approx 3.5\sigma$ for the $\ts$-[Z/H] relation and $\approx 3.5\sigma$ for the redshift evolution 
in $\ts$; see Sections~\ref{sec:ts3} and \ref{sec:tszn}), it is clear that there is no significant change 
in the results even on assuming an entirely random distribution of DLA covering factors. 

We also used the above Monte Carlo simulations to examine the effect of systematic biases in the 
covering factor. We find that if all DLAs have low (or high) covering factors, our two main results 
(the anti-correlation between $\ts$ and [Z/H] and the redshift evolution of $\ts$) remain unchanged.
The only way to reduce the statistical significance of the $\ts$-[Z/H] anti-correlation is to 
assign very low covering factors ($f < 0.1$) to DLAs with high inferred spin temperatures, so as to 
effectively reduce the derived $\ts$. This is essentially similar to the argument of \citet{curran06},
that low DLA covering factors due to geometric effects can account for the inferred high DLA 
spin temperatures. However, as discussed in Section~\ref{sec:ts_or_f}, there is statistically 
significant evidence for redshift evolution in the spin temperature even when we only consider 
the sub-sample of DLAs at $z > 1$, for which geometric effects cannot yield different DLA 
covering factors. As such, the result that the spin temperature of DLAs shows evidence of redshift 
evolution remains unaffected by the possibility that high-z DLAs are less effective at covering 
the background quasars.

Further, if low covering factors of high-$z$ DLAs are the cause of the anti-correlation between 
metallicity and spin temperature, we would expect that the anti-correlation would be detected 
at higher significance in the high-$z$ DLA sample. As discussed in Section~\ref{sec:tszn}, this 
is not the case, with the low-$z$ and high-$z$ sub-samples making roughly equal contributions
to the significance of the anti-correlation. 

We conclude that our results do not significantly depend on the assumption that the DLA covering 
factors are equal to the quasar radio core fractions. We also find no evidence that our results 
might be affected by systematic biases in the covering factor.

\subsection{DLA spin temperatures and velocity widths}

\begin{figure}
\centering
\epsfig{file=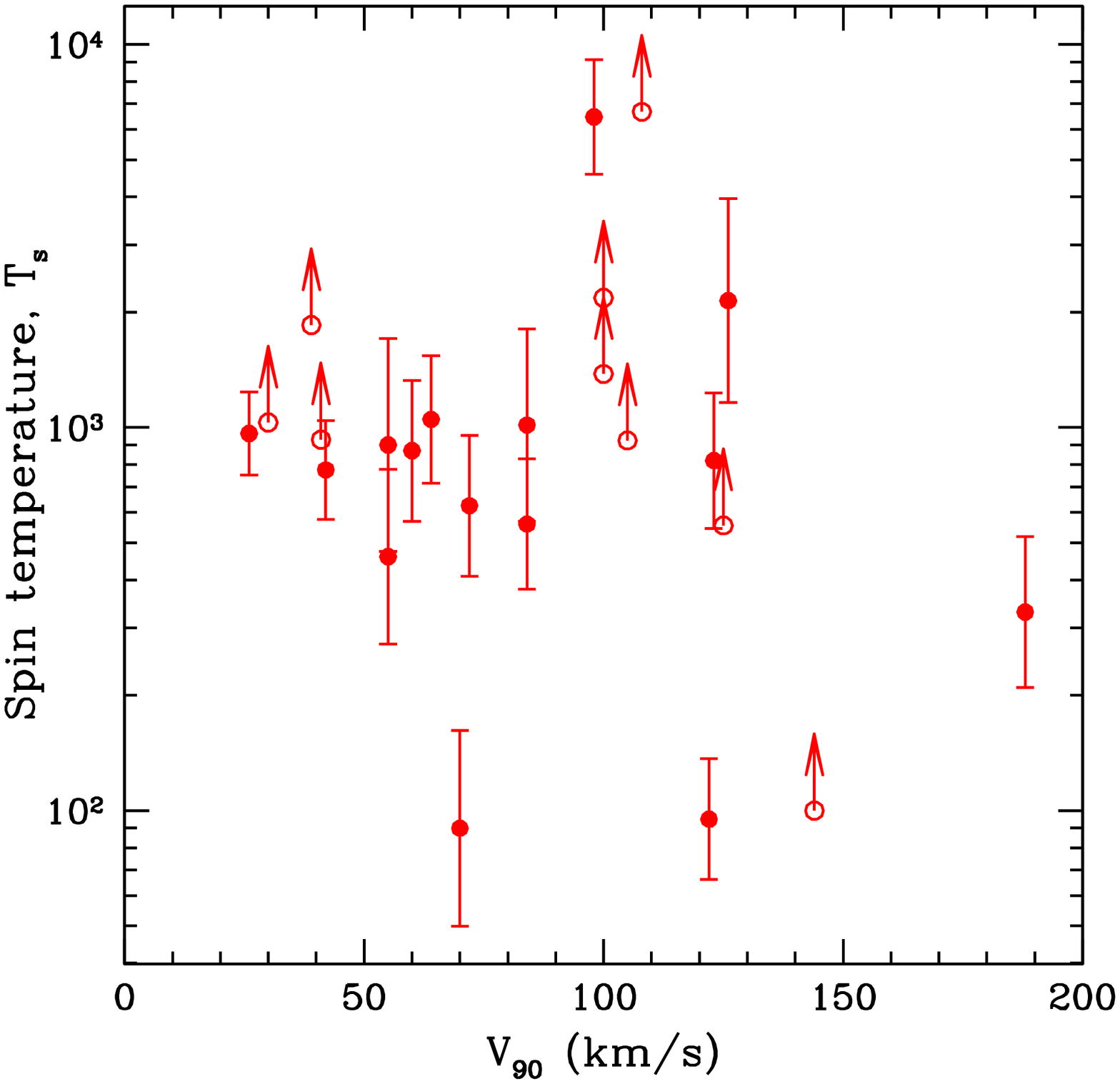,height=3.4truein,width=3.4truein}
\caption{The spin temperature $\ts$ plotted against velocity width at 90\% optical depth in 
the low-ionization metal lines, $\Delta V_{\rm 90}$, for the 25 DLAs (filled circles) and 
sub-DLAs (open star) with estimates of both quantities.  The figure shows no evidence for a 
relation between $\ts$ and $\Delta V_{\rm 90}$.}
\label{fig:tsv90}
\end{figure}

DLA metallicities have been shown to be correlated with the velocity spreads
of low-ionization metal lines \citep{wolfe98,ledoux06,prochaska08}, as well as with
the rest equivalent widths of the \mgtwo\ and \sitwo\ lines \citep{murphy07b,prochaska08}. 
We hence examined the possibility that the anti-correlation between spin temperature 
and metallicity might arise because the spin temperature depends on the velocity spread 
of the absorber. 

Our sample contains $\Delta V_{\rm 90}$ measurements for 24 DLAs, with a median 
$\Delta V_{\rm 90}$ value of 84~\kms. The detection rates of \hii\ absorption are 
$45^{+31}_{-20}$\% and $73^{+27}_{-25}$\% for systems with $\Delta V_{\rm 90}$ 
values above and below the median, respectively. We thus find no evidence that 
the detection rate of \hii\ absorption depends on the velocity spread of the low-ionization 
metal lines.

Fig.~\ref{fig:tsv90} plots the spin temperature against the 90\% optical depth velocity spread 
$\Delta V_{\rm 90}$ for the 24 DLAs with estimates of both quantities.
The figure resembles a scatter plot, with no obvious relation between the two quantities.
Similarly, the generalized Kendall-tau test (taking limits into account) finds no evidence 
of a correlation between $\ts$ and $\Delta V_{\rm 90}$ ($0.2 \sigma$ statistical significance).
We conclude that it is very unlikely that the observed correlation between $\ts$ 
and metallicity arises due to an underlying relation between the spin temperature and 
the velocity spread of the low ionization metal lines. 

The correlation between metallicity and velocity spread in DLAs has usually been interpreted 
in terms of a mass-metallicity relation \citep{ledoux06}, similar to those observed in the local 
Universe or 
in bright high-$z$ galaxies \citep[e.g.][]{tremonti04,erb06,prochaska08,neeleman13}. Since high-mass 
galaxies have higher central pressures, they are expected to have higher CNM fractions 
\citep{wolfire95}. It might then appear curious that no relation is seen between 
$\ts$ and $\Delta V_{\rm 90}$. While the sample size of systems with $\Delta V_{\rm 90}$ estimates 
is relatively small (24 systems), the anti-correlation between spin temperature and metallicity is 
detected at $3.6\sigma$ significance with a similar number of systems (29). We surmise that 
this is because the metallicity directly determines the number of cooling routes and hence 
the cold gas fraction. Conversely, it is clear that there are other contributors besides mass to 
the observed $\Delta V_{\rm 90}$ in absorption. 
For example, high mass galaxies observed at low inclinations to the line of sight can have low 
velocity spreads in absorption; DLA surveys may have some bias towards such systems, 
due to their larger surface area on the sky. There are also contributions to the velocity 
spread from merging galaxies, outflows from galactic winds, galactic infall, etc, 
which can increase $\Delta V_{\rm 90}$ for low-mass galaxies above the values expected due to 
rotation. For galaxies in the local Universe, where the mass can be estimated directly, 
\citet{zwaan08} find only a weak relation between $\Delta V_{\rm 90}$ and mass. Finally, 
the relation between [Z/H] and $\Delta V_{\rm 90}$ itself has a large spread, $\approx 1.5$~dex 
in metallicity at a given $\Delta V_{\rm 90}$ (see Fig.6[A] of \citealt{prochaska08}). It is 
hence perhaps not very surprising that the observed spin temperature does not depend strongly 
on the velocity spread of the absorbing galaxy.

\subsection{Comparisons with results from C{\sc ii}* absorption studies}
\label{sec:ciistar}

\setcounter{table}{5}
\begin{table*}
\caption{DLAs with \hii\ and \ctwostar\ or ${\rm H_2}$ absorption studies.}
\begin{center}
\begin{tabular}{|c|c|c|c|c|c|c|c|}
\hline
QSO             & $z_{\rm QSO}$ & $z_{\rm abs}$ & $\log[\nhi/{\rm cm}^{-2}]$     & $\log[l_c]/{\rm erg s^{-1} H^{-1}}$       & [Z/H]		    &  $\ts$         & log[$f_{\rm H2}$] \\
	        &               &               &                  &                   &                    &    K           &                   \\
\hline
\multicolumn{8}{|c|}{\bf DLAs with \hii\ and \ctwostar\ absorption studies} \\
\hline
0201+113     & 3.639         & 3.3869        & $21.25 \pm 0.07$ & $-26.66 \pm 0.14$ & $-1.19 \pm 0.17$   & $1050 \pm 175$ & $-4.6 \: {\rm to} \: -6.2$    \\
0336$-$017   & 3.197         & 3.0621        & $21.20 \pm 0.10$ & $-26.71 \pm 0.10$ & $-1.36 \pm 0.10$   & $> 8890$	  & $< -6.90$        \\
0458$-$020   & 2.286         & 2.0395        & $21.78 \pm 0.07$ & $> -26.37$        & $-1.269 \pm 0.072$ & $560 \pm 95$   & $< -6.40$        \\
1157+014     & 2.000         & 1.9436        & $21.80 \pm 0.07$ & $> -26.51$        & $-1.440 \pm 0.086$ & $1015 \pm 255$ & $< -6.65$        \\
1331+170     & 2.084         & 1.7764        & $21.18 \pm 0.07$ & $< -27.12$        & $-1.268 \pm 0.076$ & $625 \pm 115$  & $-1.25 \pm 0.05$ \\
1354$-$170   & 3.147         & 2.7799        & $21.30 \pm 0.15$ & $-27.06 \pm 0.16$ & $-1.83 \pm 0.16$   & $> 1030$       & $-$              \\
2342+342     & 3.053         & 2.9084        & $21.10 \pm 0.10$ & $-26.92 \pm 0.12$ & $-1.06 \pm 0.10$   & $> 2200$       & $< -6.19$        \\
\hline
\multicolumn{8}{|c|}{\bf DLAs with \hii\ and ${\rm H_2}$, but no \ctwostar, absorption studies} \\
\hline
0335$-$122   & 3.442         & 3.1799        & $20.78 \pm 0.11$ & $-$               & $-2.53 \pm 0.16$   & $> 1850$      & $< -5.10$ \\
0432$-$440   & 2.649         & 2.3023        & $20.78 \pm 0.11$ & $-$               & $-1.09 \pm 0.13$   & $> 555$       & $< -5.15$ \\
1418$-$064   & 3.689         & 3.4482        & $20.40 \pm 0.10$ & $-$               & $-1.45 \pm 0.14$   & $> 930$       & $< -5.69$ \\
\hline
\end{tabular}
\label{table:ctwo}
\end{center}
\begin{flushleft}
References: The H$_2$ fractions are from \citet{cui05}, \citet{noterdaeme08} and \citet{srianand12}, while the log[$l_c$] values are 
from \citet{wolfe08}. References for the metallicities, \hi\ column densities and spin temperatures are given in 
Table~\ref{table:main}. Note that PDLAs (e.g. the $z = 2.8111$ system towards 0528$-$250) have not been included 
in the table.
\end{flushleft}
\end{table*}

Over the last decade, it has been suggested that the strength of \ctwostar\ absorption 
provides a means of distinguishing between DLAs containing significant CNM fractions and 
ones whose neutral phase is dominated by the WNM \citep{wolfe03a,wolfe03b,wolfe04}. This is because 
the most important coolant of the neutral ISM is the [C{\sc ii}]-158$\mu$m line 
\citep[e.g.][]{pottasch79,wright91,wolfire95}. Since this arises from transitions between 
the $^2$P$_{3/2}$ and $^2$P$_{1/2}$ levels in the ground state of ionized carbon, it 
should be possible to determine the cooling rate per H atom by measuring the abundance in 
the [C{\sc ii}]~$^2$P$_{3/2}$ state from the strength of the UV \ctwostar\ line 
\citep{pottasch79}. Assuming thermal balance then allows one to infer the heating rate 
from the cooling rate: for the CNM, where the cooling is dominated by the [C{\sc ii}]-158$\mu$m 
line, the heating rate is approximately equal to the cooling rate, while, for the WNM,
the heating rate is significantly larger than the derived cooling rate. Since the heating rate 
is dominated by photoelectric emission of electrons from grains by incident far-UV radiation 
\citep{bakes94}, \citet{wolfe03a} argued that one could use this heating rate 
in conjunction with measurements of the dust-to-gas ratio in DLAs and an assumed grain photoelectric 
heating efficiency to infer the SFR per unit area in high-$z$ DLAs. This could further be used 
to probe the distribution of neutral gas between the CNM and WNM phases, by assuming two-phase 
models in pressure equilibrium \citep{field69,wolfire95,wolfire03}. 

\citet{wolfe03a} considered 
two models for the neutral ISM, a ``pure-WNM'' model with all the \hi\ in the WNM and a 
``CNM+WNM'' model in which comparable amounts of \hi\ are in each phase. They find that 
``pure-WNM'' models are ruled out for DLAs with strong \ctwostar\ absorption, as the bolometric 
luminosities of such DLAs would be far higher than observed values \citep[see also][]{wolfe03b,wolfe04}.
The ``pure-WNM'' model was found to be applicable to absorbers with low \ctwostar\ column 
densities (i.e. either non-detections of \ctwostar\ absorption or weak \ctwostar\ absorbers).
About half the DLAs in the sample of \citet{wolfe08} lie in this category. Wolfe et al. argue 
that the observed strong \ctwostar\ absorption in the remaining DLAs must arise in the CNM
and that the ``CNM+WNM'' model is applicable here. Finally, \citet{wolfe08} argue that 
the C{\sc ii}\,158$\mu$m cooling rate per hydrogen atom $l_c$ may be a separator of DLAs into 
two categories, ``high-cool'' systems with high cooling rates $l_c > 10^{-27}$~erg~s$^{-1}$~H$^{-1}$ (with 
high metallicities, large velocity spreads and significant CNM fractions) and ``low-cool'' 
systems with low cooling rates $l_c < 10^{-27}$~erg~s$^{-1}$~H$^{-1}$ (with low metallicities, low velocity
spreads and with the \hi\ mostly in the WNM phase).

With regard to \hii\ studies, \citet{wolfe03b} noted that there are two high-$z$ DLAs 
with non-detections of \hii\ absorption and high inferred lower limits on the spin temperatures, 
but which show strong \ctwostar\ absorption: these are the systems at $z = 3.0621$ towards
0336$-$017 and $z = 3.387$ towards 0201+113. \citet{wolfe03b} suggested that both systems
are likely to have high CNM fractions ($\approx 50$\%) along the sightline to the optical QSO; 
they argued that the high spin temperature limits obtained from the \hii\ absorption 
studies might arise because the relatively small ($10-20$~pc) CNM clouds do not entirely cover 
even the radio quasar core, which is significantly more extended than the optical quasar in 
both cases. Later, \citet{kanekar07} detected \hii\ absorption towards 0201+113 (see 
also \citealt{briggs97}) and also found that the radio quasar core here has a transverse size 
of $\le 286$~pc. This implies that it is possible that the CNM clouds do not fully cover the 
radio core, as suggested by \citet{wolfe03b}. However, \citet{kanekar07} also showed that the 
CNM fraction towards the radio quasar core is $\lesssim 17$\%, even if one assumes that 
the sightline towards the optical core has a far higher CNM fraction, $\approx 50$\%. Thus, 
even if the arguments of \citet{wolfe03b} are correct and the $10-20$~pc-sized CNM clouds 
do not efficiently cover the radio core, the sightline towards the radio core is still 
dominated by the WNM, with $> 80$\% of the \hi\ in this phase.

For clarity, the upper part of Table~\ref{table:ctwo} summarizes the \ctwostar\ and \hii\ results 
for the subset of DLAs with studies in both transitions and VLBI estimates of the covering factor. 
The columns in this table are: (1)~the quasar name, (2)~the quasar redshift $z_{\rm QSO}$, (3)~the 
DLA redshift, (4)~the \hi\ column density,
(5)~the C{\sc ii}\,158$\mu$m cooling rate per atom, log[$l_c$], inferred from the \ctwostar\ line strength
\citep{wolfe08}, (6)~the metallicity, [Z/H], (7)~the spin temperature $\ts$, (8)~the CNM fraction, assuming 
${\rm T_{s,CNM}} = 100$~K and ${\rm T_{s,WNM}} = 8000$~K, and (9)~the H$_2$ fraction \citep{ledoux03,noterdaeme08}.
Based on the classification of \citet{wolfe08}, the absorbers towards 1354$-$170 and 1331+170 are 
``low-cool'' systems ($\log[l_c/{\rm erg s^{-1} H^{-1}}] \lesssim -27$), while those towards 
0201+113, 0336$-$017, 0458$-$020, 1157+014 and 2342+342 are ``high-cool'' systems 
($\log[l_c/{\rm erg s^{-1} H^{-1}}] > -27$. Note that the systems towards 
1354$-$170 and 2342+342 are consistent with both categories, within $1\sigma$ errors, but are
definitely not strong \ctwostar\ absorbers.

At the outset, it is relevant to note that the \ctwostar\ and \hii\ absorption results appear
in reasonable agreement for about half the DLA sample, the systems with weak or undetected \ctwostar\ 
absorption (i.e. the ``low-cool'' systems). \citet{wolfe08} find that these systems have systematically 
lower metallicities (and dust-to-gas ratios and velocity widths) than the ``high-cool'' systems 
that have strong \ctwostar\ absorption, and conclude that the neutral gas here is predominantly 
WNM. The anti-correlation between metallicity and spin temperature would lead us to conclude that 
such DLAs have high spin temperatures and are thus dominated by the WNM; there is thus no disagreement 
between the \ctwostar\ and \hii\ results for systems with weak \ctwostar\ absorption and high spin 
temperatures. These include the DLAs at $z = 2.7799$ towards 1354$-$170 and $z = 2.9084$ towards 
2342+342, for which both the relatively weak \ctwostar\ absorption and the high spin temperatures 
are consistent with most of the \hi\ being in the WNM.


The primary difference between the results on the CNM fractions obtained from the \ctwostar\ and \hii\ 
absorption studies are for the ``high-cool'' sample of \citet{wolfe08}. However, the apparent 
discrepancy appears mainly due to the fact that \citet{wolfe03b} (and later works) only consider 
two models, ``CNM+WNM'', with half the \hi\ in the CNM phase and the other half in the WNM, and 
``pure-WNM'', with all the \hi\ in the WNM, and rule out the latter model, based on the fact 
that the \ctwostar\ absorption must arise from the CNM. But even if one accepts that all the 
\ctwostar\ absorption arises from the CNM \citep[although see][]{srianand05}, this does not imply 
that half the \hi\ must be CNM. There is no obvious problem with a model in which $10-20$\% of the 
\hi\ in DLAs with strong \ctwostar\ absorption is CNM (as suggested by Fig.~\ref{fig:cnm}), and 
with all the \ctwostar\ absorption produced by the CNM. There are three DLAs at $z > 1.7$ that 
show both \hii\ and strong \ctwostar\ absorption; these are the absorbers at $z \sim 1.9436$ 
towards 1157+014, $z \sim 2.0395$ towards 0458$-$020 and $z \sim 3.3869$ towards 0201+113. 
The three absorbers have inferred CNM fractions of $\approx 9-17$\% and intermediate metallicities, 
[Z/H]~$\approx -1.3$) (see Section~\ref{sec:cnmfrac} and Tables~\ref{table:main} and 
~\ref{table:ctwo}); these are not inconsistent with the observed strong \ctwostar\ absorption. 
In fact, given that these DLAs all have high column densities, $\nhi \ge 1.8 \times 10^{21}$~\cm,
it is possible that the \hi\ column density towards the radio core is lower than that towards 
the optical quasar by a factor of $\approx 1.5$ in these systems (see Section~\ref{sec:lmc}). 
The spin temperature estimates in these DLAs could hence be lower, and the inferred CNM fractions 
higher, by $\approx 50$\%. We emphasize that \citet{wolfe03b,wolfe04} only rule out the ``pure-WNM'' 
model for such strong \ctwostar\ absorbers, but do not rule out models with CNM fractions of 
$\approx 10-20$\% and WNM fractions of $\approx 80-90$\%. 

The sole absorber in Table~\ref{table:ctwo} for which there remains a clear discrepancy between 
the \ctwostar\ and $\ts$ data is the $z = 3.0621$ DLA towards 0336$-$017. \citet{wolfe08} classify this 
as a ``high-cool'' DLA, since $\log[l_c/{\rm erg s^{-1} H^{-1}}] = -26.71 \pm 0.10$, while we 
obtain $\ts > 8890 \times (f/0.68)$. 
One way of accounting for this difference is a fortuitously high fraction of CNM clouds towards the 
optical quasar, which do not efficiently cover the more-extended radio core \citep{wolfe03b}. Following
the argument of \citet{kanekar07} (in the context of the $z = 3.387$ DLA towards 0201+113),
 this would still imply a high WNM fraction towards the radio core of 0336$-$017, since the core 
is small enough \citep[$\lesssim 224$~pc;][]{kanekar09a} to be entirely covered by the extended 
WNM clouds. Another possibility is that the average \hi\ column density towards the quasar radio 
core is lower in the $z = 3.0621$ DLA towards 0336$-$017 than that towards the optical QSO. 
While the upper limit to the transverse core size is only $\approx 224$~pc in the $z = 3.0621$ DLA 
(i.e. lower than the median for DLAs at $z > 2$), it is possible that there is small-scale structure 
in the \hi\ distribution, so that the average \hi\ column towards the optical core is a few times 
larger than that towards the radio core. This is certainly possible along individual sightlines.
Finally, it is also possible that some of the \ctwostar\ absorption along this sightline arises 
in either the WNM or ionized gas \citep{srianand05}.

Finally, the $z = 1.7764$ DLA towards 1331+170 is a ``low-cool'' DLA, with 
$\log[l_c/{\rm erg s^{-1} H^{-1}}] < -27.12$ 
\citep{wolfe08,jorgenson10}. This has $\ts = (625 \pm 115) \times (f/0.72)$~K and a CNM fraction 
of $\lesssim 15$\%, derived from a new \hii\ absorption spectrum \citep{carswell11} and the VLBA 
covering factor of $f = 0.72$ \citep{kanekar09a}. There is no significant contradiction between the 
$\ts$ and \ctwostar\ results in this absorber, since both suggest high WNM fractions. However, it is 
difficult to reconcile the weak \ctwostar absorption with the fact that H$_2$ and C{\sc i} absorption 
have both been detected in this DLA \citep{meyer86,cui05,jorgenson10}, clearly indicating the presence 
of CNM in the absorber (as is also shown by the detected \hii\ absorption). Indeed, the H$_2$ fraction in 
the $z = 1.7764$ DLA is one of the highest in the entire DLA sample \citep{cui05,noterdaeme08}. Despite 
this, \ctwostar\ absorption is not detected (or is, at best, quite weak) in this absorber 
\citep{jorgenson10}. Similarly, the \ctwostar\ absorption is weak in the $z = 2.431$ DLA towards 
2343+125 ($\log[l_c/{\rm erg s^{-1} H^{-1}}] = -27.09 \pm 0.10$), despite the presence of 
H$_2$ absorption in this system and a high metallicity, [Z/H]~$=-0.54 \pm 0.01$ \citep{petitjean06}. 
It thus appears that, contrary to the discussion in \citet{wolfe04}, the absence of \ctwostar\ absorption 
does not imply the absence of CNM in individual DLAs, while, conversely, the presence of strong 
\ctwostar\ absorption merely indicates the presence of CNM, and not necessarily a high CNM fraction.

In summary, we find that there is no significant discrepancy between the \ctwostar\ results of 
\citet{wolfe03a,wolfe04} and our results from \hii\ absorption studies. The anti-correlation between
metallicity and spin temperature is consistent with the finding of \citet{wolfe08} that DLAs with 
high metallicities tend to be ``high-cool'' systems, while those with low metallicities have low 
cooling rates. The apparent discrepancy between the results is because \citet{wolfe03a} (and later 
papers) only considered two models, one with half the gas each in the CNM and WNM and the other with 
all the gas in the WNM. Both the \ctwostar\ and \hii\ absorption results appear to be consistent with 
a model in which (1)~high-metallicity ([Z/H]~$\gtrsim -0.5$) DLAs have high CNM fractions, $\approx 50$\%, 
(2)~intermediate-metallicity DLAs (with [Z/H]~$\approx -1$) have $10-20$\% of the neutral gas in the 
CNM, with this phase producing the \ctwostar\ and \hii\ absorption, and (3)~low-metallicity 
DLAs ([Z/H]~$\lesssim -1.5$) have most of their gas in the WNM, with low cooling rates and high 
spin temperatures.

\subsection{Comparing CNM and molecular fractions}

It is also interesting to compare the inferred CNM fractions $f_{\rm CNM}$ in DLAs 
(from Section~\ref{sec:cnmfrac} to the molecular hydrogen fraction $f_{\rm H_2} \equiv 
2 {\rm N(H_2)}/(2{\rm N(H_2)} + \nhi)$, for absorbers with estimates of both quantities. Since 
both trace the presence of cold gas, a relation between $f_{\rm CNM}$ and $f_{\rm H_2}$
might provide information on the process of molecule formation in DLAs. Searches for molecular 
hydrogen have now been carried out in nearly a hundred DLAs, mostly with high-resolution optical 
spectroscopy \citep[e.g.][]{levshakov85,petitjean00,ledoux03,cui05,petitjean06,noterdaeme08}.
However, detections of ${\rm H_2}$ absorption have been obtained in only $\approx 16$\% of 
DLAs at $z > 2$, with most absorbers showing low molecular fractions, 
$f_{\rm H_2} \lesssim 10^{-5}$ \citep[e.g.][]{ledoux03,noterdaeme08}. Detections of ${\rm H_2}$ 
absorption have been obtained in DLAs with relatively high metallicity and dust depletion, 
[X/Fe]~$> 0.4$ \citep{noterdaeme08}. This is consistent with the anti-correlation obtained 
between spin temperature and metallicity, as high-metallicity absorbers should have higher
CNM fractions and are thus more likely to form the molecular phase.

Unfortunately, there are at present only 9 DLAs with estimates of both $f_{\rm CNM}$ and 
$f_{\rm H_2}$. Most of these have also been observed in the \ctwostar\ line and are listed 
in the upper part of Table~\ref{table:ctwo}. Three additional systems with searches for 
\hii\ and ${\rm H_2}$, but not \ctwostar, absorption are listed in the lower part of the 
table. The two ${\rm H_2}$ absorbers of this sub-sample, at $z = 1.7764$ towards 1331+170 
and $z = 3.3869$ towards 0201+113, both show \hii\ absorption. However, ${\rm H_2}$ absorption
was not detected in two \hii\ absorbers, at $z = 2.0395$ towards 0458$-$020 and 
$z = 2.3476$ towards 0438$-$436. Five DLAs were detected in neither \hii\ nor ${\rm H_2}$ 
absorption, with strong upper limits on the molecular fraction (${\rm log}[f_{\rm H_2}] < -5$) 
and lower limits on the spin temperature ($\ts > 500$~K). The small sample size makes it
difficult to draw any inferences from these results, especially given that all DLAs of the
sample have high spin temperatures, $\ts \gtrsim 500$~K. Searches for ${\rm H_2}$ absorption
in DLAs with low spin temperatures would be of much interest.

\section{Summary}
\label{sec:summary}

We have carried out a deep search for redshifted \hii\ absorption in a large 
sample of DLAs at $0.68 < z < 3.44$ with the GMRT, the GBT and the WSRT. 
This has yielded detections of \hii\ absorption in two new DLAs, at $z = 3.387$ 
towards 0201+113 and $z = 2.347$ towards 0438$-$436, and confirmations of detections 
in two known \hii\ absorbers. We present evidence that the \hii\ absorption from
one of the DLAs, at $z = 2.0395$ towards 0458$-$020, has varied on timescales
of $\approx 2$~months. Similar variability in the \hii\ absorption profile has been 
earlier reported on even shorter timescales in two DLAs. 

We have also used the VLBA to obtain high spatial resolution images of a sample of 
background quasars with foreground DLAs or \hii\ absorbers 
at frequencies close to the redshifted \hii\ line frequency, in order to estimate 
the quasar core fraction. We assume that the DLA covering factor is the same as this 
core fraction. We have used our measurements of the \hii\ optical depth and the DLA 
covering factor in conjunction with the known \hi\ column density from the Lyman-$\alpha$ 
profile to estimate the spin temperature of each absorber.
Finally, we report metallicity, abundance and velocity width measurements for a set 
of DLAs with \hii\ absorption spectroscopy, from either our own observations 
or archival data.

We have combined our results with data from the literature to compile a sample 
of 37 DLAs with spin temperatures derived from a combination of the Lyman-$\alpha$ 
profile, \hii\ absorption spectroscopy and low-frequency VLBI estimates of the absorber 
covering factor. This sample does not include PDLAs, where the spin temperature could 
be influenced by the proximity of the absorber to the background quasar, or sub-DLAs, where
the spin temperature may be affected by self-shielding issues The \hi\ column density 
distribution of this sample is statistically indistinguishable from that of the much 
larger sample of all known DLAs, including absorbers detected in the SDSS.  29 DLAs of the 
sample have estimates of the gas metallicity [Z/H], 26 of the dust depletion [Z/Fe], 
and 24 of the velocity width between 90\% optical depth points, $\Delta V_{\rm 90}$. 

The main results of the paper are:

\begin{itemize}

\item{We find that the spatial extent of the quasar radio core at the foreground absorber redshift 
(as estimated from the VLBI images) is larger for high-$z$ DLAs than for low-$z$ DLAs.
However, these are {\it upper limits} to the transverse core size, due to the possibility of 
residual phase errors in the VLBA data. This is especially important for lower-frequency 
VLBI images, due to ionospheric effects; DLAs at $z > 2$ are thus the worst affected
of the sample. Despite this, the median upper limit to the transverse core size 
is $\approx 130$~pc for the full sample, while that for $z > 2$ DLAs is $\approx 350$~pc.
For all DLAs but one, the maximum value for the (upper limit to the) transverse core size is 
$\approx 1$~kpc. These are significantly lower than the size of a typical galaxy. We conclude 
that the foreground DLAs are likely to cover at least the radio cores for all the 37 DLAs 
of the sample.}

\item{We have examined whether the use of the \hi\ column density derived from the 
pencil-beam optical sightline in the equation for the \hii\ optical depth for 
the sightline towards the radio emission can cause systematic effects in the 
spin temperature estimates. This was done by smoothing a high-spatial-resolution 
(15~pc) \hii\ emission cube of the LMC to coarser resolutions, out to $\approx 1$~kpc,
to test whether the \hi\ column densities measured from the smoothed spectral cubes 
are systematically different from those at the highest spatial resolution. We find 
that the \hi\ column densities in the smoothed cube are systematically larger
than those in the 15~pc cube for $\nhi \lesssim 1 \times 10^{21}$~\cm, and systematically 
lower than those in the 15~pc cube for $\nhi \gtrsim 1 \times 10^{21}$~\cm. 
In other words, the use of the \hi\ column density from the damped Lyman-$\alpha$ profile 
in the equation for the \hii\ optical depth will tend to under-estimate the spin 
temperature for DLAs with $\nhi \lesssim 1 \times 10^{21}$~\cm, and to over-estimate
the spin temperature for DLAs with $\nhi \gtrsim 1 \times 10^{21}$~\cm\ (assuming that
the spatial distribution of neutral gas in DLAs is similar to that in the LMC). We estimated 
the magnitude of this effect at two spatial resolutions, 67~pc and 345~pc, approximately 
the median transverse size of the radio cores at the DLA redshifts for DLAs at $z < 2$ 
and $z > 2$, respectively. We find that the effect is not very significant, within a factor 
of 25\% for most of the DLAs of the sample, except for systems at low and very high 
$\nhi$ values ($\nhi < 4 \times 10^{20}$~\cm\ and $\nhi > 5.5 \times 10^{21}$~\cm, 
respectively) and with large core sizes. At these extremes of the $\nhi$ distribution, 
the spin temperature could be under-estimated by up to a factor of 2 for $\nhi \approx 2 \times 
10^{20}$~\cm, or over-estimated by up to a factor of 1.5 for $\nhi \approx 6 \times 10^{21}$~\cm.
Since there are far more DLAs at low $\nhi$ values than at extremely high ones, the 
overall effect is to give a weak bias of the spin temperature distribution of high-$z$ DLAs
towards lower values.}

\item{We have examined the full sample of DLAs for trends between different quantities, including 
the detection rate of \hii\ absorption, the spin temperature, the \hi\ column density, 
the covering factor, the metallicity, the dust depletion, the velocity width at 90\% 
optical depth, etc. Survival analysis methods, as implemented in the ASURV package, were used 
to appropriately handle censored data. 

We found no evidence that the detection rate of \hii\ absorption depends on 
the \hi\ column density, the absorber covering factor, the DLA metallicity 
or the velocity width at 90\% optical depth. For all these variables, the 
detection rates of \hii\ absorption above and below the median value were found 
to be consistent with each other within the statistical uncertainties. Similarly,
there is no statistically-significant evidence that the estimated DLA spin temperatures
depend on the absorber covering factor, the \hi\ column density or the velocity width 
at 90\% optical depth. However, we caution that the sample size of DLAs with 
\hii\ absorption studies is still quite small and the above detection rates are
dominated by Poisson errors. It is plausible that larger absorber samples will indeed show 
that the detection rates do depend on, e.g., the metallicity and the covering factor.

The detection rate of \hii\ absorption in high-$z$ DLAs (with $z > z_{med}$, where 
$z_{med} = 2.192$) is $17^{+16}_{-9}$\%, while that in low-$z$ DLAs (with $z < z_{med}$)
is $83^{+17}_{-21}$\%. The difference between the two detection rates has $\approx 
2.6 \sigma$ significance. This is tentative evidence that the probability of detecting 
\hii\ absorption increases with decreasing redshift.}

\item{We detect a difference between the spin temperature distributions of absorbers above 
and below the median redshift, $z_{med} = 2.192$, at $\approx 3.5\sigma$ significance.
The statistical significance of the difference in $\ts$ distributions is even higher,
$4.0\sigma$, if the redshift $z = 2.4$ is used to demarcate the low-$z$ and high-$z$
samples. 

We have also restricted the low-$z$ sample to absorbers with redshifts $z > 1$, 
to ensure that the angular diameter distances of the DLAs and their background quasars 
are similar for all absorbers. Again sub-dividing the sample in low-$z$ ($z < 2.683$) 
and high-$z$ ($z > 2.683$) DLAs yields a clear difference, with $3.5\sigma$ statistical
significance, between the $\ts$ distributions of the two sub-samples. Angular diameter
distances thus do not appear to play a significant role in the lower \hii\ optical depths
observed in high-$z$ DLAs. We conclude that DLA spin temperatures show clear redshift 
evolution: the high-$z$ sample contains both a smaller fraction of DLAs with low spin 
temperatures and a lower detection rate of \hii\ absorption. }

\item{We also find that the spin temperature distributions in the full sample of DLAs and in 
the Milky Way are different, at $\approx 6\sigma$ significance. This is definitive evidence 
that the neutral ISM in DLAs is significantly different from that in the Galaxy, with far 
smaller CNM fractions ($\lesssim 20$\%) in DLAs than in the Milky Way (median CNM fraction 
$\approx 27$\%).}

\item{We continue to find that the DLA metallicities and their spin temperatures are anti-correlated:
the anti-correlation between [Z/H] and $\ts$ is detected at $\approx 3.5\sigma$ significance in 
a generalized Kendall-tau test, based on 29 absorbers with estimates of metallicity, spin temperature 
and covering factor. This supports the conjecture of \citet{kanekar01a} that the low CNM fractions 
in DLAs arise due to a paucity of cooling routes in the absorbers, due to their low metallicities.  
Differences between the \hi\ column densities along the radio and optical sightlines can only
worsen any existing anti-correlation. We find only weak ($\approx 2.4\sigma$) evidence for an 
anti-correlation between $\ts$ and dust depletion, from 26~systems. It is thus unlikely that the 
relation between spin temperature and metallicity arises due to an underlying relation between 
spin temperature and dust depletion.

The DLA spin temperature estimates are based on the assumptions that the quasar core fraction can
be interpreted as the DLA covering factor and that the \hi\ column density measured towards the 
optical quasar is the same as that towards the more-extended radio emission. The metallicity estimates 
do not make either of these assumptions. If the above assumptions are violated for most of the absorbers, 
we would not expect to detect an underlying relation between $\ts$ and [Z/H]. The fact that the 
predicted anti-correlation between $\ts$ and [Z/H] is indeed detected indicates that the assumptions 
are reliable, at least in a statistical sense. }

\item{We have also examined whether the two main results of this paper, viz. the redshift evolution of 
DLA spin temperatures and the anti-correlation between spin temperature and metallicity, are affected
by the assumption that the DLA covering factor is equal to the quasar radio core fraction. This 
was done via Monte Carlo simulations with DLA spin temperatures inferred assuming (1)~an error in the 
inferred covering factor, with standard deviation ranging from $10-100$\% of the measured core fraction, 
and (2)~entirely random covering factors, selected from a uniform distribution of values between $0$ and $1$. 
For both types of errors, the above two results were obtained at $\gtrsim 3\sigma$ significance, 
after averaging over 10,000 Monte Carlo runs. Our results thus do not appear to critically depend on 
the assumption that the DLA covering factor is equal to the quasar radio core fraction.}

\item{We have examined the \ctwostar\ model of \citet{wolfe03a,wolfe04} in the light of 
our spin temperature estimates. We find that the \ctwostar\ results are entirely consistent with 
our results for at least half the DLA sample, the systems with low metallicities and undetected or 
weak \ctwostar\ absorption. The anti-correlation between metallicity and spin temperature implies
 that the neutral ISM in such systems should be predominantly WNM, as expected from the low 
cooling rates derived from the \ctwostar\ model. Similarly, DLAs with high metallicity are 
expected to have high CNM fractions based on the $\ts$-[Z/H] anti-correlation and are also expected 
to have a high cooling rate from their strong \ctwostar\ absorption; the spin temperature and 
\ctwostar\ results are thus also in agreement for such systems. Thus, the \ctwostar\ results 
are broadly consistent with the spin temperature results except for a few cases where 
line-of-sight effects may be important. Further, even for these cases, the apparent discrepancy 
between the spin temperature and \ctwostar\ results arises for systems with intermediate metallicity, 
[Z/H]~$\approx -1.3$. We find that the difference essentially arises because 
\citet{wolfe03b,wolfe04} use a CNM+WNM model with 50\% of the neutral gas in each phase; such 
a model is not required by the \ctwostar\ data. It appears possible to alleviate the apparent 
discrepancy between the \ctwostar\ and the \hii\ results by considering a model wherein $10-20$\% 
of the \hi\ in intermediate-metallicity DLAs is in the CNM, producing the majority of the 
the \ctwostar\ absorption. Such a two-phase model would yield both strong \ctwostar\ 
absorption and a high spin temperature ($\ts \approx 1000$~K). While it is beyond the 
scope of the present work to explicitly test this possibility, we conclude that there 
is no significant disagreement between the \ctwostar\ results and the spin temperature 
estimates. }

\end{itemize}

\section*{Acknowledgements}

We thank the staff of the GMRT, the GBT, the VLBA, the WSRT, the VLT, the Keck 
Observatory, the Space Telescope Science Institute and the Gemini Observatory. 
The GMRT is run by the National Centre for Radio Astrophysics of the Tata Institute 
of Fundamental Research. The WSRT is operated by ASTRON (the Netherlands Institute 
for Radio Astronomy), with support from the Netherlands Foundation for Scientific 
Research (NWO). The NRAO is a facility of the National Science Foundation operated 
under cooperative agreement by Associated Universities, Inc.. Based on observations 
obtained at the Gemini Observatory, which is operated by the Association of 
Universities for Research in Astronomy, Inc., under a cooperative agreement with 
the NSF on behalf of the Gemini partnership: the National Science Foundation (United 
States), the Science and Technology Facilities Council (United Kingdom), the National 
Research Council (Canada), CONICYT (Chile), the Australian Research Council 
(Australia), Minist\'{e}rio da Ci\^{e}ncia, Tecnologia e Inova\c{c}\~{a}o (Brazil) 
and Ministerio de Ciencia, Tecnolog\'{i}a e Innovaci\'{o}n Productiva (Argentina).
Based on observations made with the NASA/ESA Hubble Space Telescope, obtained 
from the data archive at the Space Telescope Science Institute. STScI is operated 
by the Association of Universities for Research in Astronomy, Inc. under NASA contract 
NAS~5-26555. Based on observations made with ESO Telescopes at the Paranal Observatory 
under programme IDs 67.A-0567, 68.A-0170 and 69.A-0371.
Some of the data presented herein were obtained at the W.M. Keck Observatory, 
which is operated as a scientific partnership among the California Institute 
of Technology, the University of California and the National Aeronautics and 
Space Administration. The Observatory was made possible by the generous financial 
support of the W.M. Keck Foundation. NK acknowledges support from the Department 
of Science and Technology through a Ramanujan Fellowship. Part of this work was 
carried out during a visit by NK to ESO-Santiago; he thanks ESO for hospitality.
JXP is partly supported by NSF grant AST-1109447. Basic research in radio astronomy 
at the Naval Research Laboratory is supported by 6.1 base funds. ERW acknowledges the 
support of Australian Research Council grant DP1095600. We thank Art Wolfe and an 
anonymous referee for detailed comments on earlier versions of the manuscript that 
have significantly improved its clarity.

\bibliographystyle{mn2e}
\bibliography{ms}

\label{lastpage}

\end{document}